\documentclass{tac}
\usepackage{amssymb}
\usepackage{tikz}
\usetikzlibrary{calc}

\title{A Categorical Reduction System for Linear Logic}
\author{Ryu Hasegawa}
\address{Graduate School
 of Mathematical Sciences, The University of Tokyo, Komaba 3-8-1,
 Meguro-ku, Tokyo 153-8914, Japan}

\keywords{type theory, linear logic, rewriting system}
\amsclass{03B40, 
 68N18
}

\makeatletter

\def\mor#1{\mathrel{\mathop{\longrightarrow}\limits^{\vbox
 to0pt{\vss \hbox{$\scriptstyle #1$}\kern-2pt}}}}
\def\leftmor#1{\mathrel{\mathop{\longleftarrow}\limits^{\vbox
 to0pt{\vss \hbox{$\scriptstyle #1$}\kern-2pt}}}}
\def\linearmor#1{\mathrel{\mathop{\multimap}\limits^{\vbox
 to0pt{\vss \hbox{$\scriptstyle #1$}\kern-4pt}}}}
\def\twomor#1{\mathrel{\mathop{\Rightarrow}\limits^{\vbox
 to0pt{\vss \hbox{$\scriptstyle #1$}\kern-1pt}}}}

\def\fitarr#1{\vcenter{\offinterlineskip
 \halign{\hfil ##\hfil &$##$\hfil \crcr
 \vbox to0pt{\vss \hbox{\kern2pt$\scriptstyle #1$\kern3pt}}\crcr
 \noalign{\vskip1pt}
 \leaders\hrule height.7pt depth-.3pt\hfill $\mkern2mu$
  &\vbox to0pt{\vss \llap{$\rightarrow$}\vss} \crcr}}}

\def\lmor#1{\mathrel{\fitarr{#1}}}

\newdimen\latticeUnit
\newdimen\boxMargin \newdimen\layerMargin

\newbox\tbox \newbox\ttbox
\chardef\tc=8 \chardef\tcc=9
\newdimen\tdimen \newdimen\ttdimen
\newdimen\sdimen \newdimen\ssdimen
\newtoks\ttoks \newtoks\tttoks
\newcount\tcount \newcount\ttcount
\newcount\tttcount \newcount\ttttcount

\newbox\diagbox
\newdimen\rightlim \newdimen\leftlim
\newdimen\upperlim \newdimen\lowerlim
\newdimen\dimA \newdimen\dimB
\newdimen\dimC \newdimen\dimD
\newdimen\dimE \newdimen\dimF \newdimen\dimG

\def\clist{\\}

\def\rightappend#1\to#2{\ttoks={#1\\}\tttoks=\expandafter{#2}%
 \edef#2{\the\tttoks\the\ttoks}}

{\catcode`p=12 \catcode`t=12 \gdef\gazonc#1pt{#1}}
\let\getfactor=\gazonc
\def\dimensionToNumber#1{\expandafter\getfactor\the#1}
\let\ptless=\dimensionToNumber

\def\nodebox#1{%
 \futurelet\com\nnodeboxx#1\with}

\def\nnodeboxx{%
 \ifx\com\invisible
  \let\next=\invisibleNodeBox
 \else
  \let\next=\visibleNodeBox
 \fi\next}

\def\visibleNodeBox#1\with#2#3{%
 \def\bm{#3}
 \setbox#2=\hbox{\kern\bm\vbox{\kern\bm\hbox{$\displaystyle{#1}$%
  }\kern\bm}\kern\bm}}

\def\invisible{}

\def\invisibleNodeBox#1\with#2#3{%
 \def\bm{#3}
 \setbox#2=\hbox{\kern\bm\vbox{\kern\bm\hbox{$\displaystyle{#1}$%
  }\kern\bm}\kern\bm}
 \setbox#2=\hbox{\vrule width0pt height\ht#2 depth\dp#2%
  \vbox to0pt{\hrule height0pt depth0pt width\wd#2}}}

\gdef\object(#1,#2)=#3{%
 \nodebox{#3}\tbox\boxMargin
 \tdimen=.5\wd\tbox \ttdimen=.5\ht\tbox \advance\ttdimen by.5\dp\tbox
 \edef\bbbox{(#1,#2)(\ptless{\tdimen}pt,\ptless{\ttdimen}pt)}%
 \expandafter\rightappend\bbbox\to\clist
 \advance\tdimen by-#1\latticeUnit \advance\ttdimen by-#2\latticeUnit
 \tdimen=-\tdimen \ttdimen=-\ttdimen
 \sdimen=\tdimen
  \ifdim\sdimen<\leftlim \global\leftlim=\sdimen \fi
  \advance\sdimen by\wd\tbox
  \ifdim\sdimen>\rightlim \global\rightlim=\sdimen \fi
 \sdimen=\ttdimen
  \ifdim\sdimen<\lowerlim \global\lowerlim=\sdimen \fi
  \advance\sdimen by\ht\tbox
  \ifdim\sdimen>\upperlim \global\upperlim=\sdimen \fi
 \put(\ptless{\tdimen},\ptless{\ttdimen}){\unhbox\tbox}
}

\def\refer(#1,#2)\to(#3,#4){\def\a{#1}\def\b{#2}%
 \tcount=#3 \ttcount=#4
 \expandafter\rreferr\clist\empty}
\def\rreferr\\#1{\ifx#1\empty \let\next=\relax
 \else \let\next=\rrreferrr \fi \next}
\def\rrreferrr#1,#2)(#3,#4){\def\aa{#1}\def\bb{#2}%
 \ifdim\a pt=\aa pt \ifdim\b pt=\bb pt
  \dimen\tcount=#3 \dimen\ttcount=#4
 \fi\fi \rreferr}

\gdef\edge{\let\lcommand=\line
 \morphismSwitch}

\gdef\morphism{\let\lcommand=\vector
 \morphismSwitch}

\def\morphismSwitch(#1,#2)to(#3,#4){%
 \def\mNext{\ifx\farg[\morphismBody(#1,#2)(#3,#4)%
  \else \slantMorphism(#1,#2)(#3,#4)\fi}
 \futurelet\farg\mNext}

\def\morphismBody(#1,#2)(#3,#4){%
 \slantMorphism(#1,#2)(#3,#4)%
 \expandafter\attachSwitch}

\def\abs#1{\ifdim#1<0pt-#1\else#1\fi}

\def\attachSwitch#1#2]{
 \nodebox{{\scriptstyle #2}}\tbox\boxMargin
 \rightsidefalse
 \diagonalfalse \horizontalfalse \verticalfalse
 \def\tNext{\ifx\targ[\expandafter\attachOption
  \else \attachBody\fi}
 \futurelet\targ\tNext}

\newif\ifrightside
\newif\ifhorizontal
\newif\ifvertical
\newif\ifdiagonal

\def\attachOption#1{%
 \attachOptionLoop}
\def\attachOptionLoop#1{%
 \ifx#1]
  \let\next\attachBody
 \else
  \ifx#1R \rightsidetrue \fi
  \ifx#1H \horizontaltrue \fi
  \ifx#1V \verticaltrue \fi
  \ifx#1D \diagonaltrue \fi
  \let\next\attachOptionLoop
 \fi
 \next
}

\newtoks\labelPosition
\labelPosition={.5}

\def\attachBody{%
 \dimC=\xC \advance\dimC by-\xB
 \dimD=\yC \advance\dimD by-\yB
 \dimA=\xB \advance\dimA by\the\labelPosition\dimC
 \dimB=\yB \advance\dimB by\the\labelPosition\dimD
 \ifhorizontal \ifdim0pt<\abs\dimD \attachBodyHorizontal \fi
 \else \ifvertical \ifdim0pt<\abs\dimC \attachBodyVertical \fi
 \else \ifdiagonal \attachBodyDiagonal
 \else
  \tdimen.2\dimD \tdimen=\abs\tdimen
  \ifdim\abs\dimC<\tdimen
   \attachBodyHorizontal
  \else \tdimen.2\dimC \tdimen=\abs\tdimen
  \ifdim\abs\dimD<\tdimen
   \attachBodyVertical
  \else
   \attachBodyDiagonal
  \fi\fi
 \fi\fi\fi}

\def\attachBodyHorizontal{%
 \dimG=.5\ht\tbox \advance\dimG by.5\dp\tbox
 \ifrightside \dimE=\dimD \else \dimE=-\dimD \fi
 \dimF=\dimC
 \ifdim\dimD<0.0pt
  \ifrightside \advance\dimA by-.5\wd\tbox
  \else \advance\dimA by.5\wd\tbox \fi
 \else
  \ifrightside \advance\dimA by.5\wd\tbox
  \else \advance\dimA by-.5\wd\tbox \fi
 \fi
 \displaceLabel(\ptless{\dimA},\ptless{\dimB})(\ptless{\dimE},\ptless{\dimF},\ptless{\dimG},1)
}

\def\attachBodyVertical{%
 \dimG=.5\wd\tbox
 \ifrightside \dimE=\dimC \else \dimE=-\dimC \fi
 \dimF=\dimD
 \ifdim\dimC<0.0pt
  \ifrightside \advance\dimB by.5\ht\tbox \advance\dimB by.5\dp\tbox
  \else \advance\dimB by-.5\ht\tbox \advance\dimB by-.5\dp\tbox \fi
 \else
  \ifrightside \advance\dimB by-.5\ht\tbox \advance\dimB by-.5\dp\tbox
  \else \advance\dimB by.5\ht\tbox \advance\dimB by.5\dp\tbox \fi
 \fi
 \displaceLabel(\ptless{\dimA},\ptless{\dimB})(\ptless{\dimE},\ptless{\dimF},\ptless{\dimG},-1)
}

\def\attachBodyDiagonal{%
 \ifdim\dimD<0.0pt
  \ifrightside
   \advance\dimA by-.5\wd\tbox \advance\dimA by.5\boxMargin
  \else
   \advance\dimA by.5\wd\tbox \advance\dimA by-.5\boxMargin
  \fi
 \else
  \ifrightside
   \advance\dimA by.5\wd\tbox \advance\dimA by-.5\boxMargin
  \else
   \advance\dimA by-.5\wd\tbox \advance\dimA by.5\boxMargin
  \fi
 \fi
 \ifdim\dimC<0.0pt
  \ifrightside
   \advance\dimB by.5\ht\tbox \advance\dimB by.5\dp\tbox
   \advance\dimB by-.5\boxMargin
  \else
   \advance\dimB by-.5\ht\tbox \advance\dimB by-.5\dp\tbox
   \advance\dimB by.5\boxMargin
  \fi
 \else
  \ifrightside
   \advance\dimB by-.5\ht\tbox \advance\dimB by-.5\dp\tbox
   \advance\dimB by.5\boxMargin
  \else
   \advance\dimB by.5\ht\tbox \advance\dimB by.5\dp\tbox
   \advance\dimB by-.5\boxMargin
  \fi
 \fi
 \displaceLabel(\ptless{\dimA},\ptless{\dimB})(1.0,0.0,0.0,0)
}

\def\displaceLabel(#1,#2)(#3,#4,#5,#6){%
 \put(#1,#2){%
  \dimA=#4pt
  \ifdim\dimA<0.0pt \dimA=-\dimA\fi
  \ifnum#6>0
   \begin{tikzpicture}[xscale=0.0352778, yscale=0.0352778, thin, inner sep=0]
     \path[use as bounding box] (-#5*\ptless\dimA/#3,0);
     \node at (0,0) {%
       \hbox to0pt{\hss \vbox to0pt{\vss
       \hbox{\copy\tbox}%
       \vss}\hss}};
   \end{tikzpicture}
  \else
   \begin{tikzpicture}[xscale=0.0352778, yscale=0.0352778, thin, inner sep=0]
     \path[use as bounding box] (0,#5*\ptless\dimA/#3);
     \node at (0,0) {%
       \hbox to0pt{\hss \vbox to0pt{\vss
       \hbox{\copy\tbox}%
       \vss}\hss}};
   \end{tikzpicture}
  \fi}
}

\newdimen\hd \newdimen\vd \newdimen\cd \newdimen\md
\newdimen\xB \newdimen\yB
\newdimen\xC \newdimen\yC

\def\slantMorphism(#1,#2)(#3,#4){%
 \hd=#3\latticeUnit \advance\hd by-#1\latticeUnit
 \vd=#4\latticeUnit \advance\vd by-#2\latticeUnit
 \vectorPosition(#1,#2)(#3,#4)%
 \edef\arg{{\ptless{\xB}}{\ptless{\yB}}{\ptless{\xC}}{\ptless{\yC}}}
  \ifx\lcommand\vector
   \expandafter\drawVect\arg
  \else
   \expandafter\drawLine\arg
  \fi
}

\newif\iflayer
\layerfalse

\def\drawLine#1#2#3#4{%
 \ifdim#1pt<#3pt \dimA=#1pt \else \dimA=#3pt\fi
 \ifdim#2pt<#4pt \dimB=#2pt \else \dimB=#4pt\fi
 \iflayer
  \put(\ptless\dimA,\ptless\dimB){%
    \tikz[xscale=0.0352778, yscale=0.0352778]
      \draw[line width=2..5,white] (#1,#2) -- (#3,#4);
    \tikz[xscale=0.0352778, yscale=0.0352778]
      \draw[line width=.5,black] (#1,#2) -- (#3,#4);}
 \else
  \put(\ptless\dimA,\ptless\dimB){%
    \tikz[xscale=0.0352778, yscale=0.0352778]
      \draw[line width=.5,black] (#1,#2) -- (#3,#4);}
 \fi
}
\def\drawVect#1#2#3#4{%
 \ifdim#1pt<#3pt \dimA=#1pt \else \dimA=#3pt\fi
 \ifdim#2pt<#4pt \dimB=#2pt \else \dimB=#4pt\fi
 \iflayer
  \put(\ptless\dimA,\ptless\dimB){%
    \tikz[xscale=0.0352778, yscale=0.0352778]
      \draw[>=latex,->,line width=2..5,white] (#1,#2) -- (#3,#4);
    \tikz[xscale=0.0352778, yscale=0.0352778]
      \draw[>=latex,->,line width=.5,black] (#1,#2) -- (#3,#4);}
 \else
  \put(\ptless\dimA,\ptless\dimB){%
    \tikz[xscale=0.0352778, yscale=0.0352778]
      \draw[>=latex,->,line width=.5,black] (#1,#2) -- (#3,#4);}
 \fi
}

{\catcode`p=12 \catcode`t=12 \gdef\hazonc#1.#2pt{#1}}
\def\toJnt#1{\expandafter\hazonc\the#1}
\let\getInt=\toJnt

\newdimen\xd \newdimen\yd \newdimen\xe \newdimen\ye

\def\vectorPosition(#1,#2)(#3,#4){%
 \ifdim \hd>0.0pt
  \ifdim \vd>0.0pt
   \obtainDelta(#1,#2)(#3,#4)
  \else
   \vd=-\vd
   \obtainDelta(#1,#2)(#3,#4)
   \vd=-\vd \yd=-\yd \ye=-\ye
  \fi
 \else
  \hd=-\hd
  \ifdim \vd>0.0pt
   \obtainDelta(#1,#2)(#3,#4)
   \xd=-\xd \xe=-\xe
  \else
   \vd=-\vd
   \obtainDelta(#1,#2)(#3,#4)
   \vd=-\vd \xd=-\xd \xe=-\xe \yd=-\yd \ye=-\ye
  \fi
  \hd=-\hd
 \fi
 \xB=#1\latticeUnit \advance\xB by\xd
 \yB=#2\latticeUnit \advance\yB by\yd
 \xC=#3\latticeUnit \advance\xC by\xe
 \yC=#4\latticeUnit \advance\yC by\ye
}

\def\obtainDelta(#1,#2)(#3,#4){%
 \refer(#1,#2)\to(\tc,\tcc)%
 \ifdim\hd<\vd 
  \dimA=\dimen\tcc
  \multiply\dimA by\getInt\hd
  \divide\dimA by\getInt\vd    
  \dimB=\dimen\tc
  \ifdim \dimB<\dimA
   \dimA=\dimen\tc
   \multiply\dimA by\getInt\vd
   \divide\dimA by\getInt\hd   
   \xd=\dimen\tc \yd=\dimA
  \else
   \xd=\dimA \yd=\dimen\tcc
  \fi
 \else 
  \dimA=\dimen\tc
  \multiply\dimA by\getInt\vd
  \divide\dimA by\getInt\hd    
  \dimB=\dimen\tcc
  \ifdim \dimB<\dimA
   \dimA=\dimen\tcc
   \multiply\dimA by\getInt\hd
   \divide\dimA by\getInt\vd   
   \xd=\dimA \yd=\dimen\tcc
  \else
   \xd=\dimen\tc \yd=\dimA
  \fi
 \fi
 \refer(#3,#4)\to(\tc,\tcc)%
 \ifdim\hd<\vd 
  \dimA=\dimen\tcc
  \multiply\dimA by\getInt\hd
  \divide\dimA by\getInt\vd    
  \dimB=\dimen\tc
  \ifdim \dimB<\dimA
   \dimA=\dimen\tc
   \multiply\dimA by\getInt\vd
   \divide\dimA by\getInt\hd   
   \xe=-\dimen\tc \ye=-\dimA
  \else
   \xe=-\dimA \ye=-\dimen\tcc
  \fi
 \else 
  \dimA=\dimen\tc
  \multiply\dimA by\getInt\vd
  \divide\dimA by\getInt\hd    
  \dimB=\dimen\tcc
  \ifdim \dimB<\dimA
   \dimA=\dimen\tcc
   \multiply\dimA by\getInt\hd
   \divide\dimA by\getInt\vd   
   \xe=-\dimA \ye=-\dimen\tcc
  \else
   \xe=-\dimen\tc \ye=-\dimA
  \fi
 \fi
}

\def\ptToCoord#1{%
 \tcount=\expandafter\toInt\number\ptless\latticeUnit
 \divide#1 by\tcount}

\newenvironment{diagramme}{%
  \latticeUnit=100pt \boxMargin=3pt \layerMargin=3pt
  \rightlim=0pt \leftlim=0pt \upperlim=0pt \lowerlim=0pt
  \setbox\diagbox=\hbox\bgroup
  \begin{picture}(0,0)}%
 {\end{picture}\egroup
  \tdimen=\rightlim \advance\tdimen by-\leftlim
  \ttdimen=\upperlim \advance\ttdimen by-\lowerlim
  \begin{picture}(\ptless{\tdimen},\ptless{\ttdimen})%
                 (\ptless{\leftlim},\ptless{\lowerlim})
   \put(0,0){\unhbox\diagbox}
  \end{picture}%
}

\newcount\tabsc

\def\spandiagabscoord#1#2#3#4#5#6#7{%
 \tabsc=#1 \divide\tabsc by2
 \let\ss=\scriptstyle
 \latticeUnit=1pt\boxMargin=3pt
 \object(\the\tabsc,#2)={#3}
 \object(0,0)={#4}
 \object(#1,0)={#5}
 \morphism(\the\tabsc,#2)to(0,0)[\ss{#6}][R]
 \morphism(\the\tabsc,#2)to(#1,0)[\ss{#7}]
}

\def\trianglediagabscoord#1#2#3#4#5#6#7#8{%
 \tabsc=#1 \divide\tabsc by2
 \let\ss=\scriptstyle
 \latticeUnit=1pt\boxMargin=3pt
 \object(0,#2)={#3}
 \object(#1,#2)={#4}
 \object(\the\tabsc,0)={#5}
 \morphism(0,#2)to(#1,#2)[\ss{#6}]
 \morphism(0,#2)to(\the\tabsc,0)[\ss #7][R]
 \morphism(#1,#2)to(\the\tabsc,0)[\ss #8]
}

\def\optrianglediagabscoord#1#2#3#4#5#6#7#8{%
 \tabsc=#1 \divide\tabsc by2
 \let\ss=\scriptstyle
 \latticeUnit=1pt\boxMargin=3pt
 \object(\the\tabsc,#2)={#3}
 \object(0,0)={#4}
 \object(#1,0)={#5}
 \morphism(\the\tabsc,#2)to(0,0)[\ss{#6}][R]
 \morphism(\the\tabsc,#2)to(#1,0)[\ss{#7}]
 \morphism(0,0)to(#1,0)[\ss{#8}][R]
}

\def\squarediagabscoord#1#2{%
 \def\hori{#1}\def\vert{#2}%
 \sqda}
\def\sqda#1#2#3#4#5#6#7#8{%
 \let\ss=\scriptstyle
 \latticeUnit=1pt \boxMargin=3pt
 \object(0,\vert)={#1}
 \object(\hori,\vert)={#2}
 \object(0,0)={#3}
 \object(\hori,0)={#4}
 \morphism(0,\vert)to(\hori,\vert)[\ss{#5}]
 \morphism(0,\vert)to(0,0)[\ss{#6}][R]
 \morphism(\hori,\vert)to(\hori,0)[\ss{#7}]
 \morphism(0,0)to(\hori,0)[\ss{#8}][R]
}

\newbox\tcb\setbox\tcb=\hbox{$\scriptstyle \Rightarrow$}
\newdimen\dimH \newdimen\dimI

\def\ptless#1{\expandafter\getfactor\the#1}

\newdimen\dmA \newdimen\dmB \newdimen\dmC \newdimen\dmD \newdimen\dmE

\def\lzda#1#2#3#4#5#6#7#8{%
 \let\ss=\scriptstyle
 \latticeUnit=1pt\boxMargin=3pt
 \object(0,\ptless\dmB)={#1}
 \object(-\ptless\dmA,0)={#2}
 \object(\ptless\dmA,0)={#3}
 \object(0,-\ptless\dmB)={#4}
 \morphism(0,\ptless\dmB)to(-\ptless\dmA,0)[\ss{#5}][R]
 \morphism(0,\ptless\dmB)to(\ptless\dmA,0)[\ss{#6}]
 \morphism(-\ptless\dmA,0)to(0,-\ptless\dmB)[\ss{#7}][R]
 \morphism(\ptless\dmA,0)to(0,-\ptless\dmB)[\ss{#8}]
}

\def\pentagondiagabscoord#1#2#3#4#5#6#7{%
 \def\firstobj{#3}\def\secondobj{#4}\def\thirdobj{#5}%
 \def\fourthobj{#6}\def\fifthobj{#7}%
 \argargpgda{#1}{#2}}

\def\argargpgda#1#2#3#4#5#6#7{%
  \let\ss=\scriptstyle
  \latticeUnit=1pt \boxMargin=3pt
  \pgda{#1}{#2}{#3}{#4}{#5}{#6}{#7}%
}

\def\pgda#1#2#3#4#5#6#7{%
 \dmA=#1pt\dmB=#2pt
 \dmE=\dmB
 \multiply\dmE by95\divide\dmE by100
 \multiply\dmB by59\divide\dmB by100
 \dmC=\dmA
 \multiply\dmC by81\divide\dmC by100
 \dmD=\dmA
 \multiply\dmD by19\divide\dmD by100
 \divide\dmA by2
 \object(\ptless\dmA,\ptless\dmE)={\firstobj}
 \object(0,\ptless\dmB)={\secondobj}
 \object(#1,\ptless\dmB)={\thirdobj}
 \object(\ptless\dmD,0)={\fourthobj}
 \object(\ptless\dmC,0)={\fifthobj}
 \morphism(\ptless\dmA,\ptless\dmE)to(0,\ptless\dmB)[\ss{#3}][R]
 \morphism(\ptless\dmA,\ptless\dmE)to(#1,\ptless\dmB)[\ss{#4}]
 \morphism(0,\ptless\dmB)to(\ptless\dmD,0)[\ss{#5}][R]
 \morphism(#1,\ptless\dmB)to(\ptless\dmC,0)[\ss{#6}]
 \morphism(\ptless\dmD,0)to(\ptless\dmC,0)[\ss{#7}][R]
}

\def\oppentagondiagabscoord#1#2#3#4#5#6#7{%
 \def\firstobj{#3}\def\secondobj{#4}\def\thirdobj{#5}%
 \def\fourthobj{#6}\def\fifthobj{#7}%
 \opargargpgda{#1}{#2}}

\def\opargargpgda#1#2#3#4#5#6#7{%
  \let\ss=\scriptstyle
  \latticeUnit=1pt \boxMargin=3pt
  \oppgda{#1}{#2}{#3}{#4}{#5}{#6}{#7}%
}

\def\oppgda#1#2#3#4#5#6#7{%
 \dmA=#1pt\dmB=#2pt
 \dmE=\dmB
 \multiply\dmE by95\divide\dmE by100
 \multiply\dmB by59\divide\dmB by100
 \dmC=\dmA
 \multiply\dmC by81\divide\dmC by100
 \dmD=\dmA
 \multiply\dmD by19\divide\dmD by100
 \divide\dmA by2
 \object(\ptless\dmA,-\ptless\dmE)={\fifthobj}
 \object(#1,-\ptless\dmB)={\fourthobj}
 \object(0,-\ptless\dmB)={\thirdobj}
 \object(\ptless\dmC,0)={\secondobj}
 \object(\ptless\dmD,0)={\firstobj}
 \morphism(\ptless\dmD,0)to(\ptless\dmC,0)[\ss{#3}]
 \morphism(\ptless\dmD,0)to(0,-\ptless\dmB)[\ss{#4}][R]
 \morphism(\ptless\dmC,0)to(#1,-\ptless\dmB)[\ss{#5}]
 \morphism(0,-\ptless\dmB)to(\ptless\dmA,-\ptless\dmE)[\ss{#6}][R]
 \morphism(#1,-\ptless\dmB)to(\ptless\dmA,-\ptless\dmE)[\ss{#7}]
}

\def\hexagondiagabscoord#1#2#3#4#5#6#7#8{%
 \def\firstobj{#3}\def\secondobj{#4}\def\thirdobj{#5}%
 \def\fourthobj{#6}\def\fifthobj{#7}\def\sixthobj{#8}%
 \argarghgda{#1}{#2}}

\def\argarghgda#1#2#3#4#5#6#7#8{%
  \let\ss=\scriptstyle
  \latticeUnit=1pt \boxMargin=3pt
  \hgda{#1}{#2}{#3}{#4}{#5}{#6}{#7}{#8}%
}

\def\hgda#1#2#3#4#5#6#7#8{%
 \dmA=#1pt\dmB=#2pt
 \multiply\dmB by58\divide\dmB by100
 \dmC=\dmA \divide\dmC by2
 \dmD=\dmB \divide\dmD by2
 \object(\ptless\dmC,\ptless\dmB)={\firstobj}
 \object(0,\ptless\dmD)={\secondobj}
 \object(#1,\ptless\dmD)={\thirdobj}
 \object(0,-\ptless\dmD)={\fourthobj}
 \object(#1,-\ptless\dmD)={\fifthobj}
 \object(\ptless\dmC,-\ptless\dmB)={\sixthobj}
 \morphism(\ptless\dmC,\ptless\dmB)to(0,\ptless\dmD)[\ss{#3}][R]
 \morphism(\ptless\dmC,\ptless\dmB)to(#1,\ptless\dmD)[\ss{#4}]
 \morphism(0,\ptless\dmD)to(0,-\ptless\dmD)[\ss{#5}][R]
 \morphism(#1,\ptless\dmD)to(#1,-\ptless\dmD)[\ss{#6}]
 \morphism(0,-\ptless\dmD)to(\ptless\dmC,-\ptless\dmB)[\ss{#7}][R]
 \morphism(#1,-\ptless\dmD)to(\ptless\dmC,-\ptless\dmB)[{#8}]
}

\def\putTensor(#1,#2){%
 \begin{scope}[shift={(#1,#2)}]
  \fill[black] (0,0) circle (5);
  \fill[white] (0,0) circle (4.3);
  \draw[line width=.65] (3.5, 3.5) -- (-3.5, -3.5);
  \draw[line width=.65] (-3.5, 3.5) -- (3.5, -3.5);
 \end{scope}
}

\def\putPar(#1,#2){%
 \begin{scope}[shift={(#1,#2)}]
  \fill[black] (0,0) circle (5);
  \fill[white] (0,0) circle (4.3);
  \fill[black] (0,0) circle (2.2);
  \fill[white] (0,0) circle (1.5);
 \end{scope}
}

\def\putRightDiode(#1,#2){%
 \begin{scope}[thin]
  \fill (#1+1.5, #2+0) -- (#1-2.5, #2+2.5) -- (#1-2.5, #2-2.5) -- cycle;
  \fill (#1+1.5, #2+2.5) -- (#1+1.5, #2-2.5) -- (#1+2.5, #2-2.5) -- (#1+2.5, #2+2.5) -- cycle;
 \end{scope}
}

\def\putLeftDiode(#1,#2){%
 \begin{scope}[thin]
  \fill (#1-1.5, #2+0) -- (#1+2.5, #2+2.5) -- (#1+2.5, #2-2.5) -- cycle;
  \fill (#1-1.5, #2+2.5) -- (#1-1.5, #2-2.5) -- (#1-2.5, #2-2.5) -- (#1-2.5, #2+2.5) -- cycle;
 \end{scope}
}

\newbox\tcd\setbox\tcd=\hbox{$\&$}
\def\linpar{\mathbin{\tikz[inner sep=0] \node[rotate=180] at (0,0) {$\&$};}}
\def\slinpar{\mathbin{\tikz[inner sep=0] \node[rotate=180] at (0,0) {$\scriptstyle\&$};}}

\def\scirc{\mathbin{\vcenter{\hbox{\scriptsize $\circ$}}}}

\def\llapem#1#2{\llap{\hbox to#1em{\rm #2\hss}}}
\let\tempar\par \def\par{{\tempar}}%

\def\bang{\mathord!}

\def\di#1(#2,#3)#4{%
 \dimA=#2\latticeUnit
 \dimB=#3\latticeUnit
 \put(\ptless\dimA,\ptless\dimB){\tikz[inner sep=0]
   \node[rotate=-#4] at (\ptless\dimA,\ptless\dimB)
     {\vbox to0pt{\vss\hbox to0pt{\hss #1\hss}\vss}};}}

\makeatother

\begin{document}

\maketitle

\begin{abstract}
Diagram chasing is not an easy task.
The coherence holds in a generalized sense
 if we have a mechanical method to judge
 whether given two morphisms are equal to each other.
A simple way to this end is to reform a concerned category
 into a calculus, where the instructions for the diagram chasing
 are given in the form of rewriting rules.
We apply this idea to the categorical semantics of the linear
 logic.
We build a calculus directly on the free category of the semantics.
It enables us to perform diagram chasing as essentially one-way
 computations led by the rewriting rules.
We verify the weak termination property of this calculus.
This gives the first step towards the mechanization of
 diagram chasing.
\end{abstract}

\section{Introduction}

This work started with the naive idea that
 diagram chasing in category theory may be mechanized, at least
 in specific cases.
Constructing the appropriate
 commutative diagrams to show equality of two morphisms
 is by no means straightforward.
The automation of the task would be useful and thus of interest.
Monoidal categories and symmetric monoidal
 categories admit coherence theorems ascertaining that
 any parallel morphisms are
 automatically equal to one another as far as
 they consist only of canonical isomorphisms~\cite{macl,kell}.
Autonomous categories and $\star$-autonomous categories
 do not have strict coherences but, through graphical
 presentations, checking equality of two morphisms
 can be automated~\cite{kemc,bcst,hugh}, although
 the decision procedure is intractible~\cite{heho}.

Categorical semantics of type theories provide an equivalence
 between type systems and certain categories \cite{jaco}.
A decisively classic work is the semantics of the simply
 typed lambda calculus using
 the cartesian closed category~\cite{lasc}.
It shows exact correspondence between $\beta\eta$-equal
 lambda terms and commutative diagrams.
Unfortunately, the equivalence is valid only after the process of calculation
 is ignored.
As the name suggests,
 the lambda calculus is a computational system.
Equality between two lambda terms can
 be automatically checked
 through mechanical computation by $\beta\eta$-reduction.
To ensure the equivalence between the calculus and the category, however,
 we have to identify all terms occurring during this calculation.
Thus, the dynamic content of the calculus is lost in the categorical
 semantics.

Existence of computation in the side of the lambda calculus suggests
 that the corresponding cartesian closed category
 may well be given a dynamic computational mechanism.
The $\beta\eta$-equality in the lambda calculus corresponds to
 the adjunction between product and exponential, i.e.,
 $(\hbox{--})\times B\dashv
 B\rightarrow (\hbox{--})$.
For example, the $\beta$-equivalence corresponds to
 the following commutative triangle diagram, which arises from the
 adjunction:
$$\vcenter{\small
\hbox{\begin{diagramme}
  \trianglediagabscoord{80}{40}%
   {A\times B}{(B\rightarrow A\times B)\times B\kern-20pt}{A\times B}%
   {{\rm abs}\cdot}{1}{{\rm ev}}
  \object(40,25)={\circlearrowright}
 \end{diagramme}}}
$$
where $\circlearrowright$ denotes that two legs are equal.\footnote{%
 To save space, we occasionally use a dot to signify the position where a
 suitable identity is inserted, and omit tensor and cotensor on morphisms.}
When we regard this diagram as a reduction, we modify it as
$$\vcenter{\small
\hbox{\begin{diagramme}
  \trianglediagabscoord{80}{40}%
   {A\times B}{(B\rightarrow A\times B)\times B\kern-20pt}{A\times B}%
   {{\rm abs}\cdot}{1}{{\rm ev}}
  \di{\hbox{$\Rightarrow$}}(40,25){140}
 \end{diagramme}}}
$$
 where the 2-cell double arrow $\Rightarrow$ means one-way rewriting.
The morphism $({\rm abs}\times1_B);{\rm ev}$
 contracts to $1_{A\times B}$.
Rewriting in the reverse direction is prohibited.
This idea, essentially due to Seely \cite{seel},
 looks natural but does not seem to be pursued further.
Previous works by Seely \cite{seel} and
 Jay \cite{jay} construct 2-categories
 employing ordinary lambda terms.
These are the lambda calculus presented in the style of 2-categories,
 not the rewriting system directly built on categories.

We install a calculus
 on the categorical semantics of
 the linear logic.
In the lambda calculus, when an argument of a function is accessed $n$ times,
 its $n$ copies are created to be substituted simultaneously.
The duplication process is, however, encapsulated
 in the $\beta$-reduction rule, inseparable
 from other operations.
The linear logic isolates duplication so that
 the timing and the amount of it can be controlled.
The categorical semantics of the linear logic has a symmetric monoidal
 adjunction $(\hbox{--})\otimes B\dashv
 B\multimap (\hbox{--})$ equipped with comonad $\mathord!$, such that
 $\mathord!A$ comes equipped with
 a commutative coalgebra structure on it.
The $\beta$-reduction of the linear logic becomes
$$\kern20pt\vcenter{\small
\hbox{\begin{diagramme}
  \trianglediagabscoord{80}{40}%
   {A\otimes B}{(B\multimap A\otimes B)\otimes B\kern-20pt}{A\otimes B}%
   {{\rm abs}\cdot}{1}{{\rm ev}}
  \di{\hbox{$\Rightarrow$}}(40,25){140}
 \end{diagramme}}}
$$
Alternatively, if we take a $\star$-autonomous category
 as its base,
\begingroup
\def\linpar{\mathbin{\tikz[inner sep=0] \node[rotate=180] at (0,0) {\small $\&$};}}
$$\kern20pt\vcenter{\small
\hbox{\begin{diagramme}
\oppentagondiagabscoord{70}{60}%
 {{\bf 1}\otimes A}{(A\linpar A^*)\otimes A\kern-40pt}{A}%
 {A\linpar (A^*\otimes A)\kern-15pt}{A\linpar \bot}%
 {\tau\cdot}{\sim}{\partial'}{\sim}{\cdot\gamma}
 \di{\hbox{$\Rightarrow$}}(35,-25){140}
\end{diagramme}}}$$
\endgroup
Among the defining diagrams of the categorical semantics,
 twenty-one diagrams are regarded as reduction rules.
For example,
$$\vcenter{\small
\hbox{\begin{diagramme}
\squarediagabscoord{50}{40}%
 {\mathord!A}{\mathord!\mathord!A}{\mathord!\mathord!A}%
 {\mathord!\mathord!\mathord!A}%
 {\delta}{\delta}{\delta}{\mathord!\delta}
 \di{\hbox{$\Rightarrow$}}(25,20){140}
\end{diagramme}}}
$$
 replaces one of the defining commutative diagrams of comonad.
The naturality of certain morphisms is also replaced
 with rewriting rules.
For example, naturality of the diagonal (comultiplication)
 of the comonoid gives rise to
$$\vcenter{\small
\hbox{\begin{diagramme}
\squarediagabscoord{50}{40}%
 {\mathord!A}{\mathord!B}{\kern-10pt\mathord!A\otimes \mathord!A}%
 {\mathord!B\otimes \mathord!B\kern-10pt}%
 {\mathord!f}{d}{d}{\mathord!f\otimes \mathord!f}
 \di{\hbox{$\Rightarrow$}}(25,20){140}
\end{diagramme}}}$$
 that realizes duplication.
The choice of the directions of rewriting rules is justified
 by comparison to the conversion rules in the type theory.
Details are given in section~\ref{soe19}.
These rules, in addition to $\beta\eta$-rules,
 provide twenty-three reduction rules in total.
We contend that the categorical model of the lambda
 calculus is too coarse to incorporate a computational system
 in it.
If we clearly separate copying
 from the other functions using the linear logic,
 we can directly implement rewriting on a category
 so that the obtained calculus has the desirable properties.

Our purpose is, however, not to transcribe a carbon copy of the type system
 in a category.
We build a calculus worth existing in its own right.
The linear logic allows finer control on duplication
 than the lambda calculus, yet the unit of substitutions
 is coarse.
A term is duplicated in one stroke no matter how large it is.
To improve the situation, graphical reduction systems have been
 considered \cite{gal}.
Each link in a graph can be individually duplicated so that
 optimal efficiency is attained by
 ultimate usage of sharing.
However, graphical systems have a drawback.
Arbitrary connections
 of links do not form a syntactically lawful graph in general.
Moreover, it is not obvious how to ensure that the graphs
 occurring in the process of rewriting remains meaningful.
A system by Ghani \cite{ghan} and one by Asperti \cite{aspe} are
 examples of the calculi inspired by the category theory.
The former is term rewriting and the latter is graph rewriting.

Our categorical rewriting system lies between
 term rewriting and graph rewriting.
It enables fine-grained control of resources.
We can dissect terms in order to duplicate them piece by piece.
A morphism of the form $\mathord!f$ corresponds to a box in the linear
 logic.
The linear logic has no function to split boxes,
 thus a box $\mathord!(g;f)$ must be copied
 as an assembled unit.
In contrast our system permits to decompose it into $\mathord!g;
 \mathord!f$ to activate partial duplication
 $\mathord!f;d\Rightarrow d;(\mathord!f\otimes \mathord!f)$
 by naturality of the diagonal.
A similar property is presented in a system based on
 lambda terms by Jay \cite{jay}.
For graph rewriting, in contrast, extremely fine control is enabled
 since duplication per link is allowed.
However, there is a risk that the intermediate graphs
 appearing in computation may lose semantical justification.
As our system deals with only those which are meaningful
 as morphisms, the duplicated unit $f$ always keeps its semantical meaning.
Our calculus rewrites the entities that have mathematical
 ``meaning'', whilst finer control on duplication is enabled
 than ordinary term rewriting.

Early works that view reductions as 2-cells are \cite{seel,ryst}.
Seely and Jay constructed 2-categories from lambda terms as
 mentioned above \cite{seel,jay}.
A graphical system based on the categorical semantics is
 \cite{aspe, asgu}.
The categorical abstract machine \cite{ccm} is a virtual machine
 based on categorical combinators \cite{curi}.
To the author's best knowledge, no previous works built
 computational systems directly on categories.

Our system satisfies the properties that computational
 systems require.
We verify normalizability in this paper.
Confluence will be discussed in a forthcoming paper.

\section{Linear category}

We start with the definition of categorical models of
 the linear logic.
Among several equivalent
 definitions \cite{bent,bbph,mell,mel2,scha}, we take the
 following \cite[Prop.25]{mmpr}.
A reason for this choice is that we can write down all defining conditions
 as commutative diagrams.
The reader may consult these papers for comparison between various
 models.

	\vskip1ex

\begin{definition}\label{wmz77}\rm
An (intuitionistic or classical)
 {\it linear category} is a pair of a category ${\bf C}$ and
 a functor $\mathord!:{\bf C}\rightarrow {\bf C}$ with the
 following additional structures:

	\vskip0ex
        \hangafter0\hangindent2em
        \noindent
\llapem2{(i)}%
If an intuitionisic linear category is concerned with,
 the underlying category is a symmetric monoidal closed category
 $({\bf C},\otimes,{\bf 1},\mathord\multimap)$.
If a classical linear category is concerned with,
 the underlying category is a
 $\star$-autonomous category $({\bf C},\otimes,\linpar,
 {\bf 1},\bot,(\hbox{--})^*)$.

        \noindent
\llapem2{(ii)}%
$\mathord!$ is equipped with the structure of a symmetric monoidal
 functor $(\mathord!,\tilde\varphi,\varphi_0)$.

        \noindent
\llapem2{(iii)}%
$\mathord!$ is equipped with the structure of a comonad
 $(\mathord!,\delta,\varepsilon)$ where $\delta:\mathord!
 \rightarrow \mathord!\mathord!$ and $\varepsilon:\mathord!\rightarrow{\it Id}$ are
 monoidal natural transformations.

	\noindent
\llapem2{(iv)}%
The objects of the shape $\mathord!A$ are equipped with the structure of a commutative
 comonoid $(\mathord!A,d_A,e_A)$ where collectively $d_A:\mathord!A
 \rightarrow \mathord!A\otimes \mathord!A$ and
 $e_A:\mathord!A\rightarrow {\bf 1}$ are monoidal natural transformations in $A$.

	\vskip.5ex
        \hangafter0\hangindent0pt
        \noindent
Moreover, these structures are related in the following way:

	\vskip.5ex
        \hangafter0\hangindent2em
        \noindent
\llapem2{(v)}%
Each $d_A$ and each $e_A$ give rise to coalgebra morphisms when $\mathord!A,{\bf 1}$,
 and $\mathord!A\otimes \mathord!A$ are naturally regarded as coalgebras.

	\noindent
\llapem2{(vi)}%
Each $\delta_A$ is a comonoid morphism.
\par
\end{definition}

	\vskip1ex

We give the list of all defining commutative
 diagrams, although they are absolutely standard.
In the next section, we pick up some of the
 diagrams and turn them into rewriting rules to
 build up a calculus.
So it will be instructive to give a full list
 as a preparation.

A symmetric monoidal category
 has a $2$-place functor $\otimes$ and an object ${\bf 1}$,
 and is equipped with
 natural isomorphisms $\alpha_{A,B,C}:(A\otimes B)\otimes C
\rightarrow A\otimes (B\otimes C),\,
 \lambda_A:{\bf 1}\otimes A\rightarrow A,\,\rho_A
 :A\otimes{\bf 1}\rightarrow A$, and $\sigma_{A,B}:
 A\otimes B\rightarrow B\otimes A$.
The naturality of these isomorphisms are

	\vskip2ex
\begingroup\offinterlineskip
\halign to\textwidth{\kern120pt\hfil #\hfil \tabskip1000pt plus100pt minus1000pt
 &\kern30pt\hfil #\hfil \kern120pt\tabskip0pt
\cr
$\vcenter{\footnotesize
\hbox to0pt{\hss\begin{diagramme}
\squarediagabscoord{50}{40}%
 {\kern-25pt(A\otimes B)\otimes C}{A\otimes (B\otimes C)\kern-25pt}%
 {\kern-30pt(A'\otimes B')\otimes C'}{A'\otimes (B'\otimes C')\kern-30pt}%
 {\alpha}{(f\otimes g)\otimes h}{f\otimes (g\otimes h)}{\alpha}
 \di{\hbox{$\circlearrowright$}}(25,20){0}
\end{diagramme}\hss}}$%
&
$\vcenter{\footnotesize
\hbox to0pt{\hss\begin{diagramme}
\squarediagabscoord{50}{40}%
 {{\bf 1}\otimes A}{A}{{\bf 1}\otimes A'}{A'}%
 {\lambda}{1\otimes f}{f}{\lambda}
 \di{\hbox{$\circlearrowright$}}(25,20){0}
\end{diagramme}\kern-2pt\hss}}$\kern6pt
\cr
	\noalign{\vskip4ex}
$\vcenter{\footnotesize
\hbox to0pt{\hss\begin{diagramme}
\squarediagabscoord{50}{40}%
 {A\otimes B}{B\otimes A}{A'\otimes B'}{B'\otimes A'}%
 {\sigma}{f\otimes g}{g\otimes f}{\sigma}
 \di{\hbox{$\circlearrowright$}}(25,20){0}
\end{diagramme}\hss}}$%
&
$\vcenter{\footnotesize
\hbox to0pt{\hss\begin{diagramme}
\squarediagabscoord{50}{40}%
 {A\otimes {\bf 1}}{A}{A'\otimes {\bf 1}}{A'}%
 {\rho}{f\otimes 1}{f}{\rho}
 \di{\hbox{$\circlearrowright$}}(25,20){0}
\end{diagramme}\kern-2pt\hss}}$\kern6pt
\cr
}
\endgroup

	\vskip2ex
        \noindent
 where subscripts are omitted for simplicity.
Moreover, these are subject to the following coherence conditions~\cite{kell}:

	\vskip2ex
\begingroup\offinterlineskip
\halign to\textwidth{\kern120pt\hfil #\hfil \tabskip1000pt plus100pt minus1000pt
 &\kern30pt\hfil #\hfil \kern120pt\tabskip0pt
\cr
$\vcenter{\footnotesize
 \hbox to0pt{\hss\begin{diagramme}
  \let\ss=\scriptstyle
  \latticeUnit=1pt \boxMargin=3pt
 \pentagondiagabscoord{70}{60}%
  {((A\otimes B)\otimes C)\otimes D}%
  {\kern-40pt(A\otimes (B\otimes C))\otimes D}%
  {(A\otimes B)\otimes (C\otimes D)\kern-40pt}%
  {\kern-60ptA\otimes ((B\otimes C)\otimes D)}%
  {A\otimes (B\otimes (C\otimes D))\kern-60pt}%
  {\global\labelPosition={.7}\alpha\otimes 1}%
  {\global\labelPosition={.5}\alpha}{\alpha}{\alpha}{1\otimes \alpha}%
 \di{\hbox{$\circlearrowright$}}(35,25){0}%
\end{diagramme}\hss}}$%
&
$\vcenter{\footnotesize
 \hbox to0pt{\hss\begin{diagramme}
  \let\ss=\scriptstyle
  \latticeUnit=1pt \boxMargin=3pt
  \trianglediagabscoord{50}{30}%
   {\kern-30pt(A\otimes {\bf 1})\otimes B}%
   {A\otimes ({\bf 1}\otimes B)\kern-30pt}{A\otimes B}%
   {\alpha}{\rho\otimes 1}{1\otimes\lambda}%
 \di{\hbox{$\circlearrowright$}}(25,17){0}%
 \end{diagramme}\hss}}$%
\cr
	\noalign{\vskip5ex}
$\vcenter{\footnotesize
 \hbox to0pt{\hss\begin{diagramme}
  \let\ss=\scriptstyle
  \latticeUnit=1pt \boxMargin=3pt
  \hexagondiagabscoord{80}{70}%
    {(A\otimes B)\otimes C}%
    {(B\otimes A)\otimes C}%
    {A\otimes (B\otimes C)}%
    {B\otimes (A\otimes C)}%
    {(B\otimes C)\otimes A}%
    {B\otimes (C\otimes A)}%
    {\global\labelPosition={.9}\sigma\otimes 1}%
    {\global\labelPosition={.7}\alpha}%
    {\global\labelPosition={.5}\alpha}{\sigma}%
    {\global\labelPosition={.1}1\otimes \sigma}%
    {\global\labelPosition={.3}\alpha}%
    \global\labelPosition={.5}%
 \di{\hbox{$\circlearrowright$}}(40,0){0}%
 \end{diagramme}\hss}}$%
&$\vcenter{\footnotesize
 \hbox to0pt{\hss\begin{diagramme}
  \let\ss=\scriptstyle
  \latticeUnit=1pt \boxMargin=3pt
  \trianglediagabscoord{50}{30}%
   {A\otimes B}{B\otimes A}{A\otimes B}{\sigma}{1}{\sigma}%
 \di{\hbox{$\circlearrowright$}}(25,18){0}%
 \end{diagramme}\hss}}$%
\cr
}
\endgroup

	\vskip2ex

A symmetric monoidal closed category~\cite{kemc} is further
 equipped with an adjoint
 $(\hbox{--})\otimes B\dashv B\multimap (\hbox{--})$.
We write the unit as ${\rm abs}^B_A:A\rightarrow B
 \multimap (A\otimes B)$ and the counit as ${\rm ev}^B_A:
(B\multimap A)\otimes B\rightarrow A$.
These satisfy the naturality
 as

	\vskip2ex
\begingroup\offinterlineskip
\halign to\textwidth{\kern120pt\hfil #\hfil \tabskip1000pt plus100pt minus1000pt
 &\kern30pt\hfil #\hfil \kern120pt\tabskip0pt
\cr
$\vcenter{\footnotesize
\hbox to0pt{\hss\begin{diagramme}
\squarediagabscoord{50}{40}%
  A{B\multimap (A\otimes B)\kern-30pt}{A'}{B\multimap (A'\otimes B)\kern-30pt}%
  {{\rm abs}^B_A}f{1\multimap (f\otimes 1)}{{\rm abs}^B_{A'}}
 \di{\hbox{$\circlearrowright$}}(25,20){0}
\end{diagramme}\kern-12pt\hss}}$%
&$\vcenter{\footnotesize
\hbox to0pt{\hss\begin{diagramme}
\squarediagabscoord{50}{40}%
  {\kern-30pt(B\multimap A)\otimes B}A{\kern-30pt(B\multimap A')\otimes B}{A'}%
  {{\rm ev}^B_A}{(1\multimap f)\otimes 1}f{{\rm ev}^B_{A'}}
 \di{\hbox{$\circlearrowright$}}(25,20){0}
\end{diagramme}\kern3pt\hss}}$
\cr
}
\endgroup

	\vskip2ex
        \noindent
 and render the following adjoint triangles commutative:

	\vskip2ex
\begingroup\offinterlineskip
\halign to\textwidth{\kern120pt\hfil #\hfil \tabskip1000pt plus100pt minus1000pt
 &\kern30pt\hfil #\hfil \kern120pt\tabskip0pt
\cr
$\vcenter{\footnotesize
\hbox{\hss\kern-52pt\begin{diagramme}
  \trianglediagabscoord{70}{35}%
   {A\otimes B}{(B\multimap A\otimes B)\otimes B\kern-40pt}{A\otimes B}%
   {{\rm abs}\otimes 1}{1}{{\rm ev}}
  \di{\hbox{$\circlearrowright$}}(35,20){0}
 \end{diagramme}\hss}}$%
&$\vcenter{\footnotesize
\hbox{\hss\begin{diagramme}
  \trianglediagabscoord{70}{35}%
   {B\multimap A}{B\multimap (B\multimap A)\otimes B\kern-40pt}{B\multimap A}%
   {{\rm abs}}{1}{1\multimap{\rm ev}}
  \di{\hbox{$\circlearrowright$}}(35,20){0}
 \end{diagramme}\kern-60pt\hss}}$%
\cr}
\endgroup

	\vskip2ex
        \noindent
(product has higher precedence than exponential).

A linearly distributive (or weakly distribuitve)
 category is a symmetric monidal category that has
 an additional monoidal category structure $(\linpar,\bot)$ and
 linear distribution morphisms $\partial_{A,B,C}:
 A\otimes(B\linpar C)\rightarrow (A\otimes B)\linpar C$,
 which are natural.
We write bars over the natural isomorphism of the added
 monoidal structure for distinction.
The naturality turns out to be

	\vskip2ex
\begingroup\offinterlineskip
\def\linpar{\mathbin{\tikz[inner sep=0] \node[rotate=180] at (0,0) {\footnotesize $\&$};}}
\def\slinpar{\mathbin{\tikz[inner sep=0] \node[rotate=180] at (0,0) {\tiny $\&$};}}
\halign to\textwidth{\kern120pt\hfil #\hfil \tabskip1000pt plus100pt minus1000pt
 &\kern30pt\hfil #\hfil \kern120pt\tabskip0pt
\cr
$\vcenter{\footnotesize
\hbox to0pt{\hss\begin{diagramme}
\squarediagabscoord{50}{40}%
 {\kern-25pt(A\linpar B)\linpar C}{A\linpar (B\linpar C)\kern-25pt}%
 {\kern-30pt(A'\linpar B')\linpar C'}{A'\linpar (B'\linpar C')\kern-30pt}%
 {\bar\alpha}{(f\slinpar g)\slinpar h}{f\slinpar (g\slinpar h)}{\bar\alpha}
 \di{\hbox{$\circlearrowright$}}(25,20){0}
\end{diagramme}\hss}}$%
&
$\vcenter{\footnotesize
\hbox to0pt{\hss\begin{diagramme}
\squarediagabscoord{50}{40}%
 {\bot\linpar A}{A}{\bot\linpar A'}{A'}%
 {\bar\lambda}{1\slinpar f}{f}{\bar\lambda}
 \di{\hbox{$\circlearrowright$}}(25,20){0}
\end{diagramme}\kern12pt\hss}}$%
\cr
	\noalign{\vskip4ex}
$\vcenter{\footnotesize
\hbox to0pt{\hss\begin{diagramme}
\squarediagabscoord{50}{40}%
 {A\linpar B}{B\linpar A}{A'\linpar B'}{B'\linpar A'}%
 {\bar\sigma}{f\slinpar g}{g\slinpar f}{\bar\sigma}
 \di{\hbox{$\circlearrowright$}}(25,20){0}
\end{diagramme}\hss}}$%
&
$\vcenter{\footnotesize
\hbox to0pt{\hss\begin{diagramme}
\squarediagabscoord{50}{40}%
 {A\linpar \bot}{A}{A'\linpar \bot}{A'}%
 {\bar\rho}{f\slinpar 1}{f}{\bar\rho}
 \di{\hbox{$\circlearrowright$}}(25,20){0}
\end{diagramme}\kern12pt\hss}}$%
\cr
	\noalign{\vskip4ex}
$\vcenter{\footnotesize
\hbox to0pt{\hss\begin{diagramme}
\squarediagabscoord{50}{40}%
 {\kern-25pt A\otimes (B\linpar C)}{(A\otimes B)\linpar C\kern-25pt}%
 {\kern-30pt A'\otimes (B'\linpar C')}{(A'\otimes B')\linpar C'\kern-30pt}%
 {\partial}{f\otimes(g\slinpar h)}{(f\otimes g)\slinpar h}{\partial}
 \di{\hbox{$\circlearrowright$}}(25,20){0}
\end{diagramme}\hss}}$%
\cr
}
\endgroup

	\vskip2ex
        \noindent
The coherence conditions for the added monoidal structure are

	\vskip2ex
\begingroup\offinterlineskip
\halign to\textwidth{\kern120pt\hfil #\hfil \tabskip1000pt plus100pt minus1000pt
 &\kern30pt\hfil #\hfil \kern120pt\tabskip0pt
\cr
$\vcenter{\footnotesize
\def\linpar{\mathbin{\tikz[inner sep=0] \node[rotate=180] at (0,0) {\footnotesize $\&$};}}
\def\slinpar{\mathbin{\tikz[inner sep=0] \node[rotate=180] at (0,0) {\tiny $\&$};}}
 \hbox to0pt{\hss\begin{diagramme}
  \let\ss=\scriptstyle
  \latticeUnit=1pt \boxMargin=3pt
 \pentagondiagabscoord{70}{60}%
  {((A\linpar B)\linpar C)\linpar D}%
  {\kern-40pt(A\linpar (B\linpar C))\linpar D}%
  {(A\linpar B)\linpar (C\linpar D)\kern-40pt}%
  {\kern-60ptA\linpar ((B\linpar C)\linpar D)}%
  {A\linpar (B\linpar (C\linpar D))\kern-60pt}%
  {\global\labelPosition={.7}\bar\alpha\slinpar 1}%
  {\global\labelPosition={.5}\bar\alpha}{\bar\alpha}{\bar\alpha}{1\slinpar \bar\alpha}
 \di{\hbox{$\circlearrowright$}}(35,25){0}
\end{diagramme}\hss}}$%
&$\vcenter{\footnotesize
\def\linpar{\mathbin{\tikz[inner sep=0] \node[rotate=180] at (0,0) {\footnotesize $\&$};}}
\def\slinpar{\mathbin{\tikz[inner sep=0] \node[rotate=180] at (0,0) {\tiny $\&$};}}
 \hbox to0pt{\hss\begin{diagramme}
  \let\ss=\scriptstyle
  \latticeUnit=1pt \boxMargin=3pt
  \trianglediagabscoord{50}{30}%
   {\kern-30pt(A\linpar {\bf 1})\linpar B}%
   {A\linpar ({\bf 1}\linpar B)\kern-30pt}{A\linpar B}%
   {\bar\alpha}{\bar\rho\slinpar 1}{1\slinpar\bar\lambda}
 \di{\hbox{$\circlearrowright$}}(25,17){0}
 \end{diagramme}\kern-6pt\hss}}$
\cr
	\noalign{\vskip4ex}
$\vcenter{\footnotesize
\def\linpar{\mathbin{\tikz[inner sep=0] \node[rotate=180] at (0,0) {\footnotesize $\&$};}}
\def\slinpar{\mathbin{\tikz[inner sep=0] \node[rotate=180] at (0,0) {\tiny $\&$};}}
 \hbox to0pt{\hss\begin{diagramme}
  \let\ss=\scriptstyle
  \latticeUnit=1pt \boxMargin=3pt
  \hexagondiagabscoord{80}{70}%
    {(A\linpar B)\linpar C}%
    {(B\linpar A)\linpar C}%
    {A\linpar (B\linpar C)}%
    {B\linpar (A\linpar C)}%
    {(B\linpar C)\linpar A}%
    {B\linpar (C\linpar A)}%
    {\global\labelPosition={.7}\bar\sigma\slinpar 1}%
    {\global\labelPosition={.5}\bar\alpha}{\bar\alpha}{\bar\sigma}%
    {\global\labelPosition={.3}1\slinpar \bar\sigma}%
    {\global\labelPosition={.5}\bar\alpha}
 \di{\hbox{$\circlearrowright$}}(40,0){0}
 \end{diagramme}\hss}}$%
&$\vcenter{\footnotesize
\def\linpar{\mathbin{\tikz[inner sep=0] \node[rotate=180] at (0,0) {\footnotesize $\&$};}}
\def\slinpar{\mathbin{\tikz[inner sep=0] \node[rotate=180] at (0,0) {\tiny $\&$};}}
 \hbox to0pt{\hss\begin{diagramme}
  \let\ss=\scriptstyle
  \latticeUnit=1pt \boxMargin=3pt
  \trianglediagabscoord{50}{30}%
   {A\linpar B}{B\linpar A}{A\linpar B}{\bar\sigma}{1}{\bar\sigma}
 \di{\hbox{$\circlearrowright$}}(25,18){0}
 \end{diagramme}\kern-6pt\hss}}$
\cr
}
\endgroup

	\vskip2ex
        \noindent
The following are the coherences of linear distribution morphisms \cite{cose}.
Here $\partial'_{ABC}:(A\linpar B)\otimes C\rightarrow A\linpar (B\otimes C)$ is
 induced from $\partial_{ABC}$ by symmetry of tensor and cotensor.
The label $\sim$ represents appropriate structural isomorphisms.

	\vskip2ex
\begingroup\offinterlineskip
\halign to\textwidth{\kern120pt\hfil #\hfil \tabskip1000pt plus100pt minus1000pt
 &\kern30pt\hfil #\hfil \kern120pt\tabskip0pt
\cr
$\vcenter{\footnotesize
\def\linpar{\mathbin{\tikz[inner sep=0] \node[rotate=180] at (0,0) {\footnotesize $\&$};}}
\def\slinpar{\mathbin{\tikz[inner sep=0] \node[rotate=180] at (0,0) {\tiny $\&$};}}
\hbox to0pt{\hss\begin{diagramme}
\trianglediagabscoord{50}{30}%
 {\kern-20pt {\bf 1}\otimes (A\linpar B)}%
 {({\bf 1}\otimes A)\linpar B\kern-20pt}{A\linpar B}%
 {\partial}{\sim}{\sim}
 \di{\hbox{$\circlearrowright$}}(25,18){0}
\end{diagramme}\hss}}$%
&$\vcenter{\footnotesize
\def\linpar{\mathbin{\tikz[inner sep=0] \node[rotate=180] at (0,0) {\footnotesize $\&$};}}
\def\slinpar{\mathbin{\tikz[inner sep=0] \node[rotate=180] at (0,0) {\tiny $\&$};}}
\hbox to0pt{\hss\begin{diagramme}
\pentagondiagabscoord{70}{60}%
 {A\otimes (B\linpar (C\linpar D))}%
 {\kern-40ptA\otimes ((B\linpar C)\linpar D)}%
 {(A\otimes B)\linpar (C\linpar D)\kern-40pt}%
 {\kern-60pt(A\otimes (B\linpar C))\linpar D}%
 {((A\otimes B)\linpar C)\linpar D\kern-60pt}%
 {\sim}{\partial}{\partial}{\sim}{\partial\cdot}
 \di{\hbox{$\circlearrowright$}}(35,25){0}
\end{diagramme}\hss}}$%
\cr
	\noalign{\vskip3ex}
$\vcenter{\footnotesize
\def\linpar{\mathbin{\tikz[inner sep=0] \node[rotate=180] at (0,0) {\footnotesize $\&$};}}
\def\slinpar{\mathbin{\tikz[inner sep=0] \node[rotate=180] at (0,0) {\tiny $\&$};}}
\hbox to0pt{\hss\begin{diagramme}
\pentagondiagabscoord{70}{60}%
 {(A\linpar B)\otimes (C\linpar D)}%
 {\kern-40ptA\linpar (B\otimes (C\linpar D))}%
 {((A\linpar B)\otimes C)\linpar D\kern-40pt}%
 {\kern-60ptA\linpar ((B\otimes C)\linpar D)}%
 {(A\linpar (B\otimes C))\linpar D\kern-60pt}%
 {\partial'}{\partial}{\cdot\partial}%
 {\global\labelPosition={.45}\partial'\cdot}
 {\global\labelPosition={.5}\sim}
 \di{\hbox{$\circlearrowright$}}(35,25){0}
\end{diagramme}\hss}}$%
\cr
}

\endgroup

	\vskip2ex
        \noindent
 and their duals:

	\vskip2ex
        \noindent
\begingroup\offinterlineskip
\halign to\textwidth{\kern120pt\hfil #\hfil \tabskip1000pt plus100pt minus1000pt
 &\kern30pt\hfil #\hfil \kern120pt\tabskip0pt
\cr
$\vcenter{\footnotesize
\def\linpar{\mathbin{\tikz[inner sep=0] \node[rotate=180] at (0,0) {\footnotesize $\&$};}}
\def\slinpar{\mathbin{\tikz[inner sep=0] \node[rotate=180] at (0,0) {\tiny $\&$};}}
\hbox to0pt{\hss\begin{diagramme}
\optrianglediagabscoord{50}{30}%
 {A\otimes B}{\kern-20pt A\otimes(B\linpar \bot)}%
 {(A\otimes B)\linpar\bot\kern-20pt}%
 {\sim}{\sim}{\partial}
 \di{\hbox{$\circlearrowright$}}(25,12){0}
\end{diagramme}\hss}}$%
&$\kern40pt\vcenter{\footnotesize
\def\linpar{\mathbin{\tikz[inner sep=0] \node[rotate=180] at (0,0) {\footnotesize $\&$};}}
\def\slinpar{\mathbin{\tikz[inner sep=0] \node[rotate=180] at (0,0) {\tiny $\&$};}}
\hbox to0pt{\hss\begin{diagramme}
\oppentagondiagabscoord{70}{60}%
 {\kern-60pt A\otimes (B\otimes (C\linpar D))}%
 {A\otimes ((B\otimes C)\linpar D)\kern-60pt}%
 {\kern-40pt (A\otimes B)\otimes (C\linpar D)}%
 {(A\otimes (B\otimes C))\linpar D\kern-40pt}%
 {((A\otimes B)\otimes C)\linpar D}%
 {\cdot\partial}{\sim}{\partial}%
 {\global\labelPosition={.1}\partial}%
 {\global\labelPosition={.5}\sim}
 \di{\hbox{$\circlearrowright$}}(35,-25){0}
\end{diagramme}\hss}}$%
\cr
	\noalign{\vskip3ex}
$\vcenter{\footnotesize
\def\linpar{\mathbin{\tikz[inner sep=0] \node[rotate=180] at (0,0) {\footnotesize $\&$};}}
\def\slinpar{\mathbin{\tikz[inner sep=0] \node[rotate=180] at (0,0) {\tiny $\&$};}}
\hbox to0pt{\hss\begin{diagramme}
\oppentagondiagabscoord{70}{60}%
 {\kern-60pt A\otimes((B\linpar C)\otimes D)}%
 {(A\otimes(B\linpar C))\otimes D\kern-60pt}%
 {\kern-40pt A\otimes(B\linpar (C\otimes D))}%
 {((A\otimes B)\linpar C)\otimes D\kern-40pt}%
 {(A\otimes B)\linpar (C\otimes D)}%
 {\sim}{\cdot\partial'}{\partial\cdot}%
 {\global\labelPosition={.1}\partial}%
 {\global\labelPosition={.0}\partial'}
 \global\labelPosition={.5}
 \di{\hbox{$\circlearrowright$}}(35,-25){0}
\end{diagramme}\hss}}$%
\cr
}
\endgroup

	\vskip2ex
        \noindent
These three diagrams are obtained if we upend a transparent sheet
 on which the former three are written so that the back surface
 comes to the front, reverse the direction of the arrow,
 and interchange tensor and cotensor.

A $*$-autonomous category is a linearly distributive
 category equipped with a map $A\mapsto A^*$
 on objects as well as
 two families of morphisms: $\tau_A:{\bf 1}
 \rightarrow A\linpar A^*$ and
 $\gamma_A:A^*\otimes A\rightarrow\bot$.
We need to demand neither that $(\hbox{-})^*$ be
 functorial, nor that $\tau_A$ and $\gamma_A$ be natural~\cite{cose}.
We distinguish between $A^{**}$ and $A$, which are naturally isomorphic.
The coherence conditions for these are

	\vskip2ex

\begingroup\offinterlineskip
\halign to\textwidth{\kern120pt\hfil #\hfil \tabskip1000pt plus100pt minus1000pt
 &\kern30pt\hfil #\hfil \kern120pt\tabskip0pt
\cr
$\vcenter{\footnotesize
\def\linpar{\mathbin{\tikz[inner sep=0] \node[rotate=180] at (0,0) {\footnotesize $\&$};}}
\def\slinpar{\mathbin{\tikz[inner sep=0] \node[rotate=180] at (0,0) {\tiny $\&$};}}
\hbox to0pt{\hss\begin{diagramme}
\oppentagondiagabscoord{70}{60}%
 {\kern-5pt{\bf 1}\otimes A}{(A\linpar A^*)\otimes A\kern-35pt}{A}%
 {A\linpar (A^*\otimes A)\kern-15pt}{A\linpar \bot}%
 {\tau\otimes 1}{\sim}{\partial'}{\sim}{1\slinpar\gamma}
 \di{\hbox{$\circlearrowright$}}(35,-25){0}
\end{diagramme}\kern-16pt\hss}}$%
&$\vcenter{\footnotesize
\def\linpar{\mathbin{\tikz[inner sep=0] \node[rotate=180] at (0,0) {\footnotesize $\&$};}}
\def\slinpar{\mathbin{\tikz[inner sep=0] \node[rotate=180] at (0,0) {\tiny $\&$};}}
\hbox to0pt{\hss\begin{diagramme}
\oppentagondiagabscoord{70}{60}%
 {\kern-5ptA^*\otimes{\bf 1}}{A^*\otimes(A\linpar A^*)\kern-40pt}{A^*}%
 {(A^*\otimes A)\linpar A^*\kern-15pt}{\bot\linpar A^*}%
 {1\otimes\tau}{\sim}{\partial}{\sim}{\gamma\slinpar 1}
 \di{\hbox{$\circlearrowright$}}(35,-25){0}
\end{diagramme}\kern-16pt\hss}}$%
\cr
}
\endgroup
 
	\vskip2ex

The condition (ii) of Def.~\ref{wmz77} requires that $\mathord!:{\bf C}
 \rightarrow {\bf C}$ is
 a symmetric monoidal functor.
This means that the functor $\mathord!$ is
 equipped with
 a natural transformation $\tilde\varphi_{A,B}:\mathord!A
\otimes\mathord!B\rightarrow\mathord!(A\otimes B)$
 and a morphism $\varphi_0:{\bf 1}\rightarrow
 \mathord!{\bf 1}$.
The tilde is added so that these are
 dinstinguished if the subscripts are omitted.
The naught signifies it to be nullary.
The naturality of $\tilde\varphi$ is

	\vskip2ex

\begingroup\offinterlineskip
\halign to\textwidth{\kern120pt\hfil #\hfil \tabskip1000pt plus100pt minus1000pt
 &\kern30pt\hfil #\hfil \kern120pt\tabskip0pt
\cr
$\vcenter{\footnotesize
\hbox to0pt{\hss\begin{diagramme}
\squarediagabscoord{50}{40}%
  {\kern-5pt\mathord!A\otimes \mathord!B}{\mathord!(A\otimes B)\kern-5pt}
  {\kern-10pt\mathord!A'\otimes \mathord!B'}{\mathord!(A'\otimes B')\kern-10pt}
  {\tilde\varphi}{\mathord!f\otimes\mathord!g}%
  {\mathord!(f\otimes g)}{\tilde\varphi}
 \di{\hbox{$\circlearrowright$}}(25,20){0}
\end{diagramme}\hss}}$\kern-4pt
\cr
}
\endgroup
 
	\vskip2ex
        \noindent
 and the coherence conditions are

	\vskip2ex
\begingroup\offinterlineskip
\halign to\textwidth{\kern120pt\hfil #\hfil \tabskip1000pt plus100pt minus1000pt
 &\kern30pt\hfil #\hfil \kern120pt\tabskip0pt
\cr
$\vcenter{\footnotesize
 \hbox to0pt{\hss\begin{diagramme}
  \let\ss=\scriptstyle
  \latticeUnit=1pt \boxMargin=3pt
  \hexagondiagabscoord{80}{70}%
    {(\mathord!A\otimes \mathord!B)\otimes \mathord!C}%
    {\kern-10pt\mathord!A\otimes (\mathord!B\otimes \mathord!C)}%
    {\mathord!(A\otimes B)\otimes \mathord!C\kern-10pt}%
    {\kern-10pt\mathord!A\otimes \mathord!(B\otimes C)}
    {\mathord!((A\otimes B)\otimes C)\kern-10pt}%
    {\mathord!(A\otimes (B\otimes C))}%
     {\global\labelPosition={.9}\alpha}{\tilde\varphi\otimes 1}%
     {\global\labelPosition={.5}1\otimes \tilde\varphi}%
     {\tilde\varphi}%
     {\global\labelPosition={-.2}\tilde\varphi}{\mathord!\alpha}%
 \global\labelPosition={.5}%
 \di{\hbox{$\circlearrowright$}}(40,0){0}%
 \end{diagramme}\hss}}$
&$\vcenter{\footnotesize
\hbox to0pt{\hss\begin{diagramme}
\squarediagabscoord{50}{40}%
  {\kern-5pt\mathord!A\otimes \mathord!B}{\mathord!(A\otimes B)\kern-5pt}
  {\kern-5pt\mathord!B\otimes \mathord!A}{\mathord!(B\otimes A)\kern-5pt}%
  {\tilde\varphi}{\sigma}{\mathord!\sigma}{\tilde\varphi}
 \di{\hbox{$\circlearrowright$}}(25,20){0}
\end{diagramme}\kern-9pt\hss}}$
\cr
	\noalign{\vskip3ex}
$\vcenter{\footnotesize
\hbox to0pt{\hss\begin{diagramme}
\squarediagabscoord{50}{40}%
 {{\bf 1}\otimes\bang A}{\bang{\bf 1}\otimes\bang A}%
 {\bang A}{\bang({\bf 1}\otimes A)}%
 {\varphi_0\otimes 1}{\sim}{\tilde\varphi}{\sim}
 \di{\hbox{$\circlearrowright$}}(25,20){0}
\end{diagramme}\hss}}$%
\cr
}
\endgroup

	\vskip2ex
        \noindent
 where $\sim$ denotes an appropriate structural isomorphism.

The condition (iii) of the linear category is the requirement
 that the functor $\mathord!$ is endowed with a comonad structure.
Namely, two natural transformations $\delta_A:\mathord!A
 \rightarrow \mathord!\mathord!A$ and $\varepsilon_A:
 \mathord!A\rightarrow A$ are associated.
The naturality is

	\vskip2ex
\begingroup\offinterlineskip
\halign to\textwidth{\kern120pt\hfil #\hfil \tabskip1000pt plus100pt minus1000pt
 &\kern30pt\hfil #\hfil \kern120pt\tabskip0pt
\cr
$\vcenter{\footnotesize
\hbox to0pt{\hss\begin{diagramme}
\squarediagabscoord{50}{40}%
 {\mathord!A}{\mathord!B}{\mathord!\mathord!A}{\mathord!\mathord!B}%
 {\mathord!f}{\delta}{\delta}{\mathord!\mathord!f}
 \di{\hbox{$\circlearrowright$}}(25,20){0}
\end{diagramme}\hss}}$%
&$\vcenter{\footnotesize
\hbox to0pt{\hss\begin{diagramme}
\squarediagabscoord{50}{40}%
 {\mathord!A}{\mathord!B}{A}{B}%
 {\mathord!f}{\varepsilon}{\varepsilon}{f}
 \di{\hbox{$\circlearrowright$}}(25,20){0}
\end{diagramme}\hss}}$%
\cr
}
\endgroup

	\vskip2ex
        \noindent
 and the coherence conditions are

	\vskip2ex
\begingroup\offinterlineskip
\halign to\textwidth{\kern120pt\hfil #\hfil \tabskip1000pt plus100pt minus1000pt
 &\kern30pt\hfil #\hfil \kern120pt\tabskip0pt
\cr
$\vcenter{\footnotesize
\hbox to0pt{\hss\begin{diagramme}
\squarediagabscoord{50}{40}%
 {\mathord!A}{\mathord!\mathord!A}{\mathord!\mathord!A}%
 {\mathord!\mathord!\mathord!A}%
 {\delta}{\delta}{\delta}{\mathord!\delta}
 \di{\hbox{$\circlearrowright$}}(25,20){0}
\end{diagramme}\hss}}$%
&$\vcenter{\footnotesize
\hbox to0pt{\hss\begin{diagramme}
\trianglediagabscoord{50}{30}%
 {\mathord!A}{\mathord!\mathord!A}{\mathord!A}%
 {\delta}{1}{\varepsilon}%
 \di{\hbox{$\circlearrowright$}}(25,18){0}
\end{diagramme}\hss}}$%
\cr
	\noalign{\vskip5ex}
$\vcenter{\footnotesize
\hbox to0pt{\hss\begin{diagramme}
\trianglediagabscoord{50}{30}%
 {\mathord!A}{\mathord!\mathord!A}{\mathord!A}%
 {\delta}{1}{\mathord!\varepsilon}%
 \di{\hbox{$\circlearrowright$}}(25,18){0}
\end{diagramme}\hss}}$%
\cr
}
\endgroup

	\vskip2ex
        \noindent
Moreover, $\delta$ and $\varepsilon$ are required to be
 monoidal natural transformations.
In general, for monoidal functors $F$ and $G$,
 a natural transformation $\nu:F\rightarrow G$ is monoidal
 if it commutes with $(\tilde\varphi,\varphi_0)$ for
 $F$ and $G$ in an obvious sense.
Whenever $\mathord!$ is a monoidal functor, so is
 $\mathord!\mathord!$.
An identity functor is
 always a monoidal functor.
Hence, it makes sense to require
 $\delta:\mathord!\rightarrow
\mathord!\mathord!$ and $\varepsilon:\mathord!
\rightarrow {\it Id}$ to be monoidal.
This amounts to the following four diagrams:

	\vskip2ex
\begingroup\offinterlineskip
\halign to\textwidth{\kern120pt\hfil #\hfil \tabskip1000pt plus100pt minus1000pt
 &\kern30pt\hfil #\hfil \kern120pt\tabskip0pt
\cr
$\vcenter{\footnotesize
\hbox to0pt{\hss\begin{diagramme}
\pentagondiagabscoord{70}{60}%
 {\mathord!A\otimes\mathord!B}%
 {\mathord!\mathord!A\otimes \mathord!\mathord!B}%
 {\mathord!(A\otimes B)}%
 {\kern-20pt\mathord!(\mathord!A\otimes \mathord!B)}%
 {\mathord!\mathord!(A\otimes B)\kern-20pt}%
  {\delta\otimes\delta}{\tilde\varphi}{\tilde\varphi}
  {\delta}{\mathord!\tilde\varphi}%
 \di{\hbox{$\circlearrowright$}}(35,25){0}
\end{diagramme}\hss}}$%
&$\vcenter{\footnotesize
\hbox to0pt{\hss\begin{diagramme}
\trianglediagabscoord{50}{30}%
 {\kern-10pt\mathord!A\otimes \mathord!B}%
 {\mathord!(A\otimes B)\kern-10pt}{A\otimes B}%
 {\tilde\varphi}{\varepsilon\otimes\varepsilon}{\varepsilon}%
 \di{\hbox{$\circlearrowright$}}(25,18){0}
\end{diagramme}\hss}}$%
\cr
	\noalign{\vskip5ex}
$\vcenter{\footnotesize
\hbox to0pt{\hss\begin{diagramme}
\squarediagabscoord{50}{40}%
 {{\bf 1}}{\mathord!{\bf 1}}{\mathord!{\bf 1}}{\mathord!\mathord!{\bf 1}}%
 {\varphi_0}{\varphi_0}{\delta}{\mathord!\varphi_0}
 \di{\hbox{$\circlearrowright$}}(25,20){0}
\end{diagramme}\hss}}$%
&$\vcenter{\footnotesize
\hbox to0pt{\hss\begin{diagramme}
\trianglediagabscoord{50}{30}%
 {{\bf 1}}{\mathord!{\bf 1}}{\bf 1}%
 {\varphi_0}{1}{\varepsilon}
 \di{\hbox{$\circlearrowright$}}(25,18){0}
\end{diagramme}\hss}}$%
\cr
}
\endgroup

	\vskip2ex

The condition (iv) requires that
 the objects of the form $\mathord!A$ have the
 structure of commutative comonoids.
Namely, there are family of morphisms
 $d_A:\mathord!A\rightarrow \mathord!A\otimes \mathord!A$
 and $e_A:\mathord!A\rightarrow{\bf 1}$, rendering the
 following diagrams commutative:

	\vskip2ex
\begingroup\offinterlineskip
\halign to\textwidth{\kern120pt\hfil #\hfil \tabskip1000pt plus100pt minus1000pt
 &\kern30pt\hfil #\hfil \kern120pt\tabskip0pt
\cr
$\vcenter{\footnotesize
\hbox to0pt{\hss\begin{diagramme}
\pentagondiagabscoord{70}{60}%
 {\bang A}{\bang A\otimes \bang A}{\bang A\otimes \bang A}%
 {\kern-40pt(\bang A\otimes \bang A)\otimes \bang A}%
 {\bang A\otimes (\bang A\otimes \bang A)\kern-40pt}%
 dd{d\otimes 1}{1\otimes d}{\alpha}
 \di{\hbox{$\circlearrowright$}}(35,25){0}
\end{diagramme}\hss}}$%
&$\vcenter{\footnotesize
\hbox to0pt{\hss\begin{diagramme}
\trianglediagabscoord{50}{30}%
   {\mathord!A}{\mathord!A\otimes \mathord!A}{\mathord!A\otimes \mathord!A}%
   dd{\sigma}
 \di{\hbox{$\circlearrowright$}}(25,18){0}
\end{diagramme}\kern-16pt\hss}}$
\cr
	\noalign{\vskip3ex}
$\vcenter{\footnotesize
\hbox to0pt{\hss\kern12pt\begin{diagramme}
\trianglediagabscoord{50}{30}%
 {\mathord!A}{\mathord!A\otimes \mathord!A}{{\bf 1}\otimes \mathord!A}%
 {d}{\sim}{e\otimes 1}
 \di{\hbox{$\circlearrowright$}}(25,18){0}
\end{diagramme}\hss}}$%
\cr
}
\endgroup

	\vskip2ex
        \noindent
Moreover $d_A$ and $e_A$ must be monoidal natural
 transformations.
Consequently, $\mathord!f$ is always a comonoid morphism.
The naturality is given as

	\vskip2ex
\begingroup\offinterlineskip
\halign to\textwidth{\kern120pt\hfil #\hfil \tabskip1000pt plus100pt minus1000pt
 &\kern30pt\hfil #\hfil \kern120pt\tabskip0pt
\cr
$\vcenter{\footnotesize
\hbox to0pt{\hss\begin{diagramme}
\squarediagabscoord{50}{40}%
 {\mathord!A}{\mathord!B}{\kern-10pt\mathord!A\otimes \mathord!A}%
 {\mathord!B\otimes \mathord!B\kern-10pt}%
 {\mathord!f}{d}{d}{\mathord!f\otimes \mathord!f}
 \di{\hbox{$\circlearrowright$}}(25,20){0}
\end{diagramme}\hss}}$%
&$\vcenter{\footnotesize
\hbox to0pt{\hss\begin{diagramme}
\trianglediagabscoord{50}{30}%
 {\mathord!A}{\mathord!B}{{\bf 1}}%
 {\mathord!f}{e}{e}
 \di{\hbox{$\circlearrowright$}}(25,18){0}
\end{diagramme}\hss}}$%
\cr
}
\endgroup

	\vskip2ex
        \noindent
 and requiring them to be monoidal amounts to

	\vskip2ex
\begingroup\offinterlineskip
\halign to\textwidth{\kern120pt\hfil #\hfil \tabskip1000pt plus100pt minus1000pt
 &\kern30pt\hfil #\hfil \kern120pt\tabskip0pt
\cr
$\vcenter{\footnotesize
\hbox to0pt{\hss\begin{diagramme}
\pentagondiagabscoord{70}{60}%
 {\mathord!A\otimes \mathord!B}%
 {\kern-40pt
    (\mathord!A\otimes \mathord!A)\otimes (\mathord!B\otimes \mathord!B)}%
 {\mathord!(A\otimes B)}%
 {\kern-70pt
    (\mathord!A\otimes \mathord!B)\otimes (\mathord!A\otimes \mathord!B)}%
 {\mathord!(A\otimes B)\otimes \mathord!(A\otimes B)\kern-70pt}%
  {d\otimes d}{\tilde\varphi}{\sim}%
  {d}{\tilde\varphi\otimes\tilde\varphi}
 \di{\hbox{$\circlearrowright$}}(35,25){0}
\end{diagramme}\kern8pt\hss}}$%
&$\vcenter{\footnotesize
\hbox to0pt{\hss\begin{diagramme}
\squarediagabscoord{50}{40}%
  {\kern-5pt\mathord!A\otimes\mathord!B}%
  {\mathord!(A\otimes B)\kern-5pt}%
  {{\bf 1}\otimes{\bf 1}}{{\bf 1}}%
  {\tilde\varphi}{e\otimes e}{e}{\sim}
 \di{\hbox{$\circlearrowright$}}(25,20){0}
\end{diagramme}\hss}}$%
\cr
	\noalign{\vskip5ex}
$\vcenter{\footnotesize
\hbox to0pt{\hss\begin{diagramme}
\squarediagabscoord{50}{40}%
 {{\bf 1}}{\mathord!{\bf 1}}{{\bf 1}\otimes{\bf 1}}%
 {\bang{\bf 1}\otimes\bang{\bf 1}}
 {\varphi_0}{\sim}{d}{\varphi_0\otimes\varphi_0}
 \di{\hbox{$\circlearrowright$}}(25,20){0}
\end{diagramme}\hss}}$%
&$\vcenter{\footnotesize
\hbox to0pt{\hss\begin{diagramme}
\trianglediagabscoord{50}{30}%
  {{\bf 1}}{\mathord!{\bf 1}}{{\bf 1}}%
  {\varphi_0}{1}{e}
 \di{\hbox{$\circlearrowright$}}(25,18){0}
\end{diagramme}\hss}}$%
\cr
}
\endgroup

	\vskip2ex
        \noindent
Note that $A\mapsto \mathord!A\otimes\mathord!A$
 and the constant ${\bf 1}$ are
 monoidal functors.

The condition (v) is concerned with coalgebra morphisms.
The notion of coalgebras is associated with comonads.
If $X\mor f\mathord!X$ and
 $Y\mor g\mathord!Y$ are coalgebras, a coalgebra morphism between
 these is simply a morphism $X\mor kY$ satisfying
 $f;\mathord!k=k;g$.
The morphism $\mathord!A\mor{\delta_A}\mathord!\mathord!A$
 gives rise to a coalgebra, often called a free coalgebra.
Similarly $\mathord!A\otimes \mathord!A$ and ${\bf 1}$ have
 natural coalgebra structures on them.
The linear category requires that
 $d_A:\mathord!A\rightarrow \mathord!A\otimes \mathord!A$
 and $e_A:\mathord!A\rightarrow{\bf 1}$ are coalgebra
 morphisms.
Namely, the following diagrams are commutative:

	\vskip2ex
\begingroup\offinterlineskip
\halign to\textwidth{\kern120pt\hfil #\hfil \tabskip1000pt plus100pt minus1000pt
 &\kern30pt\hfil #\hfil \kern120pt\tabskip0pt
\cr
$\vcenter{\footnotesize
\hbox to0pt{\hss\begin{diagramme}
\pentagondiagabscoord{70}{60}%
 {\mathord!A}{\mathord!A\otimes \mathord!A}{\mathord!\mathord!A}%
 {\kern-20pt\mathord!\mathord!A\otimes \mathord!\mathord!A}%
 {\mathord!(\mathord!A\otimes \mathord!A)\kern-20pt}
 {d}{\delta}{\delta\otimes\delta}{\mathord!d}{\tilde\varphi} 
 \di{\hbox{$\circlearrowright$}}(35,25){0}
\end{diagramme}\kern12pt\hss}}$%
&$\vcenter{\footnotesize
\hbox to0pt{\hss\begin{diagramme}
\squarediagabscoord{50}{40}%
 {\mathord!A}{\mathord!\mathord!A}{\bf 1}{\mathord!{\bf 1}}%
 {\delta}{e}{\mathord!e}{\varphi_0}
 \di{\hbox{$\circlearrowright$}}(25,20){0}
\end{diagramme}\hss}}$%
\cr
}
\endgroup

	\vskip2ex

Finally, the condition (vi) says that
 $\delta_A:\mathord!A\rightarrow \mathord!
\mathord!A$ is a comonoid morphism.
Namely, it is commutative to the comonoid structure:

	\vskip2ex
\begingroup\offinterlineskip
\halign to\textwidth{\kern120pt\hfil #\hfil \tabskip1000pt plus100pt minus1000pt
 &\kern30pt\hfil #\hfil \kern120pt\tabskip0pt
\cr
$\vcenter{\footnotesize
\hbox to0pt{\hss\begin{diagramme}
 \squarediagabscoord{50}{40}%
 {\mathord!A}{\mathord!\mathord!A}%
 {\kern-5pt\mathord!A\otimes \mathord!A}%
 {\mathord!\mathord!A\otimes \mathord!\mathord!A\kern-5pt}%
 {\delta}{d}{d}{\delta \otimes \delta}
 \di{\hbox{$\circlearrowright$}}(25,20){0}
\end{diagramme}\hss}}$%
&$\vcenter{\footnotesize
\hbox to0pt{\hss\begin{diagramme}
\trianglediagabscoord{50}{30}%
 {\mathord!A}{\mathord!\mathord!A}{{\bf 1}}%
 {\delta}{e}{e}%
 \di{\hbox{$\circlearrowright$}}(25,18){0}
\end{diagramme}\hss}}$%
\cr
}
\endgroup

	\vskip2ex
        \noindent
As a consequence, every coalgebra morphism between free coalgebras
 turn out to be a comonoid morphism.

\section{A categorical reduction system}

Suppose that a set of atomic objects is given.
A {\it free} (intuitionistic or
 classical) linear category is naturally defined.
It is obtained by freely generating by the constructors
 and taking quotients with regard to
 the due conditions in Def.~\ref{wmz77}.
Since all conditions are given as commutative diagrams,
 taking quotients makes sense.

We regard the free linear category as a syntactic structure.
Our goal is to develop a dynamic calculus installed directly on the
 category.
In analogy to type systems, objects of the category correspond
 to types, while morphisms correspond to terms.
As type systems are designed as rewriting calculi of terms,
 our categorical system is realized as a rewriting calculus of morphisms.

If we consider the free intuitionistic linear category, the base of which is
 a symmetric monoidal closed
 category, the set of objects are generated by

	\vskip2ex
        \noindent\kern5em
$A\ \ \mathrel{::=}\ \ X\ \ |\ \ {\bf 1}\ \ |\ \ A\otimes A\ \ |\ \ A\multimap A
 \ \ |\ \ \mathord!A$

	\vskip2ex
        \noindent
 where $X$ ranges over a given set of atomic objects.
Atomic morphisms are identities, with {\it structural isomorphisms}:

	\vskip2ex
        \noindent\kern5em
\begingroup\small
$A\mor{1_A}A$
\endgroup

	\vskip2ex
        \noindent\kern5em
\begingroup\small
\vbox{\halign{&$#$\hfil\kern3em\cr
 (A\otimes B)\otimes C\lmor{\alpha_{ABC}}A\otimes (B\otimes C)
 & A\otimes (B\otimes C)\lmor{\alpha_{ABC}^{-1}}(A\otimes B)\otimes C\cr
 {\bf 1}\otimes A\lmor{\lambda_A}A
 & A\lmor{\lambda_A^{-1}}{\bf 1}\otimes A\cr
 A\otimes {\bf 1}\lmor{\rho_A}A
 & A\lmor{\rho_A^{-1}}A\otimes {\bf 1}\cr
 A\otimes B\lmor{\sigma_{AB}}B\otimes A
 & B\otimes A\lmor{\sigma_{AB}^{-1}}A\otimes B,\cr
}}
\endgroup

	\vskip2ex
        \noindent
 the units and the counits of the adjunction:

	\vskip2ex
        \noindent\kern5em
\begingroup\small
\vbox{\halign{&$#$\hfil\kern5em\cr
 A\lmor{{\rm abs}^B_A}B\multimap A\otimes B
 & (B\multimap A)\otimes B\lmor{{\rm ev}^B_A}A\cr
}}
\endgroup

	\vskip2ex
        \noindent
 together with the morphisms given in Def.~\ref{wmz77}:

	\vskip2ex
        \noindent\kern5em
\begingroup\small
\vbox{\halign{&$#$\hfil\kern5em\cr
 \mathord!A\otimes \mathord!B\lmor{\tilde\varphi_{AB}}\mathord!(A\otimes B)
 & \mathord!A\mor{\delta_A}\mathord!\mathord!A\cr
 {\bf 1}\mor{\varphi_0}\mathord!{\bf 1}
 & \mathord!A\mor{\varepsilon_A}A\cr
 & \mathord!A\mor{d_A}\mathord!A\otimes \mathord!A\cr
 & \mathord!A\mor{e_A}{\bf 1}\cr
}}
\endgroup

	\vskip2ex
        \noindent
For future reference, the last six atomic morphisms are called
 {\it algebraic morphisms}.
The set of morphisms is generated from the atomic morphisms by
 (diagramatic order) composition $f;g$
 and the functorial operations $f\otimes g,\>\mathord!f$,
 and $1_B\multimap f$.
Subscripts will often be omitted.
Our system will be designed so that subscripts have no significance.
It is analogous to ordinary type systems where rules depend only
 on the shape of terms, not on types.

We introduce a congruence relation over morphisms.
Two morphisms that are equivalent under this congruence
 are understood to be able to be rewritable from one another.
First, we have the axioms of categories and the elementary property
 of functors:

	\vskip2ex
        \noindent\kern5em
\begingroup\small
\vbox{\halign{&$#$\hfil\kern5em\cr
 f;1=f=1;f & F1=1\cr
 (f;g);h=f;(g;h) & F(f;g)=Ff;Fg\cr
}}
\endgroup

	\vskip2ex
        \noindent
 where $F$ is one of either $(\hbox{--})\otimes(\hbox{--}),\>B\multimap(\hbox{--})$,
 or $\mathord!(\hbox{--})$.
For the tensor product, we appropriately reform
 the equality as it is a 2-place functor.
Second, each structural isomorphism and its inverse are actual inverses:

	\vskip2ex
        \noindent\kern5em
\begingroup\small
\vbox{\halign{&$#$\hfil\kern5em\cr
 \alpha;\alpha^{-1}=1 & \alpha^{-1};\alpha=1 & \cdots\cr
}}
\endgroup

	\vskip2ex

Next, we consider the case where the base category is $\star$-autonomous.
Among several equivalent definitions known for $\star$-autonomous
 categories \cite{barr,hugh}, we adopt the one using the linearly
 distributive category \cite{cose}.
The set of objects is generated by

	\vskip2ex
        \noindent\kern5em
$A\ \ \mathrel{::=}\ \ X\ \ |\ \ {\bf 1}\ \ |\ \ \bot\ \ |
 \ \ A\otimes A\ \ |\ \ A\linpar A\ \ |\ \ A^*\ \ |\ \ \mathord!A$

	\vskip2ex
        \noindent
In addition to the structural isomorphisms for $\otimes$,
 we include the isomorphisms giving the symmetric monoidal structure
 on $\linpar$:

	\vskip2ex
        \noindent\kern5em
\begingroup\small
\def\linpar{\mathbin{\tikz[inner sep=0] \node[rotate=180] at (0,0) {\small $\&$};}}
\vbox{\halign{&$#$\hfil\kern3em\cr
 (A\linpar B)\linpar C\lmor{\bar\alpha_{ABC}}A\linpar (B\linpar C)
 & A\linpar (B\linpar C)\lmor{\bar\alpha_{ABC}^{-1}}(A\linpar B)\linpar C\cr
 \bot\linpar A\lmor{\bar\lambda_A}A
 & A\lmor{\bar\lambda_A^{-1}}\bot\linpar A\cr
 A\linpar \bot\lmor{\bar\rho_A}A
 & A\lmor{\bar\rho_A^{-1}}A\linpar \bot\cr
 A\linpar B\lmor{\bar\sigma_{AB}}B\linpar A
 & B\linpar A\lmor{\bar\sigma_{AB}^{-1}}A\linpar B\cr
}}
\endgroup

	\vskip2ex
        \noindent
Moreover the linear distribution morphisms

	\vskip2ex
        \noindent\kern5em
\begingroup\small
\def\linpar{\mathbin{\tikz[inner sep=0] \node[rotate=180] at (0,0) {\small $\&$};}}
\vbox{\halign{&$#$\hfil\kern5em\cr
 A\otimes (B\linpar C)\lmor{\partial_{ABC}}(A\otimes B)\linpar C\cr
}}
\endgroup

	\vskip2ex
        \noindent
 are added.
The morphisms ${\rm abs}$ and ${\rm ev}$ are removed.
In place, we add duality morphisms:

	\vskip2ex
        \noindent\kern5em
\begingroup\small
\def\linpar{\mathbin{\tikz[inner sep=0] \node[rotate=180] at (0,0) {\small $\&$};}}
\vbox{\halign{&$#$\hfil\kern5em\cr
 {\bf 1}\mor{\tau_A}A\linpar A^*
 & A^*\otimes A\mor{\gamma_A}\bot\cr
}}
\endgroup

	\vskip2ex
        \noindent
The six algebraic morphisms are the same.
The set of morphisms is generated from the atomic morphisms above by
 composition $f;g$ and the functorial operations $f\otimes g,\>f\linpar g$,
 and $\mathord!f$.
We note that $(\hbox{--})^*$ is not regarded as a contravariant functor \cite{cose}.
We distinguish $A^{**}$ from $A$.

The core of our calculus lies in the orientation
 of diagrams, which we shortly provide.
Among the commutative diagrams listed in the
 previous section, twenty-three diagrams are selected
 and reformed into rewriting rules.
The rest remain to be equivalences.
The orientation of rewriting is denoted by a double arrow.
We can rewrite only in the designated direction, whilst
 between equivalent morphisms we allow rewriting in either direction.
In other words, we give a rewriting system modulo congruence.
The selected diagrams comprise of one diagram in (ii) of Def.~\ref{wmz77},
 all of (iii), all save two diagrams of (iv), all of (v), and all of (vi).
Moreover, the adjoint triangles of monoidal closedness and the
 defining diagrams for $*$-autonomy are turned to rewriting rules.
Our tactic is to select diagrams as much as the resulting
 calculus keeps the desirable properties.
It is hopeless to reform all diagrams to rewriting rules.
For example, consider the following diagram,  which is one of (iv):
$$\vcenter{\footnotesize
\hbox{\hss\begin{diagramme}
\pentagondiagabscoord{70}{60}%
 {\bang A}{\bang A\otimes \bang A}{\bang A\otimes \bang A}%
 {\kern-40pt(\bang A\otimes \bang A)\otimes \bang A}%
 {\bang A\otimes (\bang A\otimes \bang A)\kern-40pt}%
 dd{d\otimes 1}{1\otimes d}{\alpha}
 \di{\hbox{$\circlearrowright$}}(35,25){0}
\end{diagramme}\hss}}$$
This diagram has symmetry.
Enforcing a rule so that 
only one-way rewriting is permitted
 would yield a useless calculus.

The following are the list of the twenty-three rewriting rules.
The first twenty-one diagrams are common for both the symmetric monoidal closed base
 and the $\star$-autonomous base.
The last two depend on the selected base.
In the diagrams, the label $\sim$ denotes appropriate structural isomorphisms
 and $f$ is an arbitrary morphism.

	\vskip2ex
\begingroup\offinterlineskip
\halign to\textwidth{\kern60pt\hfil #\hfil \tabskip1000pt plus100pt minus1000pt
 &\kern10pt$\cdots$\kern5pt{\footnotesize #}\kern20pt\tabskip0pt
 &\kern20pt\hfil #\hfil \tabskip1000pt plus100pt minus1000pt
 &\kern10pt$\cdots$\kern5pt{\footnotesize #}\tabskip0pt\cr
$\vcenter{\footnotesize
\hbox{\begin{diagramme}
\squarediagabscoord{50}{40}%
 {\mathord!A}{\mathord!\mathord!A}{\mathord!\mathord!A}%
 {\mathord!\mathord!\mathord!A}%
 {\delta}{\delta}{\delta}{\mathord!\delta}
 \di{\hbox{$\Rightarrow$}}(25,20){140}
\end{diagramme}}}$%
&(1)%
&$\vcenter{\footnotesize
\hbox{\begin{diagramme}
\trianglediagabscoord{50}{30}%
 {\mathord!A}{\mathord!\mathord!A}{\mathord!A}%
 {\delta}{1}{\varepsilon}%
 \di{\hbox{$\Rightarrow$}}(25,18){140}
\end{diagramme}}}$%
&(2)%
\cr
	\noalign{\vskip5ex}
$\vcenter{\footnotesize
\hbox{\begin{diagramme}
\trianglediagabscoord{50}{30}%
 {\mathord!A}{\mathord!\mathord!A}{\mathord!A}%
 {\delta}{1}{\mathord!\varepsilon}%
 \di{\hbox{$\Rightarrow$}}(25,18){140}
\end{diagramme}}}$%
&(3)%
&$\vcenter{\footnotesize
\hbox{\begin{diagramme}
 \squarediagabscoord{50}{40}%
 {\mathord!A}{\mathord!\mathord!A}%
 {\kern-5pt\mathord!A\otimes \mathord!A}%
 {\mathord!\mathord!A\otimes \mathord!\mathord!A\kern-5pt}%
 {\delta}{d}{d}{\delta \otimes \delta}
 \di{\hbox{$\Rightarrow$}}(25,20){140}
\end{diagramme}}}$%
&(4)%
\cr
	\noalign{\vskip5ex}
$\kern-9pt\vcenter{\footnotesize
\hbox{\begin{diagramme}
\pentagondiagabscoord{70}{60}%
 {\mathord!A}{\mathord!A\otimes \mathord!A}{\mathord!\mathord!A}%
 {\kern-20pt\mathord!\mathord!A\otimes \mathord!\mathord!A}%
 {\mathord!(\mathord!A\otimes \mathord!A)\kern-20pt}
 {d}{\delta}{\delta\otimes\delta}{\mathord!d}{\tilde\varphi} 
 \di{\hbox{$\Rightarrow$}}(35,25){140}
\end{diagramme}}}$%
&(5)%
&$\vcenter{\footnotesize
\hbox{\begin{diagramme}
\trianglediagabscoord{50}{30}%
 {\mathord!A}{\mathord!\mathord!A}{{\bf 1}}%
 {\delta}{e}{e}%
 \di{\hbox{$\Rightarrow$}}(25,18){140}
\end{diagramme}}}$%
&(6)%
\cr
	\noalign{\vskip5ex}
$\vcenter{\footnotesize
\hbox{\begin{diagramme}
\squarediagabscoord{50}{40}%
 {\mathord!A}{\mathord!\mathord!A}{\bf 1}{\mathord!{\bf 1}}%
 {\delta}{e}{\mathord!e}{\varphi_0}
 \di{\hbox{$\Rightarrow$}}(25,20){140}
\end{diagramme}}}$%
&(7)%
&$\kern11pt\vcenter{\footnotesize
\hbox{\begin{diagramme}
\trianglediagabscoord{50}{30}%
 {\mathord!A}{\mathord!A\otimes \mathord!A}{{\bf 1}\otimes \mathord!A}%
 {d}{\sim}{e\otimes 1}
 \di{\hbox{$\Rightarrow$}}(25,18){140}
\end{diagramme}}}$%
&(8)%
\cr
	\noalign{\vskip5ex}
$\vcenter{\footnotesize
\hbox{\begin{diagramme}
\pentagondiagabscoord{70}{60}%
 {\mathord!A\otimes\mathord!B}%
 {\mathord!\mathord!A\otimes \mathord!\mathord!B}%
 {\mathord!(A\otimes B)}%
 {\kern-20pt\mathord!(\mathord!A\otimes \mathord!B)}%
 {\mathord!\mathord!(A\otimes B)\kern-20pt}%
  {\delta\otimes\delta}{\tilde\varphi}{\tilde\varphi}
  {\delta}{\mathord!\tilde\varphi}%
 \di{\hbox{$\Rightarrow$}}(35,25){140}
\end{diagramme}}}$%
&(9)%
&$\vcenter{\footnotesize
\hbox{\begin{diagramme}
\trianglediagabscoord{50}{30}%
 {\kern-10pt\mathord!A\otimes \mathord!B}%
 {\mathord!(A\otimes B)\kern-10pt}{A\otimes B}%
 {\tilde\varphi}{\varepsilon\otimes\varepsilon}{\varepsilon}%
 \di{\hbox{$\Rightarrow$}}(25,18){140}
\end{diagramme}}}$%
&(10)%
\cr
	\noalign{\vskip5ex}
$\kern-8pt\vcenter{\footnotesize
\hbox{\begin{diagramme}
\pentagondiagabscoord{70}{60}%
 {\mathord!A\otimes \mathord!B}%
 {\kern-40pt
    (\mathord!A\otimes \mathord!A)\otimes (\mathord!B\otimes \mathord!B)}%
 {\mathord!(A\otimes B)}%
 {\kern-70pt
    (\mathord!A\otimes \mathord!B)\otimes (\mathord!A\otimes \mathord!B)}%
 {\mathord!(A\otimes B)\otimes \mathord!(A\otimes B)\kern-70pt}%
  {d\otimes d}{\tilde\varphi}{\sim}%
  {d}{\tilde\varphi\otimes\tilde\varphi}
 \di{\hbox{$\Rightarrow$}}(35,25){140}
\end{diagramme}}}$%
&(11)%
&$\kern5pt\vcenter{\footnotesize
\hbox{\begin{diagramme}
\squarediagabscoord{50}{40}%
  {\kern-5pt\mathord!A\otimes\mathord!B}%
  {\mathord!(A\otimes B)\kern-5pt}%
  {{\bf 1}\otimes{\bf 1}}{{\bf 1}}%
  {\tilde\varphi}{e\otimes e}{e}{\sim}
 \di{\hbox{$\Rightarrow$}}(25,20){140}
\end{diagramme}}}$%
&(12)%
\cr
	\noalign{\vskip5ex}
$\vcenter{\footnotesize
\hbox{\begin{diagramme}
\squarediagabscoord{50}{40}%
 {{\bf 1}}{\mathord!{\bf 1}}{\mathord!{\bf 1}}{\mathord!\mathord!{\bf 1}}%
 {\varphi_0}{\varphi_0}{\delta}{\mathord!\varphi_0}
 \di{\hbox{$\Rightarrow$}}(25,20){140}
\end{diagramme}}}$%
&(13)%
&$\vcenter{\footnotesize
\hbox{\begin{diagramme}
\trianglediagabscoord{50}{30}%
 {{\bf 1}}{\mathord!{\bf 1}}{\bf 1}%
 {\varphi_0}{1}{\varepsilon}
 \di{\hbox{$\Rightarrow$}}(25,18){140}
\end{diagramme}}}$%
&(14)%
\cr
	\noalign{\vskip5ex}
$\vcenter{\footnotesize
\hbox{\begin{diagramme}
\squarediagabscoord{50}{40}%
 {{\bf 1}}{\mathord!{\bf 1}}{{\bf 1}\otimes{\bf 1}}%
 {\bang{\bf 1}\otimes\bang{\bf 1}}
 {\varphi_0}{\sim}{d}{\varphi_0\otimes\varphi_0}
 \di{\hbox{$\Rightarrow$}}(25,20){140}
\end{diagramme}}}$%
&(15)%
&$\vcenter{\footnotesize
\hbox{\begin{diagramme}
\trianglediagabscoord{50}{30}%
  {{\bf 1}}{\mathord!{\bf 1}}{{\bf 1}}%
  {\varphi_0}{1}{e}
 \di{\hbox{$\Rightarrow$}}(25,18){140}
\end{diagramme}}}$%
&(16)%
\cr
	\noalign{\vskip5ex}
$\kern4pt\vcenter{\footnotesize
\hbox{\begin{diagramme}
\squarediagabscoord{50}{40}%
 {{\bf 1}\otimes\bang A}{\bang{\bf 1}\otimes\bang A}%
 {\bang A}{\bang({\bf 1}\otimes A)}%
 {\varphi_0\otimes 1}{\sim}{\tilde\varphi}{\sim}
 \di{\hbox{$\Rightarrow$}}(25,20){140}
\end{diagramme}}}$%
&(17)%
&$\vcenter{\footnotesize
\hbox{\begin{diagramme}
\squarediagabscoord{50}{40}%
 {\mathord!A}{\mathord!B}{\mathord!\mathord!A}{\mathord!\mathord!B}%
 {\mathord!f}{\delta}{\delta}{\mathord!\mathord!f}
 \di{\hbox{$\Rightarrow$}}(25,20){140}
\end{diagramme}}}$%
&(18)%
\cr
	\noalign{\vskip5ex}
$\vcenter{\footnotesize
\hbox{\begin{diagramme}
\squarediagabscoord{50}{40}%
 {\mathord!A}{\mathord!B}{A}{B}%
 {\mathord!f}{\varepsilon}{\varepsilon}{f}
 \di{\hbox{$\Rightarrow$}}(25,20){140}
\end{diagramme}}}$%
&(19)%
&$\vcenter{\footnotesize
\hbox{\begin{diagramme}
\squarediagabscoord{50}{40}%
 {\mathord!A}{\mathord!B}{\kern-10pt\mathord!A\otimes \mathord!A}%
 {\mathord!B\otimes \mathord!B\kern-10pt}%
 {\mathord!f}{d}{d}{\mathord!f\otimes \mathord!f}
 \di{\hbox{$\Rightarrow$}}(25,20){140}
\end{diagramme}}}$%
&(20)%
\cr
	\noalign{\vskip5ex}
$\vcenter{\footnotesize
\hbox{\begin{diagramme}
\trianglediagabscoord{50}{30}%
 {\mathord!A}{\mathord!B}{{\bf 1}}%
 {\mathord!f}{e}{e}
 \di{\hbox{$\Rightarrow$}}(25,18){140}
\end{diagramme}}}$%
&(21)%
\cr
}
\endgroup

	\vskip3ex
        \noindent
The remaining two rules are interchanged depending
 on which base category is adopted.
If we choose the symmetric monoidal closed category \cite{kemc},

	\vskip2ex
\begingroup\offinterlineskip
\halign to\textwidth{\kern40pt\hfil #\hfil \tabskip1000pt plus100pt minus1000pt
 &\kern10pt$\cdots$\kern5pt{\footnotesize #}\kern10pt\tabskip0pt
 &\kern10pt\hfil #\hfil \tabskip1000pt plus100pt minus1000pt
 &\kern10pt$\cdots$\kern5pt{\footnotesize #}\tabskip0pt\cr
$\kern5pt\vcenter{\footnotesize
\hbox{\begin{diagramme}
  \trianglediagabscoord{70}{35}%
   {A\otimes B}{(B\multimap A\otimes B)\otimes B\kern-40pt}{A\otimes B}%
   {{\rm abs}\otimes 1}{1}{{\rm ev}}
  \di{\hbox{$\Rightarrow$}}(35,20){140}
 \end{diagramme}}}\kern20pt$%
&(22)%
&$\kern5pt\vcenter{\footnotesize
\hbox{\begin{diagramme}
  \trianglediagabscoord{70}{35}%
   {B\multimap A}{B\multimap (B\multimap A)\otimes B\kern-40pt}{B\multimap A}%
   {{\rm abs}}{1}{1\multimap{\rm ev}}
  \di{\hbox{$\Rightarrow$}}(35,20){140}
 \end{diagramme}}}\kern20pt$%
&(23)%
\cr}
\endgroup

	\vskip3ex
        \noindent
If we select the $\star$-autonomous category we replace the above by
 the following two, where
 $\partial'_{ABC}:(A\linpar B)\otimes C\rightarrow A\linpar (B\otimes C)$ is
 induced from $\partial_{ABC}$ by the symmetry of tensor and cotensor,

	\vskip2ex
\begingroup\offinterlineskip
\halign to\textwidth{\kern15pt\hfil #\hfil \tabskip1000pt plus100pt minus1000pt
 &\kern10pt$\cdots$\kern5pt{\footnotesize #}\kern10pt\tabskip0pt
 &\kern10pt\hfil #\hfil \tabskip1000pt plus100pt minus1000pt
 &\kern10pt$\cdots$\kern5pt{\footnotesize #}\tabskip0pt\cr
$\kern30pt\vcenter{\footnotesize
\def\linpar{\mathbin{\tikz[inner sep=0] \node[rotate=180] at (0,0) {\footnotesize $\&$};}}
\def\slinpar{\mathbin{\tikz[inner sep=0] \node[rotate=180] at (0,0) {\tiny $\&$};}}
\hbox{\begin{diagramme}
\oppentagondiagabscoord{70}{60}%
 {\kern-5pt{\bf 1}\otimes A}{(A\linpar A^*)\otimes A\kern-35pt}{A}%
 {A\linpar (A^*\otimes A)\kern-15pt}{A\linpar \bot}%
 {\tau\otimes 1}{\sim}{\partial'}{\sim}{1\slinpar\gamma}
 \di{\hbox{$\Rightarrow$}}(35,-25){140}
\end{diagramme}}}\kern15pt$%
&(22)%
&$\kern10pt\vcenter{\footnotesize
\def\linpar{\mathbin{\tikz[inner sep=0] \node[rotate=180] at (0,0) {\footnotesize $\&$};}}
\def\slinpar{\mathbin{\tikz[inner sep=0] \node[rotate=180] at (0,0) {\tiny $\&$};}}
\hbox{\begin{diagramme}
\oppentagondiagabscoord{70}{60}%
 {\kern-5ptA^*\otimes{\bf 1}}{A^*\otimes(A\linpar A^*)\kern-40pt}{A^*}%
 {(A^*\otimes A)\linpar A^*\kern-15pt}{\bot\linpar A^*}%
 {1\otimes\tau}{\sim}{\partial}{\sim}{\gamma\slinpar 1}
 \di{\hbox{$\Rightarrow$}}(35,-25){140}
\end{diagramme}}}\kern15pt$%
&(23)%
\cr
}
\endgroup

	\vskip3ex
        \noindent
If we define $X\multimap Y$ with $Y\linpar X^*$,
 ${\rm abs}^B_A$ with $A\mor\sim A\otimes{\bf 1}\mor{\cdot\tau_B}A\otimes
 (B\linpar B^*)\mor\partial (A\otimes B)\linpar B^*$ and
 ${\rm ev}^B_A$ with $(A\linpar B^*)\otimes B\mor{\partial'}
 A\linpar (B^*\otimes B)\mor{\cdot\gamma_B}A\linpar \bot\mor\sim A$,
 then rule (22) and (23) for the symmetric monoidal closed base are
 a consequence of the corresponding rules for the $\star$-autonomous
 base.

In place of referring to the rules by numbers, we call them
 by the shape of their redexes.
For example, rule (1) is called $(\delta;\delta)$-type,
 rule (5) $(\delta;\mathord!d)$-type, and
 rule (17) $(\varphi_0;\tilde\varphi)$-type.
We call (1) through (7) collectively $\delta$-type
 as the redexes start with $\delta$.
Likewise we call (9) through (12) $\tilde\varphi$-type,
 and (13) through (17) $\varphi_0$-type.
If we collectively deal with (1)
 through (17) starting with one of $\delta,d,\tilde\varphi,
 \varphi_0$, we call a reduction in the
 group an {\it algebraic reduction}.
We call (18) through (21) {\it naturality reductions}\footnote{%
The naturality of $\tilde\varphi$ is taken to be equivalence.
We comment that if we turn the naturality into a
 rewriting rule in either orientation, the local confluence discussed
 in the next section fails.}.
Rule (22) and (23) are called $\beta$ and $\eta$ reductions
 respectively.

\section{Example: local confluence}

We give several examples of computation in our calculus.
We consider a few cases of local confluence.
Global confluence will be discussed in a forthcoming paper.

First, let us consider $\varphi_0;\delta;\varepsilon$.
If we contract $\delta;\varepsilon$ by rule (1), we obtain

	\vskip3ex
	\noindent\hfill
$$\vcenter{\footnotesize
\hbox{\begin{diagramme}
 \let\ss=\scriptstyle
 \latticeUnit=.84pt \boxMargin=3pt
 \object(0,80)={\bf 1}
 \object(0,0)={\bang{\bf 1}}
 \object(80,80)={\bang{\bf 1}}
 \object(80,0)={\bang\bang{\bf 1}}
 \morphism(0,80)to(80,80)[\varphi_0]
 \morphism(80,80)to(80,0)[\delta]
 \morphism(80,0)to(0,0)[\varepsilon]
 \morphism(80,80)to(0,0)[1]
 \morphism(0,80)to(0,0)[\varphi_0][R]
 \di{\hbox{$\Rightarrow$}}(55,25){180}
 \di{\hbox{$\circlearrowright$}}(25,55){0}
\end{diagramme}}}$$

	\vskip3ex
	\noindent
while, if we contract $\varphi_0;\delta$ by rule (13), then
 we have a sequence of contractions as follows:

$$\vcenter{\footnotesize
\hbox{\begin{diagramme}
 \let\ss=\scriptstyle
 \latticeUnit=.84pt \boxMargin=3pt
 \object(0,80)={\bf 1}
 \object(0,40)={\bf 1}
 \object(0,0)={\bang{\bf 1}}
 \object(40,40)={\bang{\bf 1}}
 \object(80,80)={\bang{\bf 1}}
 \object(80,0)={\bang\bang{\bf 1}}
 \morphism(0,80)to(80,80)[\varphi_0]
 \morphism(80,80)to(80,0)[\delta]
 \morphism(80,0)to(0,0)[\varepsilon]
 \morphism(0,80)to(40,40)[\varphi_0]
 \morphism(40,40)to(80,0)[\global\labelPosition{.2}\bang\varphi_0]
 \morphism(40,40)to(0,40)[\global\labelPosition{.5}\varepsilon]
 \morphism(0,80)to(0,40)[1][R]
 \morphism(0,40)to(0,0)[\varphi_0][R]
 \di{\hbox{$\Rightarrow$}}(55,55){180}
 \di{\hbox{$\Rightarrow$}}(12,52){180}
 \di{\hbox{$\Rightarrow$}}(25,20){180}
\end{diagramme}}}$$

	\vskip3ex
        \noindent
 where (14) and (19) are used in addition.
The leftmost vertical arrows in the two diagrams are
 both equal to $\varphi_0$.

Second, let us consider $\delta;\mathord!d;\delta$.
If we contract $\mathord!d;\delta$ first by a naturality reduction,

	\vskip3ex
	\noindent\hfill
$$\vcenter{\footnotesize
\hbox{\begin{diagramme}
 \let\ss=\scriptstyle
 \latticeUnit=.84pt \boxMargin=3pt
 \object(-38,92)={\bang A}
 \object(38,92)={\bang\bang A}
 \object(-92,38)={\bang A\otimes\bang A}
 \object(92,38)={\bang (\bang A\otimes\bang A)}
 \object(-92,-38)={\kern-10pt\bang\bang A\otimes \bang\bang A}
 \object(92,-38)={\bang\bang (\bang A\otimes\bang A)\kern-10pt}
 \object(-38,-92)={\kern-10pt\bang(\bang A\otimes\bang A)}
 \object(38,-92)={\bang (\bang\bang A\otimes \bang\bang A)\kern-10pt}
 \object(-20,0)={\bang\bang A}
 \object(34,10)={\bang\bang\bang A}
 \morphism(-38,92)to(38,92)[\delta]
 \morphism(-38,92)to(-92,38)[d][R]
 \morphism(38,92)to(92,38)[\bang d]
 \morphism(-92,38)to(-92,-38)[\delta\otimes\delta][R]
 \morphism(92,38)to(92,-38)[\delta]
 \morphism(-92,-38)to(-38,-92)[\tilde\varphi][R]
 \morphism(-38,-92)to(38,-92)[\bang(\delta\otimes\delta)][R]
 \morphism(38,-92)to(92,-38)[\bang\tilde\varphi][R]
 \morphism(-38,92)to(-20,0)[\delta]
 \morphism(-20,0)to(-38,-92)[\bang d]
 \morphism(38,92)to(34,10)[\delta]
 \morphism(-20,0)to(34,10)[\bang \delta]
 \morphism(34,10)to(92,-38)[\bang\bang d][R]
 \di{\hbox{$\Rightarrow$}}(6,44){140}
 \di{\hbox{$\Rightarrow$}}(65,12){140}
 \di{\hbox{$\Rightarrow$}}(-58,-3){140}
 \di{\hbox{$\Rightarrow$}}(22,-45){140}
\end{diagramme}}}$$\hfill\rlap{}

	\vskip3ex
	\noindent
If we contract $\delta;\mathord!d$ first,

	\vskip3ex
	\noindent\hfill
$\vcenter{\footnotesize
\hbox{\begin{diagramme}
 \let\ss=\scriptstyle
 \latticeUnit=.84pt \boxMargin=3pt
 \object(-38,92)={\bang A}
 \object(38,92)={\bang\bang A}
 \object(-92,38)={\bang A\otimes\bang A}
 \object(92,38)={\bang (\bang A\otimes\bang A)}
 \object(-92,-38)={\kern-10pt\bang\bang A\otimes \bang\bang A}
 \object(92,-38)={\bang\bang (\bang A\otimes\bang A)\kern-10pt}
 \object(-38,-92)={\kern-10pt\bang(\bang A\otimes\bang A)}
 \object(38,-92)={\bang (\bang\bang A\otimes \bang\bang A)\kern-10pt}
 \object(10,20)={\bang\bang A\otimes\bang\bang A}
 \object(0,-34)={\bang\bang\bang A\otimes\bang\bang\bang A}
 \morphism(-38,92)to(38,92)[\delta]
 \morphism(-38,92)to(-92,38)[d][R]
 \morphism(38,92)to(92,38)[\bang d]
 \morphism(-92,38)to(-92,-38)[\delta\otimes\delta][R]
 \morphism(92,38)to(92,-38)[\delta]
 \morphism(-92,-38)to(-38,-92)[\tilde\varphi][R]
 \morphism(-38,-92)to(38,-92)[\bang(\delta\otimes\delta)][R]
 \morphism(38,-92)to(92,-38)[\bang\tilde\varphi][R]
 \morphism(-92,38)to(10,20)[\delta\otimes\delta]
 \morphism(10,20)to(92,38)[\tilde\varphi][R]
 \morphism(10,20)to(0,-34)[\delta\otimes\delta]
 \morphism(-92,-38)to(0,-34)[\bang\delta\otimes\bang\delta][R]
 \morphism(0,-34)to(38,-92)[\tilde\varphi]
 \di{\hbox{$\Rightarrow$}}(5,60){140}
 \di{\hbox{$\Rightarrow$}}(55,-15){140}
 \di{\hbox{$\Rightarrow$}}(-42,-5){140}
 \di{\hbox{$\circlearrowright$}}(-18,-65){0}
\end{diagramme}}}$\hfill\rlap{}

	\vskip3ex
	\noindent
The next example of critical pairs starts with
 $\tilde\varphi;\delta;\bang d$.
If we contract $\delta;\mathord!d$ first,

	\vskip3ex
	\noindent\hfill
$\kern40pt\vcenter{\footnotesize
\hbox{\begin{diagramme}
 \let\ss=\scriptstyle
 \latticeUnit=.96pt \boxMargin=3pt
 \object(0,100)={\bang A\otimes \bang B}
 \object(-64,77)={\kern-25pt\bang A\otimes \bang A
    \otimes\bang B\otimes \bang B}
 \object(64,77)={\bang(A\otimes B)}
 \object(-98,17)={\kern-40pt\bang\bang A\otimes \bang\bang A
    \otimes\bang\bang B\otimes \bang\bang B}
 \object(98,17)={\bang\bang(A\otimes B)}
 \object(-86,-50)={\kern-40pt\bang\bang A\otimes \bang\bang B
    \otimes\bang\bang A\otimes \bang\bang B}
 \object(86,-50)={\bang(\bang (A\otimes B)
    \otimes \bang (A\otimes B))\kern-40pt}
 \object(-34,-94)={\kern-60pt\bang(\bang A\otimes\bang B)
    \otimes\bang(\bang A\otimes\bang B)}
 \object(34,-94)={\bang(\bang A\otimes \bang B
    \otimes\bang A\otimes \bang B)\kern-60pt}
 \object(-45,40)={\bang A\otimes \bang B
    \otimes\bang A\otimes \bang B\kern-30pt}
 \object(5,10)={\bang(A\otimes B)\otimes\bang(A\otimes B)\kern-10pt}
 \object(0,-25)={\bang\bang(A\otimes B)
    \otimes\bang\bang(A\otimes B)\kern-20pt}
 \morphism(0,100)to(-64,77)[d\otimes d][R]
 \morphism(0,100)to(64,77)[\tilde\varphi]
 \morphism(-64,77)to(-98,17)[\delta\otimes\delta\otimes\delta\otimes\delta][R]
 \morphism(64,77)to(98,17)[\delta]
 \morphism(-98,17)to(-86,-50)[\sim][R]
 \morphism(98,17)to(86,-50)[\bang d]
 \morphism(-86,-50)to(-34,-94)[\tilde\varphi\otimes\tilde\varphi][R]
 \morphism(34,-94)to(86,-50)[\bang(\tilde\varphi\otimes\tilde\varphi)][R]
 \morphism(-34,-94)to(34,-94)[\tilde\varphi][R]
 \morphism(-64,77)to(-45,40)[\sim]
 \morphism(64,77)to(5,10)[d]
 \morphism(-45,40)to(-86,-50)[\delta\otimes\delta\otimes\delta\otimes\delta]
 \morphism(-45,40)to(5,10)[\tilde\varphi\otimes\tilde\varphi]
 \morphism(5,10)to(0,-25)[\delta\otimes\delta]
 \morphism(0,-25)to(86,-50)[\tilde\varphi][R]
 \morphism(-34,-94)to(0,-25)[\bang\tilde\varphi\otimes\bang\tilde\varphi][R]
 \di{\hbox{$\Rightarrow$}}(60,-5){140}
 \di{\hbox{$\Rightarrow$}}(0,65){140}
 \di{\hbox{$\Rightarrow$}}(-40,-42){140}
 \di{\hbox{$\circlearrowright$}}(-80,-5){0}
 \di{\hbox{$\circlearrowright$}}(31,-65){0}
\end{diagramme}}}$\hfill\rlap{}

	\vskip3ex
	\noindent
If we contract $\tilde\varphi;\delta$ first

	\vskip3ex
	\noindent\hfill
$\kern40pt\vcenter{\footnotesize
\hbox{\begin{diagramme}
 \let\ss=\scriptstyle
 \latticeUnit=.96pt \boxMargin=3pt
 \object(0,100)={\bang A\otimes \bang B}
 \object(-64,77)={\kern-25pt\bang A\otimes \bang A
    \otimes\bang B\otimes \bang B}
 \object(64,77)={\bang(A\otimes B)}
 \object(-98,17)={\kern-40pt\bang\bang A\otimes \bang\bang A
    \otimes\bang\bang B\otimes \bang\bang B}
 \object(98,17)={\bang\bang(A\otimes B)}
 \object(-86,-50)={\kern-40pt\bang(\bang A\otimes \bang A)
    \otimes\bang(\bang B\otimes \bang B)}
 \object(86,-50)={\bang(\bang (A\otimes B)
    \otimes \bang (A\otimes B))\kern-40pt}
 \object(-34,-94)={\kern-60pt\bang(\bang A\otimes\bang A
    \otimes\bang B\otimes\bang B)}
 \object(34,-94)={\bang(\bang A\otimes\bang B
    \otimes\bang A\otimes\bang B)\kern-60pt}
 \object(-10,30)={\bang\bang A\otimes\bang\bang B}
 \object(10,-30)={\bang(\bang A\otimes \bang B)}
 \morphism(0,100)to(-64,77)[d\otimes d][R]
 \morphism(0,100)to(64,77)[\tilde\varphi]
 \morphism(-64,77)to(-98,17)[\delta\otimes\delta\otimes\delta\otimes\delta][R]
 \morphism(64,77)to(98,17)[\delta]
 \morphism(-98,17)to(-86,-50)[\tilde\varphi\otimes\tilde\varphi][R]
 \morphism(98,17)to(86,-50)[\bang d]
 \morphism(-86,-50)to(-34,-94)[\tilde\varphi][R]
 \morphism(34,-94)to(86,-50)[\bang(\tilde\varphi\otimes\tilde\varphi)][R]
 \morphism(-34,-94)to(34,-94)[\sim][R]
 \morphism(0,100)to(-10,30)[\delta\otimes\delta]
 \morphism(-10,30)to(10,-30)[\tilde\varphi]
 \morphism(-10,30)to(-86,-50)[\bang d\otimes\bang d]
 \morphism(10,-30)to(98,17)[\bang\tilde\varphi]
 \morphism(10,-30)to(-34,-94)[\bang(d\otimes d)]
 \di{\hbox{$\Rightarrow$}}(42,35){140}
 \di{\hbox{$\Rightarrow$}}(32,-52){140}
 \di{\hbox{$\circlearrowright$}}(-35,-45){0}
 \di{\hbox{$\Rightarrow$}}(-50,35){140}
\end{diagramme}}}$\hfill\rlap{}

	\vskip2ex
        \noindent
The obtained sequences of morphisms are not exactly equal.
By the coherence theorem of symmetric monoidal functors, however,
 they are equivalent.

\section{Comparison to a type theory}\label{soe19}

We briefly discuss the relation of our calculus
 to a type system of intuitionistic linear logic.
It is intended
 to justify the design of the calculus.
The following comparison shows that
 our categorical calculus is a refinement of a term calculus.
Furthermore, it explains why twenty-three diagrams are
 oriented in that way.

We use the dual intuitionistic linear logic due to Barber \cite{barb},
 modified slightly.
The modality $\mathord!A$ is used to decorate types while
 $\mathop\sharp M$ is used instead to decorate
 terms\footnote{%
Simply for immediate viewability.
Barber uses $\mathord!$ for both.}.
In the type environment, we use
 $\mathop\sharp x\mathbin:A$ and $x\mathbin:A$ to distinguish
 the intuitionistic part and the linear part, instead of
 partitioning by a semicolon as in the original.
The modality $\sharp$ means that $x$ is in the intuitionistic part.
So an environment $\Gamma$ is a finite sequence of 
 $\mathop\sharp x_i\mathbin:A_i$ or $x_i\mathbin:A_i$ where
 the order has no significance.
In the original system the intuitionistic part
 is strictly separated from the linear part.
Instead, we use lifting to change the modality of a variable:

	\vskip2ex
	\noindent\kern5em
\vtop{\small\offinterlineskip
\halign{\strut $#$\hfil\cr
 \Gamma,\>x\mathop:A\vdash M\mathop:B\cr
 \noalign{\hrule}
 \Gamma,\>\sharp x\mathop:A\vdash M\mathop:B\cr}}

	\vskip2ex
        \noindent
Accordingly, we limit the axiom to the shape
 of $x\mathbin:A\vdash x\mathbin:A$.
Weakening and contraction are given explicitly:

	\vskip2ex
	\noindent\kern5em
\vtop{\small\offinterlineskip
\halign{\strut $#$\hfil\cr
 \Gamma\vdash M\mathop:B\cr
 \noalign{\hrule}
 \Gamma,\>\sharp x\mathop:A\vdash M\mathop:B\cr}}\kern5em
\vtop{\small\offinterlineskip
\halign{\strut $#$\hfil\cr
 \Gamma,\>\sharp x'\mathop:A,\>\sharp x''\mathop:A\vdash M\mathop:B\cr
 \noalign{\hrule}
 \Gamma,\>\sharp x\mathop:A\vdash M[x/x'x'']\mathop:B\cr}}

	\vskip2ex
	\noindent
In the contraction rule $M[x/x'x'']$ denotes the operation to
 substitute $x$ simultaneously for $x'$ and $x''$.

For terms we use postfix notation $M\{\mathop\sharp x\mapsto N\}$
 in place of the prefix let-operator.
It replaces ${\sf let}\ \mathord!x\ {\sf be}\ N\ {\sf in}\ M$
 in Barber's system.
The $\beta$-rule for $\sharp$ is given as
 $M\{\mathop\sharp x\mapsto\mathop\sharp N\}
 \Rightarrow M[N/x]$, and the $\eta$-rule
 as $\mathop\sharp x\{
 \mathop\sharp x\mapsto M\}\Rightarrow M$.

The type system is interpreted in the free intuitionistic
 linear category in a standard way.
A type judgement $\Gamma\vdash M\mathbin:B$ corresponds
 to a morphism $f:\Gamma\rightarrow B$ where
 $\Gamma$ denotes the sequence of
 $\mathord!A_i$ or $A_i$ connected by $\otimes$.
If the type environment contains $\mathop\sharp x_i\mathbin:A_i$
 we use $\mathord!A_i$, and if it contains $x_i\mathbin:A_i$
 we use $A_i$.
Here we associate a morphism with a derivation tree, rather
 than with a term,
It is necessary for a fine analysis of the relation between
 the type system and our calculus.

First, let us justify the $\beta$-reduction for $\sharp$:

	\vskip2ex
	\noindent\kern5em
\begingroup\small
$\vcenter{\offinterlineskip
 \halign{\strut\hfil $#$\hfil\cr
  \vbox{\halign{\strut\hfil $#$\hfil &&\qquad \hfil $#$\hfil\cr
   \vbox{\halign{\strut\hfil $#$\hfil\cr
    \vdots\raise5pt\rlap{\scriptsize$\pi_1$}\cr
    \Gamma,\,\sharp x\mathbin:A\ \vdash\ M\mathbin:B\cr}}%
   &
   \vbox{\halign{\strut\hfil $#$\hfil\cr
    \vdots\raise5pt\rlap{\scriptsize$\pi_2$}\cr
    \sharp\Delta\ \vdash\ K\mathbin:A\cr
    \noalign{\hrule}
    \sharp\Delta\ \vdash\ \sharp K\mathbin:\mathord!A\cr}}\cr}}\cr
  \noalign{\hrule}
  \Gamma,\,\sharp\Delta\ \vdash\ M\{\sharp x\mapsto\sharp K\}:B\cr}}
\kern3em\Rightarrow\kern3em
\vcenter{\offinterlineskip
 \halign{\strut\hfil $#$\hfil\cr
  \vdots\raise5pt\rlap{\scriptsize$\pi_1{\sharp^x}\pi_2$}\cr
  \Gamma,\,\sharp\Delta\ \vdash\ M[K/x]\mathbin:B\cr}}$
\endgroup

	\vskip2ex
	\noindent
 where $\mathop\sharp\Delta$ denotes that all type assignments
 in the environment have the shape of $\mathop\sharp x_i:A_i$.
The derivation $\pi_1{\sharp^x}\pi_2$ is obtained by connecting $\pi_2$
 at the place of the axiom involving $x$ in $\pi_1$.

We split cases according to the last rule involving
 the variable $x$ in $\pi_1$.
If the last inference is lifting:

	\vskip2ex
	\noindent\kern5em
\begingroup\small
\vtop{\offinterlineskip
\halign{\strut\hfil $#$\hfil\cr
 \vdots\raise5pt\rlap{\scriptsize $\rho_1$}\cr
 \Gamma,\,x\mathbin:A\ \vdash\ M\mathbin:B\cr
 \noalign{\hrule}
 \Gamma,\,\sharp x\mathbin:A\ \vdash\ M\mathbin:B\cr}}
\endgroup

	\vskip2ex
	\noindent
 then let $\rho_1$ be interpreted by $\Gamma\otimes A\mor fB$
 and $\pi_2$ by $\mathord!\Delta \mor hA$.
For example, if $\Delta$ consists of two types $C_1$ and $C_2$,
 the derivation before rewriting is interpreted as
 the counterclockwise sequence of arrows from
$\Gamma\otimes\mathord!C_1\otimes\mathord!C_2$ to $B$ in
 the following diagram, and the one after reduction
 is the other extreme going clockwise.
The contraction of derivations is
 realized by computation in our calculus as
$$\vcenter{\footnotesize
\hbox{\begin{diagramme}
 \let\ss=\scriptstyle
 \latticeUnit=1.1pt \boxMargin=3pt
 \object(0,-80)={\Gamma\otimes \bang A}
 \object(50,-40)={\Gamma\otimes A}
 \object(100,-40)={B}
 \object(0,-40)={\kern-30pt\Gamma\otimes \bang(\bang C_1\otimes \bang C_2)}
 \object(50,0)={\Gamma\otimes \bang C_1\otimes\bang C_2\kern-30pt}
 \object(0,0)={\kern-25pt\Gamma\otimes\bang\bang C_1\otimes\bang\bang C_2}
 \object(0,40)={\kern-20pt\Gamma\otimes\bang C_1\otimes\bang C_2}
 \morphism(0,-80)to(50,-40)[\cdot\varepsilon][R]
 \morphism(50,-40)to(100,-40)[f][R]
 \morphism(0,-40)to(50,0)[\cdot\varepsilon][R]
 \morphism(0,-40)to(0,-80)[\cdot\bang h][R]
 \morphism(50,0)to(50,-40)[\cdot h][L]
 \morphism(0,0)to(0,-40)[\cdot\tilde \varphi][R]
 \morphism(0,0)to(50,0)[\cdot\varepsilon\varepsilon][R]
 \morphism(0,40)to(0,0)[\cdot\delta\delta][R]
 \morphism(0,40)to(50,0)[1][L]
 \di{\hbox{$\Rightarrow$}}(27,-43){-40}
 \di{\hbox{$\Rightarrow$}}(20,-16){-40}
 \di{\hbox{$\Rightarrow$}}(20,10){-40}
\end{diagramme}}}$$
The rules used here are (2), (10), and (19).
If $\Delta$ is empty, rule (14) is used since $\varphi_0$ is
 employed in place of $\tilde\varphi$.
If the last inference of $\pi_1$ is contraction:

	\vskip2ex
	\noindent\kern5em
\begingroup\small
\vtop{\offinterlineskip
\halign{\strut\hfil $#$\hfil\cr
 \vdots\raise5pt\rlap{\scriptsize $\rho_1$}\cr
 \Gamma,\,\sharp x'\mathbin:A,\,\sharp x''\mathbin:A\ \vdash\ M\mathbin:B\cr
 \noalign{\hrule}
 \Gamma,\,\sharp x\mathbin:A\ \vdash\ M[x/x'x'']\mathbin:B\cr}}
\endgroup

	\vskip2ex
	\noindent
 a similar analysis shows that the
 rules (4), (11), (15), and (20) are utilized.
If the last inference of $\pi_1$ is weakening

	\vskip2ex
	\noindent\kern5em
\begingroup\small
\vtop{\offinterlineskip
\halign{\strut\hfil $#$\hfil\cr
 \vdots\raise5pt\rlap{\scriptsize $\rho_1$}\cr
 \Gamma\ \vdash\ M\mathbin:B\cr
 \noalign{\hrule}
 \Gamma,\,\sharp x\mathbin:A\ \vdash\ M\mathbin:B\cr}}
\endgroup

	\vskip2ex
	\noindent
 the rules (6), (12), (16), and (21) are used.
If the last inference of $\pi_1$ is $\sharp$-introduction,

	\vskip2ex
	\noindent\kern5em
\begingroup\small
\vtop{\offinterlineskip
\halign{\strut\hfil $#$\hfil\cr
 \vdots\raise5pt\rlap{\scriptsize $\rho_1$}\cr
 \sharp\Gamma,\,\sharp x\mathbin:A\ \vdash\ M\mathbin:B\cr
 \noalign{\hrule}
 \sharp\Gamma,\,\sharp x\mathbin:A\ \vdash\ \sharp M\mathbin:\mathord!B\cr}}
\endgroup

	\vskip2ex
	\noindent
 the rules (1), (9), (13), and (18) are used.
When $\Delta$ is empty, rule (17) is also used.

Second, we consider the $\eta$-reduction for $\sharp$ modality:

	\vskip2ex
	\noindent\kern5em
\begingroup\small
$\vcenter{%
 \halign{\strut\hfil $#$\hfil\cr
  \vbox{\halign{\strut\hfil $#$\hfil &&\qquad \hfil $#$\hfil\cr
   \vbox{\halign{\strut\hfil $#$\hfil\cr
    \noalign{\hrule}
    x\mathbin:A\ \vdash x\mathbin:A\cr
    \noalign{\hrule}
    \sharp x\mathbin:A\ \vdash\ x\mathbin:A\cr
    \noalign{\hrule}
    \sharp x\mathbin:A\ \vdash\ \sharp x\mathbin:\mathord!A\cr}}%
   &
   \vbox{\halign{\strut\hfil $#$\hfil\cr
    \vdots\raise5pt\rlap{\scriptsize $\pi$}\cr
    \Delta\ \vdash\ K\mathbin:\mathord!A\cr}}\cr}}\cr
  \noalign{\hrule}
  \Delta\ \vdash\ \sharp x\{\sharp x\mapsto K\}\mathbin:\mathord!A\cr}}
 \qquad\Rightarrow\qquad
\vcenter{\halign{\strut\hfil $#$\hfil\cr
 \vdots\raise5pt\rlap{\scriptsize $\pi$}\cr
 \Delta\ \vdash\ K\mathbin:\mathord!A\cr}}
$
\endgroup

	\vskip2ex
	\noindent
The left-hand side is interpreted by
$\Delta\mor h\mathord!A\mor{\delta}
\mathord!\mathord!A\mor{\mathord!\varepsilon}
\mathord!A$, and
 rule (3) is used.

If the type assignment introduced by weakening is deleted by
 contraction, both are superfluous.
Hence we can have the following simplifying rule:

	\vskip2ex
	\noindent\kern5em
\begingroup\small
$\vcenter{\halign{\strut\hfil $#$\hfil\cr
 \vdots\cr
 \Gamma,\,\sharp x'\mathbin:A\ \vdash M\mathbin:B\cr
 \noalign{\hrule}
 \Gamma,\,\sharp x'\mathbin:A,\,\sharp x''\mathbin:A\ \vdash\ M\mathbin:B\cr
 \noalign{\hrule}
 \Gamma,\,\sharp x\mathbin:A\ \vdash\ M[x/x']\mathbin:B\cr}}
 \qquad\Rightarrow\qquad
\vcenter{\halign{\strut\hfil $#$\hfil\cr
 \vdots\cr
 \Gamma,\,\sharp x\mathbin:A\ \vdash M[x/x']\mathbin:B\cr}}$
\endgroup

	\vskip2ex
	\noindent
 where rule (8) is used.

The rest are the interchanging rule of the $\sharp$ modality with
 weakening and contraction:

	\vskip2ex
	\noindent\kern5em
\begingroup\small
$\vcenter{\halign{\strut\hfil $#$\hfil\cr
 \vdots\cr
 \sharp \Gamma,\,\sharp x'\mathbin:A,\,\sharp x''\mathbin:A\ \vdash\ M\mathbin:B\cr
 \noalign{\hrule}
 \sharp \Gamma,\,\sharp x\mathbin:A\ \vdash\ M[x/x'x'']\mathbin:B\cr
 \noalign{\hrule}
 \sharp \Gamma,\,\sharp x\mathbin:A\ \vdash\ \sharp (M[x/x'x''])\mathbin:\mathord!B\cr}}
 \qquad\Rightarrow\qquad
\vcenter{\halign{\strut\hfil $#$\hfil\cr
 \vdots\cr
 \sharp \Gamma,\,\sharp x'\mathbin:A,\,\sharp x''\mathbin:A\ \vdash\ M\mathbin:B\cr
 \noalign{\hrule}
 \sharp \Gamma,\,\sharp x'\mathbin:A,\,\sharp x''\mathbin:A\ \vdash\sharp M\mathbin:\mathord!B\cr
 \noalign{\hrule}
 \sharp \Gamma,\,\sharp x\mathbin:A\ \vdash\ (\sharp M)[x/x'x'']\mathbin:\mathord!B\cr}}
$
\endgroup

	\vskip2ex
	\noindent\kern5em
\begingroup\small
$\vcenter{\halign{\strut\hfil $#$\hfil\cr
 \vdots\cr
 \sharp \Gamma\ \vdash\ M  \mathbin:B\cr
 \noalign{\hrule}
 \sharp \Gamma,\,\sharp x\mathbin:A\ \vdash\ M\mathbin:B\cr
 \noalign{\hrule}
 \sharp \Gamma,\,\sharp x\mathbin:A\ \vdash\ \sharp M\mathbin:\mathord!B\cr}}
 \qquad\Rightarrow\qquad
\vcenter{\halign{\strut\hfil $#$\hfil\cr
 \vdots\cr
 \sharp \Gamma\ \vdash\ M  \mathbin:B\cr
 \noalign{\hrule}
 \sharp\Gamma\ \vdash\ \sharp M\mathbin:\mathord!B\cr
 \noalign{\hrule}
 \sharp \Gamma,\,\sharp x\mathbin:A\ \vdash\ \sharp M\mathbin:\mathord!B\cr}}
$
\endgroup

	\vskip2ex
	\noindent
The former uses rule (5) and the latter uses rules (7)
 and (17).

Finally the $\beta$-reduction for abstraction
 $(\lambda x.\,M)K\Rightarrow M[K/x]$ corresponds to rule (22)
 and the $\eta$-reduction $\lambda x.\,Mx\Rightarrow M$ to rule (23).

Every rule, save rule (17), is used exactly once, as observed from the
 analysis above.
Each rule has an intrinsic role.
The rewriting orientation of the diagrams is determined so that
 the contractions of derivation trees are simulated by
 our calculus.
In addition, a single rewriting step of terms is realized by
 several steps of categorical rewriting.
Therefore we conclude that the categorical calculus is a refinement
 of the term calculus.
Furthermore, as mentioned in the introduction, we permit the
 rewriting $\mathord!(f;g)\rightsquigarrow \mathord!f
 ;\mathord!g$.
So our system incarnates a mechanism to decompose a term and substitute only a subterm
 obtained by decomposition.
Our calculus is also a refinement in this sense.

\section{Normalizability}

We show the weak termination of the categorical reduction system.
Hereafter we consider the system with base a $\star$-autonomous category.
A reason for the choice is that naturality of
${\rm abs}_A$ and ${\rm ev}_A$ in the symmetric monoidal
 closed category is awkward and cumbersome to handle.
Moreover, the latter is simulated by the former.

\begin{definition}\label{gbh44}\rm
A {\it normal form} is a morphism that is equivalent to the shape that has
 no redexes.
\end{definition}

	\vskip1ex

A redex may, however, be created from none as a consequence of congruence.
For example, the obvious normal
 form $\mathord!A\mor{\varepsilon_A}A$ is, by composing
 $1_{\mathord!A}=\mathord!1_A$, equivalent to
 $\mathord!A\mor{\mathord!1_A}\mathord!A\mor{\varepsilon_A}A$ that has a
 naturality redex.

	\vskip1ex

\begin{definition}\label{mmu38}\rm
A {\it reversible} reduction is one of the naturality rules
 (18) through (21) where $f$ is an identity, a structural isomorphism or
 its inverse, or their compositions.
\end{definition}

	\vskip1ex
        
We can cancel reversible reductions.
For example, suppose that $\mathord!A\mor{\mathord!f}\mathord!B\mor{d_B}
\mathord!B\otimes \mathord!B$ is contracted to $\mathord!A\mor{d_A}\mathord!A
\otimes \mathord!A\mor{\mathord!f\otimes \mathord!f}
\mathord!B\otimes \mathord!B$.
Then, by attaching $1_{\mathord!A}=\mathord!f;\mathord!f^{-1}$ in front and
 transferring $\mathord!f^{-1}$ by naturality, we restore a morphism
 that is equivalent to the original.
Reversible reductions are regarded to be inessential.

	\vskip1ex

\begin{lemma}\label{qxw29}
Only reversible reductions occur in a reduction sequence from a normal form.
\end{lemma}

\proof
The morphisms equivalent to identities are written as
 the composition of structural isomorphisms.
\endproof

	\vskip1ex
        \noindent
Once a morphism reaches a normal form,
 we can have only inessential reductions afterwards.
In the following, we ignore reversible redexes.
We assume they are removed by contraction implicitly.

	\vskip1ex

\begin{definition}\label{ftu80}\rm
A morphism (weakly) {\it terminates} if there
 is a finite reduction sequence ending with a normal
 form\footnote{%
In this paper, strong termination scarcely occurs.
Hence we omit ``weakly'' for simplicity.}.
\end{definition}

	\vskip0ex

We are not motivated by constructing a graph reduction system,
 yet it is helpful to use graphs to avoid a nuisance
 incurred by structural isomorphisms and their coherences.
In this paper, we only modestly use graphs, that
 are introduced informally to enhance intuitive understanding.
To discuss confluence in a forthcoming paper, we will rely on full
 graphical visualization.
We will not intend to construct a graph reduction system, though.

As in \cite{bcst}, we represent a morphism
$f:A_1\otimes A_2\otimes\cdots\otimes A_m
 \rightarrow B_1\linpar B_2\linpar\cdots\linpar B_n$ in the $\star$-autonomous
 category as a figure
$$
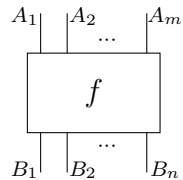
\begin{tikzpicture}[xscale=0.0352778, yscale=0.0352778, thin, inner sep=0]
  \draw (25,-15) rectangle (-25,15);
  \draw (-10,15) -- (-10,30)  (-20,15) -- (-20,30)  (20,15) -- (20,30);
  \draw (-10,-15) -- (-10,-30)  (-20,-15) -- (-20,-30)  (20,-15) -- (20,-30);
  \node at (0,0) {$f$};
  \node at (-26,29) {$\scriptstyle A_1$};
  \node at (-4,29) {$\scriptstyle A_2$};
  \node at (27,29) {$\scriptstyle A_m$};
  \node at(5,20) {$\scriptstyle \cdots$};
  \node at (-26,-29) {$\scriptstyle B_1$};
  \node at (-4,-29) {$\scriptstyle B_2$};
  \node at (27,-29) {$\scriptstyle B_n$};
  \node at(5,-20) {$\scriptstyle \cdots$};
\end{tikzpicture}
$$

	\vskip2ex
        \noindent
Each of tensor and cotensor has two gates

	\vskip2ex
        \noindent
\centerline{%
$\vcenter
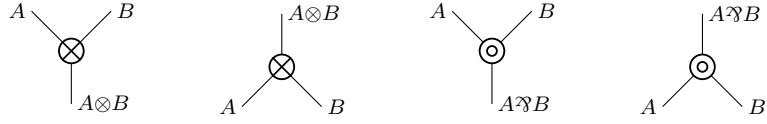
{\hbox{\begin{tikzpicture}[xscale=0.0352778, yscale=0.0352778, thin, inner sep=0]
  \def\p#1#2{%
    \ifcase #1
      \or \ifx#2x  0       \else  0 \fi
    \fi}
  \def\x#1{\p#1x}
  \def\y#1{\p#1y}
  \path[use as bounding box] (26,17) rectangle (-26,-23);
  \draw (\x1,\y1) -- (\x1-15,\y1+15);
  \draw (\x1,\y1) -- (\x1+15,\y1+15);
  \draw (\x1,\y1) -- (\x1,\y1-20);
  \putTensor(\x1,\y1)
  \node[left] at (\x1-17,\y1+15) {$\scriptstyle A$};
  \node[right] at (\x1+17,\y1+15) {$\scriptstyle B$};
  \node[right] at (\x1+2,\y1-20) {$\scriptstyle A\otimes B$};
\end{tikzpicture}}}$
\qquad
$\vcenter{\hbox{\begin{tikzpicture}[xscale=0.0352778, yscale=0.0352778, thin, inner sep=0]
  \def\p#1#2{%
    \ifcase #1
      \or \ifx#2x  0       \else  0 \fi
    \fi}
  \def\x#1{\p#1x}
  \def\y#1{\p#1y}
  \path[use as bounding box] (26,23) rectangle (-26,-17);
  \draw (\x1,\y1) -- (\x1-15,\y1-15);
  \draw (\x1,\y1) -- (\x1+15,\y1-15);
  \draw (\x1,\y1) -- (\x1,\y1+20);
  \putTensor(\x1,\y1)
  \node[left] at (\x1-17,\y1-15) {$\scriptstyle A$};
  \node[right] at (\x1+17,\y1-15) {$\scriptstyle B$};
  \node[right] at (\x1+2,\y1+20) {$\scriptstyle A\otimes B$};
\end{tikzpicture}}}$
\qquad
$\vcenter{\hbox{\begin{tikzpicture}[xscale=0.0352778, yscale=0.0352778, thin, inner sep=0]
  \def\p#1#2{%
    \ifcase #1
      \or \ifx#2x  0       \else  0 \fi
    \fi}
  \def\x#1{\p#1x}
  \def\y#1{\p#1y}
  \path[use as bounding box] (26,17) rectangle (-26,-23);
  \draw (\x1,\y1) -- (\x1-15,\y1+15);
  \draw (\x1,\y1) -- (\x1+15,\y1+15);
  \draw (\x1,\y1) -- (\x1,\y1-20);
  \putPar(\x1,\y1)
  \node[left] at (\x1-17,\y1+15) {$\scriptstyle A$};
  \node[right] at (\x1+17,\y1+15) {$\scriptstyle B$};
  \node[right] at (\x1+2,\y1-20) {$\scriptstyle A\slinpar B$};
\end{tikzpicture}}}$
\qquad
$\vcenter{\hbox{\begin{tikzpicture}[xscale=0.0352778, yscale=0.0352778, thin, inner sep=0]
  \def\p#1#2{%
    \ifcase #1
      \or \ifx#2x  0       \else  0 \fi
    \fi}
  \def\x#1{\p#1x}
  \def\y#1{\p#1y}
  \path[use as bounding box] (26,23) rectangle (-26,-17);
  \draw (\x1,\y1) -- (\x1-15,\y1-15);
  \draw (\x1,\y1) -- (\x1+15,\y1-15);
  \draw (\x1,\y1) -- (\x1,\y1+20);
  \putPar(\x1,\y1)
  \node[left] at (\x1-17,\y1-15) {$\scriptstyle A$};
  \node[right] at (\x1+17,\y1-15) {$\scriptstyle B$};
  \node[right] at (\x1+2,\y1+20) {$\scriptstyle A\slinpar B$};
\end{tikzpicture}}}$%
}

	\vskip2ex
        \noindent
We use a double circle in place of $\linpar$ as the latter symbol is
 not symmetric under a vertical flip.
The linear distribution morphism
 $\partial:A\otimes (B\linpar C)\rightarrow (A\otimes
 B)\linpar C$, for example, corresponds to
$$
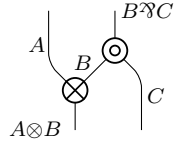
\begin{tikzpicture}[xscale=0.0352778, yscale=0.0352778, thin, inner sep=0]
  \def\p#1#2{%
    \ifcase #1
      \or \ifx#2x  0       \else  0 \fi
      \or \ifx#2x  \x1     \else  \y1+22 \fi
      \or \ifx#2x  \x2+15  \else  \y2+15 \fi
      \or \ifx#2x  \x3+10  \else  \y1 \fi
      \or \ifx#2x  \x3     \else  \y3+22 \fi
      \or \ifx#2x  \x2-10  \else  \y5 \fi
    \fi}
  \def\x#1{\p#1x}
  \def\y#1{\p#1y}
 \path[use as bounding box] (25,65) rectangle (-10,0);
 \draw (\x1,\y1+7) -- (\x2,\y2);
 \draw (\x2,\y2) -- (\x3,\y3);
 \draw (\x3,\y3) -- (\x5,\y5-7);
 \draw (\x2,\y2)[rounded corners=4] -- (\x6,\y2+10) -- (\x6,\y6-7);
 \draw (\x3,\y3)[rounded corners=4] -- (\x4,\y3-10) -- (\x4,\y4+7);
 \putTensor(\x2,\y2)
 \putPar(\x3,\y3)
 \node[left] at (\x6-1, \y6-20) {$\scriptstyle A$};
 \node[left] at ($ .5*(\x2,\y2)+.5*(\x3,\y3)+(-1,3) $) {$\scriptstyle B$};
 \node[right] at (\x4+2, \y4+20) {$\scriptstyle C$};
 \node[right] at (\x3+2, \y3+15) {$\scriptstyle B\slinpar C$};
 \node[left] at (\x2-5, \y2-15) {$\scriptstyle A\otimes B$};
\end{tikzpicture}
$$
Duality morphisms $\tau_A$ and $\gamma_A$ are represented by bends:

	\vskip2ex
        \noindent
\centerline{%
$\vcenter
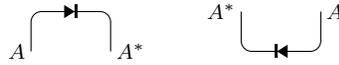
{\hbox{\begin{tikzpicture}[xscale=0.0352778, yscale=0.0352778, thin, inner sep=0]
  \def\p#1#2{%
    \ifcase #1
      \or \ifx#2x  0       \else  0 \fi
    \fi}
  \def\x#1{\p#1x}
  \def\y#1{\p#1y}
  \path[use as bounding box] (26,8) rectangle (-26,-23);
  \draw (\x1-15,\y1-15)[rounded corners=4] -- (\x1-15,\y1) -- (\x1+15,\y1) -- (\x1+15,\y1-15);
  \putRightDiode(\x1,\x2)
  \node[left] at (\x1-17,\y1-15) {$\scriptstyle A$};
  \node[right] at (\x1+17,\y1-15) {$\scriptstyle A^*$};
\end{tikzpicture}}}$
\qquad
$\vcenter{\hbox{\begin{tikzpicture}[xscale=0.0352778, yscale=0.0352778, thin, inner sep=0]
  \def\p#1#2{%
    \ifcase #1
      \or \ifx#2x  0       \else  0 \fi
    \fi}
  \def\x#1{\p#1x}
  \def\y#1{\p#1y}
  \path[use as bounding box] (-26,-8) rectangle (26,23);
  \draw (\x1-15,\y1+15)[rounded corners=4] -- (\x1-15,\y1) -- (\x1+15,\y1) -- (\x1+15,\y1+15);
  \putLeftDiode(\x1,\x2)
  \node[left] at (\x1-17,\y1+15) {$\scriptstyle A^*$};
  \node[right] at (\x1+17,\y1+15) {$\scriptstyle A$};
\end{tikzpicture}}}$%
}

	\vskip2ex
        \noindent
We add a diode-like symbol to signify which side has the duality star,
 so that it is restored if the labels attached to wires are omitted.
The duality $\beta\eta$-reduction corresponds to the operation
 straightening double bends:

	\vskip2ex
        \noindent\kern5em
$\vcenter
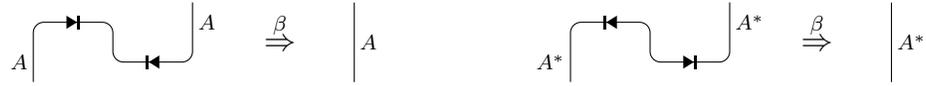
{\hbox{\begin{tikzpicture}[xscale=0.0352778, yscale=0.0352778, thin, inner sep=0]
  \def\p#1#2{%
    \ifcase #1
      \or \ifx#2x  0       \else  0 \fi
      \or \ifx#2x  \x1+30  \else  \y1-15 \fi
    \fi}
  \def\x#1{\p#1x}
  \def\y#1{\p#1y}
  \path[use as bounding box] (57,8) rectangle (-27,-23);
  \draw (\x1-15,\y2-7.5)[rounded corners=4] -- (\x1-15,\y1) -- (\x1+15,\y1)
     -- (\x1+15,\y2) -- (\x2+15,\y2) -- (\x2+15,\y1+7.5);
  \putRightDiode(\x1,\y1)
  \putLeftDiode(\x2,\y2)
  \node[left] at (\x1-17,\y2) {$\scriptstyle A$};
  \node[right] at (\x2+17,\y1) {$\scriptstyle A$};
\end{tikzpicture}}}
\quad\twomor\beta\quad
\vcenter{\hbox{\begin{tikzpicture}[xscale=0.0352778, yscale=0.0352778, thin, inner sep=0]
  \def\p#1#2{%
    \ifcase #1
      \or \ifx#2x  0       \else  0 \fi
      \or \ifx#2x  \x1     \else  \y1-30 \fi
    \fi}
  \def\x#1{\p#1x}
  \def\y#1{\p#1y}
  \path[use as bounding box] (-7,5) rectangle (7,-35);
  \draw (\x1,\y1) -- (\x2,\y2);
  \node[right] at ($ .5*(\x1,\y1)+.5*(\x1,\y2)+(2,0) $) {$\scriptstyle A$};
\end{tikzpicture}}}$
	\kern5em
$\vcenter{\hbox{\begin{tikzpicture}[xscale=0.0352778, yscale=0.0352778, thin, inner sep=0]
  \def\p#1#2{%
    \ifcase #1
      \or \ifx#2x  0       \else  0 \fi
      \or \ifx#2x  \x1+30  \else  \y1-15 \fi
    \fi}
  \def\x#1{\p#1x}
  \def\y#1{\p#1y}
  \path[use as bounding box] (57,8) rectangle (-27,-23);
  \draw (\x1-15,\y2-7.5)[rounded corners=4] -- (\x1-15,\y1) -- (\x1+15,\y1)
     -- (\x1+15,\y2) -- (\x2+15,\y2) -- (\x2+15,\y1+7.5);
  \putLeftDiode(\x1,\y1)
  \putRightDiode(\x2,\y2)
  \node[left] at (\x1-17,\y2) {$\scriptstyle A^*$};
  \node[right] at (\x2+17,\y1) {$\scriptstyle A^*$};
\end{tikzpicture}}}
\quad\twomor\beta\quad
\vcenter{\hbox{\begin{tikzpicture}[xscale=0.0352778, yscale=0.0352778, thin, inner sep=0]
  \def\p#1#2{%
    \ifcase #1
      \or \ifx#2x  0       \else  0 \fi
      \or \ifx#2x  \x1     \else  \y1-30 \fi
    \fi}
  \def\x#1{\p#1x}
  \def\y#1{\p#1y}
  \path[use as bounding box] (-7,5) rectangle (7,-35);
  \draw (\x1,\y1) -- (\x2,\y2);
  \node[right] at ($ .5*(\x1,\y1)+.5*(\x1,\y2)+(2,0) $) {$\scriptstyle A^*$};
\end{tikzpicture}}}$

	\vskip2ex
        \noindent
For $f:A\rightarrow B$, its dual $f^*:B^*\rightarrow A^*$ is
 depicted as
$$
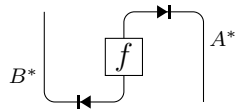
\vcenter{\hbox{\begin{tikzpicture}[xscale=0.0352778, yscale=0.0352778, thin, inner sep=0]
  \def\p#1#2{%
    \ifcase #1
      \or \ifx#2x  0       \else  0 \fi
      \or \ifx#2x  \x1+15  \else  \y1+17 \fi
      \or \ifx#2x  \x1-15  \else  \y1-17 \fi
    \fi}
  \def\x#1{\p#1x}
  \def\y#1{\p#1y}
  \path[use as bounding box] (43,25) rectangle (-43,-25);
  \draw (\x1,\y1+7)[rounded corners=4] -- (\x1,\y2) -- (\x1+30,\y2) -- (\x1+30,\y3);
  \draw (\x1,\y1-7)[rounded corners=4] -- (\x1,\y3) -- (\x1-30,\y3) -- (\x1-30,\y2);
  \draw (\x1+7,\y1+7) rectangle (\x1-7,\y1-7);
  \putRightDiode(\x2,\y2)
  \putLeftDiode(\x3,\y3)
  \node at (\x1,\y1) {$f$};
  \node[right] at (\x1+32, \y1+7) {$\scriptstyle A^*$};
  \node[left] at (\x1-32, \y1-7) {$\scriptstyle B^*$};
\end{tikzpicture}}}
$$

	\vskip2ex

The verification of the weak termination is based on the
 standard reducibility method.
The following proof strategy is inspired by \cite{gira}.

	\vskip1ex

\begin{definition}\rm\label{rby51}
A {\it positive funnel} on object $A$ is a set ${\cal S}$ of morphisms
 $X\mor fA$ for varied $X$ satisfying the following
 two conditions:

	\vskip.5ex
        \hangafter0\hangindent2em
        \noindent
\llapem2{(i)}%
The identity $A\mor{1}A$ is a member of ${\cal S}$.

	\vskip0ex
        \noindent
\llapem2{(ii)}%
Each $f\in {\cal S}$ terminates.

	\vskip.5ex
        \hangafter0\hangindent0pt
        \noindent
A {\it negative funnel} on $A$ is a set ${\cal S}$ of morphisms
 $A\mor fX$ for varied $X$ satisfying the same conditions
 (i) and (ii).

	\vskip.5ex
        \hangafter0\hangindent0em
        \noindent
When we simply say a funnel, it
 means either a positive funnel or a negative funnel.
\end{definition}

	\vskip1ex

\begin{definition}\rm\label{jbv29}
Given an object $A$, let
 ${\cal S}$ be a set of morphisms of the form $X
 \mor fA$ (resp.~$A\mor fX$).
The {\it complement} ${\cal S}^\bot$ is
 the set of all morphisms $A\mor gY$
 (resp.~$Y\mor gA$) subject to the condition that
 $f;g$ (resp.~$g;f$) terminates for all $f\in {\cal S}$.
\end{definition}

	\vskip0ex

\begin{lemma}\label{rnr91}
If ${\cal S}$ is a positive (negative) funnel on $A$,
 the complement ${\cal S}^\bot$ is a negative (resp.~positive) funnel.
\end{lemma}

\proof
Condition (i) of ${\cal S}^\bot$ follows from (ii) of ${\cal S}$.
Condition (ii) of ${\cal S}^\bot$ follows from (i) of ${\cal S}$.
\endproof

	\vskip1ex

\begin{lemma}\label{rsh62}
Let ${\cal R}$ and ${\cal S}$ be a set of morphisms of the form $X
 \mor fA$ or of the form $A\mor fX$.

	\vskip.5ex
        \hangafter0\hangindent2em
        \noindent
\llapem2{(i)}%
If ${\cal R}\subseteq {\cal S}$, then
 ${\cal S}^\bot\subseteq {\cal R}^\bot$.

	\vskip0ex
        \noindent
\llapem2{(ii)}%
${\cal S}\subseteq {\cal S}^{\bot\bot}$.

	\vskip0ex
        \noindent
\llapem2{(iii)}%
${\cal S}^\bot={\cal S}^{\bot\bot\bot}$.
	\par
\end{lemma}

\proof
Standard.
\endproof

	\vskip1ex

\begin{definition}\rm\label{yjd37}
We define sets ${\cal R}\otimes{\cal S}$, ${\cal R}\linpar
 {\cal S}$, $\mathord!{\cal S}$, and ${\cal S}^*$
 for positive (or negative) funnels ${\cal R}$ and ${\cal S}$.

	\vskip2ex
        \noindent\kern5em
\vbox{\halign{$#$\hfil &${}\ =\ #$\hfil\cr
 {\cal R}\otimes {\cal S}
 & \{f\otimes g\,|\>f\in {\cal R},\;g\in {\cal S}\}\cr
 {\cal R}\linpar {\cal S}
 & \{f\linpar g\,|\>f\in {\cal R},\;g\in {\cal S}\}\cr
 \mathord!{\cal S} & \{\mathord!f\,|\>f\in {\cal S}\}\cr
 {\cal S}^* & \{f^*\,|\>f\in {\cal S}\}.\cr
}}

	\vskip2ex
        \noindent
The first three sets
 ${\cal R}\otimes {\cal S}$, ${\cal R}\linpar {\cal S}$,
 and $\mathord!{\cal S}$ are clearly positive (or negative) funnels.
However, the last set ${\cal S}^*$ is not a funnel.
In fact, identity $1_{A^*}$ is not a member of ${\cal S}^*$,
 since it does not equal $(1_A)^*$, as $(\hbox{-})^*$ is not
 a contravariant functor.
\end{definition}

	\vskip1ex

\begin{lemma}\label{oou03}
Let $A_1\otimes A_2\otimes\cdots\otimes A_m
 \mor fB_1\linpar B_2\linpar\cdots\linpar B_n$ be
 a morphism.
Then

	\vskip2ex
        \noindent\kern5em
$\vcenter
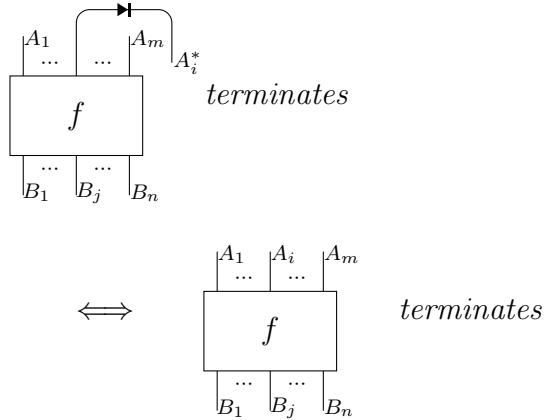
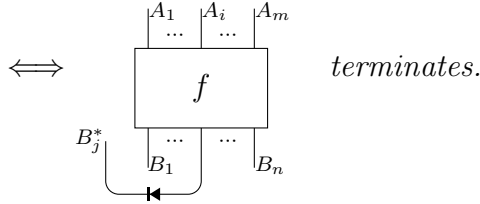
{\hbox{\begin{tikzpicture}[xscale=0.0352778, yscale=0.0352778, thin, inner sep=0]
  \path[use as bounding box] (44,45) rectangle (-24,-30);
  \draw (25,-15) -- (25,15) -- (-25,15) -- (-25,-15) -- cycle;
  \draw (0,15)[rounded corners=5] -- (0,40) -- (36,40) -- (36,20);
  \putRightDiode(18,40)
  \draw (-20,15) --  (-20,30);
  \draw (20,15) -- (20,30);
  \draw (0,-15) -- (0,-30);
  \draw (-20,-15) -- (-20,-30);
  \draw (20,-15) -- (20,-30);
  \node at (0,0) {$f$};
  \node at (-14.5,29) {$\scriptstyle A_1$};
  \node at (42,20) {$\scriptstyle A_i^*$};
  \node at (27,29) {$\scriptstyle A_m$};
  \node at (-10,20) {$\scriptstyle \cdots$};
  \node at (10,20) {$\scriptstyle \cdots$};
  \node at (-15,-29) {$\scriptstyle B_1$};
  \node at (5,-29) {$\scriptstyle B_j$};
  \node at (26,-29) {$\scriptstyle B_n$};
  \node at (-10,-20) {$\scriptstyle \cdots$};
  \node at (10,-20) {$\scriptstyle \cdots$};
\end{tikzpicture}}}$ terminates

	\vskip1ex
        \noindent\kern5em
$\qquad\Longleftrightarrow\qquad
\vcenter{\hbox{\begin{tikzpicture}[xscale=0.0352778, yscale=0.0352778, thin, inner sep=0]
  \path[use as bounding box] (44,45) rectangle (-24,-30);
  \draw (25,-15) -- (25,15) -- (-25,15) -- (-25,-15) -- cycle;
  \draw (0,15) -- (0,30);
  \draw (-20,15) --  (-20,30);
  \draw (20,15) -- (20,30);
  \draw (0,-15) -- (0,-30);
  \draw (-20,-15) -- (-20,-30);
  \draw (20,-15) -- (20,-30);
  \node at (0,0) {$f$};
  \node at (-14.5,29) {$\scriptstyle A_1$};
  \node at (5,29) {$\scriptstyle A_i$};
  \node at (27,29) {$\scriptstyle A_m$};
  \node at (-10,20) {$\scriptstyle \cdots$};
  \node at (10,20) {$\scriptstyle \cdots$};
  \node at (-15,-29) {$\scriptstyle B_1$};
  \node at (5,-29) {$\scriptstyle B_j$};
  \node at (26,-29) {$\scriptstyle B_n$};
  \node at (-10,-20) {$\scriptstyle \cdots$};
  \node at (10,-20) {$\scriptstyle \cdots$};
\end{tikzpicture}}}$ terminates

	\vskip1ex
        \noindent\kern5em
$\qquad\Longleftrightarrow\qquad
\vcenter{\hbox{\begin{tikzpicture}[xscale=0.0352778, yscale=0.0352778, thin, inner sep=0]
  \path[use as bounding box] (44,45) rectangle (-24,-30);
  \putLeftDiode(-18,-40)
  \draw (25,-15) -- (25,15) -- (-25,15) -- (-25,-15) -- cycle;
  \draw (0,15) -- (0,30);
  \draw (-20,15) --  (-20,30);
  \draw (20,15) -- (20,30);
  \draw (0,-15)[rounded corners=5] -- (0,-40) -- (-36,-40) -- (-36,-20);
  \draw (-20,-15) -- (-20,-30);
  \draw (20,-15) -- (20,-30);
  \node at (0,0) {$f$};
  \node at (-14.5,29) {$\scriptstyle A_1$};
  \node at (5,29) {$\scriptstyle A_i$};
  \node at (27,29) {$\scriptstyle A_m$};
  \node at (-10,20) {$\scriptstyle \cdots$};
  \node at (10,20) {$\scriptstyle \cdots$};
  \node at (-15,-29) {$\scriptstyle B_1$};
  \node at (-42,-20) {$\scriptstyle B_j^*$};
  \node at (26,-29) {$\scriptstyle B_n$};
  \node at (-10,-20) {$\scriptstyle \cdots$};
  \node at (10,-20) {$\scriptstyle \cdots$};
\end{tikzpicture}}}$ terminates.
\end{lemma}

	\vskip2ex

\proof
As they are symmetric, we will verify only the first equivalence.
Suppose that $f$ with a bend terminates.
If it leads to a normal form where the bend is intact,
 there is an obvious terminating reduction sequence from $f$.
Otherwise, the bend vanishes by $\eta$-reduction as in

	\vskip2ex
        \noindent\kern5em
$\vcenter
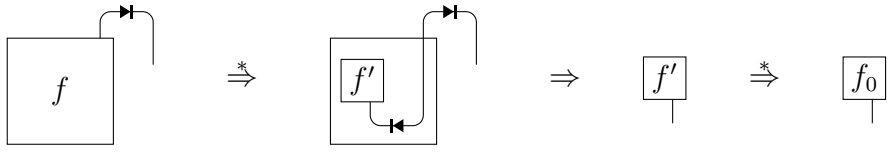
{\hbox{\begin{tikzpicture}[xscale=0.0352778, yscale=0.0352778, thin, inner sep=0]
  \def\p#1#2{%
    \ifcase #1
      \or \ifx#2x  0       \else  0 \fi
      \or \ifx#2x  \x1+25  \else  \y1+30 \fi
    \fi}
  \def\x#1{\p#1x}
  \def\y#1{\p#1y}
  \path[use as bounding box] (39,35) rectangle (-20,-25);
  \draw (\x1+20,\y1+20) rectangle (\x1-20,\y1-20);
  \draw (\x1+15,\y1+20)[rounded corners=4] -- (\x1+15,\y2) -- (\x2+10,\y2)
     -- (\x2+10,\y2-20);
  \putRightDiode(\x2,\y2)
  \node at (\x1,\y1){$f$};
\end{tikzpicture}}}$%
	\qquad $\twomor*$\qquad
$\vcenter{\hbox{\begin{tikzpicture}[xscale=0.0352778, yscale=0.0352778, thin, inner sep=0]
  \def\p#1#2{%
    \ifcase #1
      \or \ifx#2x  0       \else  0 \fi
      \or \ifx#2x  \x1+25  \else  \y1+30 \fi
      \or \ifx#2x  \x1+5   \else  \y1-13 \fi
      \or \ifx#2x  \x1-8   \else  \y1+4 \fi
    \fi}
  \def\x#1{\p#1x}
  \def\y#1{\p#1y}
  \path[use as bounding box] (39,35) rectangle (-24,-25);
  \draw (\x1+20,\y1+20) rectangle (\x1-20,\y1-20);
  \draw (\x4+8,\y4+8) rectangle (\x4-8,\y4-8);
  \draw (\x1-5,\y4-8)[rounded corners=4] -- (\x1-5,\y1-13) -- (\x1+15,\y1-13)
     -- (\x1+15,\y1+20) -- (\x1+15,\y2) -- (\x2+10,\y2) -- (\x2+10,\y2-20);
  \putRightDiode(\x2,\y2)
  \putLeftDiode(\x3,\y3)
  \node at (\x4,\y4){$f'$};
\end{tikzpicture}}}$%
	\qquad $\twomor{}$\qquad
$\vcenter{\hbox{\begin{tikzpicture}[xscale=0.0352778, yscale=0.0352778, thin, inner sep=0]
  \def\p#1#2{%
    \ifcase #1
      \or \ifx#2x  0       \else  0 \fi
      \or \ifx#2x  \x1+3   \else  \y1-8 \fi
    \fi}
  \def\x#1{\p#1x}
  \def\y#1{\p#1y}
  \draw[use as bounding box] (\x1+8,\y1+8) rectangle (\x1-8,\y1-8);
  \draw (\x2,\y2) -- (\x2,\y2-9);
  \node at (\x1,\y1){$f'$};
\end{tikzpicture}}}$%
	\qquad $\twomor*$\qquad
$\vcenter{\hbox{\begin{tikzpicture}[xscale=0.0352778, yscale=0.0352778, thin, inner sep=0]
  \def\p#1#2{%
    \ifcase #1
      \or \ifx#2x  0       \else  0 \fi
      \or \ifx#2x  \x1+3   \else  \y1-8 \fi
    \fi}
  \def\x#1{\p#1x}
  \def\y#1{\p#1y}
  \draw[use as bounding box] (\x1+8,\y1+8) rectangle (\x1-8,\y1-8);
  \draw (\x2,\y2) -- (\x2,\y2-9);
  \node at (\x1,\y1){$f_0$};
\end{tikzpicture}}}$%

	\vskip2ex
        \noindent
 where $f_0$ is a normal form.
Then we have

	\vskip2ex
        \noindent\kern5em
$\vcenter
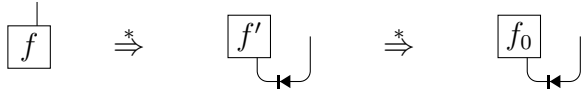
{\hbox{\begin{tikzpicture}[xscale=0.0352778, yscale=0.0352778, thin, inner sep=0]
  \def\p#1#2{%
    \ifcase #1
      \or \ifx#2x  0       \else  0 \fi
      \or \ifx#2x  \x1+3   \else  \y1+8 \fi
    \fi}
  \def\x#1{\p#1x}
  \def\y#1{\p#1y}
  \draw[use as bounding box] (\x1+8,\y1+8) rectangle (\x1-8,\y1-8);
  \draw (\x2,\y2) -- (\x2,\y2+9);
  \node at (\x1,\y1){$f$};
\end{tikzpicture}}}$%
	\qquad $\twomor*$\qquad
$\vcenter{\hbox{\begin{tikzpicture}[xscale=0.0352778, yscale=0.0352778, thin, inner sep=0]
  \def\p#1#2{%
    \ifcase #1
      \or \ifx#2x  0       \else  0 \fi
      \or \ifx#2x  \x1+25  \else  \y1+30 \fi
      \or \ifx#2x  \x1+5   \else  \y1-13 \fi
      \or \ifx#2x  \x1-8   \else  \y1+4 \fi
    \fi}
  \def\x#1{\p#1x}
  \def\y#1{\p#1y}
  \path[use as bounding box] (19,25) rectangle (-24,-25);
  \draw (\x4+8,\y4+8) rectangle (\x4-8,\y4-8);
  \draw (\x1-5,\y4-8)[rounded corners=4] -- (\x1-5,\y1-13) -- (\x1+15,\y1-13)
     -- (\x1+15,\y4);
  \putLeftDiode(\x3,\y3)
  \node at (\x4,\y4){$f'$};
\end{tikzpicture}}}$%
	\qquad $\twomor*$\qquad
$\vcenter{\hbox{\begin{tikzpicture}[xscale=0.0352778, yscale=0.0352778, thin, inner sep=0]
  \def\p#1#2{%
    \ifcase #1
      \or \ifx#2x  0       \else  0 \fi
      \or \ifx#2x  \x1+25  \else  \y1+30 \fi
      \or \ifx#2x  \x1+5   \else  \y1-13 \fi
      \or \ifx#2x  \x1-8   \else  \y1+4 \fi
    \fi}
  \def\x#1{\p#1x}
  \def\y#1{\p#1y}
  \path[use as bounding box] (19,25) rectangle (-24,-25);
  \draw (\x4+8,\y4+8) rectangle (\x4-8,\y4-8);
  \draw (\x1-5,\y4-8)[rounded corners=4] -- (\x1-5,\y1-13) -- (\x1+15,\y1-13)
     -- (\x1+15,\y4);
  \putLeftDiode(\x3,\y3)
  \node at (\x4,\y4){$f_0$};
\end{tikzpicture}}}$%

	\vskip2ex
        \noindent
The last is a normal form unless the bend is connected to another bend,
 forming a $\beta$-redex.
If so the contraction of the redex leads to a normal form.
The converse is straightforward.
\endproof

	\vskip1ex

\begin{lemma}\label{daf47}
If ${\cal S}$ is a negative (or positive) funnel, then
  $({\cal S}^*)^\bot$ is a negative (or positive) funnel.
\end{lemma}

\proof
Suppose ${\cal S}$ is a negative funnel.
Since every morphism in ${\cal S}^*$ terminates by Lem.~\ref{oou03},
 $({\cal S}^*)^\bot$ contains an identity.
We verify that each morphism $f\in({\cal S}^*)^\bot$ terminates.
As ${\cal S}$ contains an identity,
$\vcenter{\hbox{\begin{tikzpicture}[xscale=0.0352778, yscale=0.0352778, thin, inner sep=0]
 \draw (-20,3)[rounded corners=3] -- (-20,-3) --(0,-3) -- (0,3) -- (20,3) -- (20,-3);
 \putLeftDiode(-10,-3)
 \putRightDiode(10,3)
\end{tikzpicture}}}$
 is a member of ${\cal S}^*$.
Hence
$$
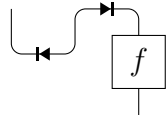
\vcenter{\hbox{\begin{tikzpicture}[xscale=0.0352778, yscale=0.0352778, thin, inner sep=0]
  \def\p#1#2{%
    \ifcase #1
      \or \ifx#2x  0       \else  0 \fi
      \or \ifx#2x  \x1-12  \else  \y1+20 \fi
      \or \ifx#2x  \x2-24  \else  \y2-17 \fi
    \fi}
  \def\x#1{\p#1x}
  \def\y#1{\p#1y}
  \path[use as bounding box] (14,28) rectangle (-65,-25);
  \draw (\x1+10,\y1+10) rectangle (\x1-10,\y1-10);
  \draw (\x1,\y1-20) -- (\x1,\y1-10);
  \draw (\x1,\y1+10)[rounded corners=4] -- (\x1,\y2) -- (\x2,\y2);
  \draw (\x2,\y2)[rounded corners=4] -- ($ .5*(\x2,\y2)+.5*(\x3,\y2) $)
    -- ($ .5*(\x2,\y3)+.5*(\x3,\y3) $) -- (\x3,\y3);
  \draw (\x3,\y3)[rounded corners=4] -- (\x3-12,\y3) -- (\x3-12,\y2);
  \putRightDiode(\x2,\y2)
  \putLeftDiode(\x3,\y3)
  \node at (\x1,\y1){$f$};
\end{tikzpicture}}}
$$
 terminates.
By Lem.~\ref{oou03}, we can eliminate the leftmost bend.
With a similar argument, we can eliminate the other bend as well.
\endproof

	\vskip1ex

\begin{definition}\rm\label{iwu15}
For each object $A$, we define
 a negative funnel $R^-(A)$ and a positive funnel $R^+(A)$ by
 induction on the construction of $A$.

	\vskip1ex
        \noindent\kern5em
\vbox{\halign{$#$\hfil&${}\ =\ #$\hfil\cr
 R^-(A) & \{1_A\}^\bot\qquad\hbox{if $A$ is atomic}\cr
 R^-({\bf 1}) & \{1_{\bf 1}\}^\bot\cr
 R^+(\bot) & \{1_{\bot}\}^\bot\cr
 R^-(A\otimes B) & (R^+(A)\otimes R^+(B))^\bot\cr
 R^+(A\linpar B) & (R^-(A)\linpar R^-(B))^\bot\cr
 R^-(A^*) & ((R^-(A))^*)^\bot \cr
 R^-(\mathord!A) & (\mathord!R^+(A))^\bot\cr
}}

	\vskip2ex
        \noindent
Only one of $R^-(A)$ and $R^+(A)$ is listed above.
The other is defined as its complement $(\hbox{--})^\bot$.
For example, $R^+(\mathord!A)$ is
 $(R^-(\mathord!A))^\bot$.
Evidently, $R^+(A)$ and $R^-(A)$ are the complements of each other.
\end{definition}

	\vskip1ex

\begin{lemma}\label{aun57}
Let $A\mor fB$ and $C\mor gD$ be morphisms.
The following implications hold:

	\vskip2ex
        \noindent\kern5em
\vbox{\halign{$#$\hfil &$\qquad\Longrightarrow\qquad #$\hfil\cr
 f\in R^-(A) & \mathord!f\in R^-(\mathord!A)\cr
 f\in R^+(B) & \mathord!f\in R^+(\mathord!B)\cr}}
	\vskip1ex
        \noindent\kern5em
\vbox{\halign{$#$\hfil &$\qquad\Longrightarrow\qquad #$\hfil\cr
 f\in R^-(A),\ g\in R^-(C) & f\otimes g\in R^-(A\otimes C)\cr
 f\in R^+(B),\ g\in R^+(D) & f\otimes g\in R^+(B\otimes D)\cr
	\noalign{\vskip1ex}
 f\in R^-(A),\ g\in R^-(C) & f\linpar g\in R^-(A\linpar C)\cr
 f\in R^+(B),\ g\in R^+(D) & f\linpar g\in R^+(B\linpar D)\cr}}
	\vskip1ex
        \noindent\kern5em
\vbox{\halign{$#$\hfil &$\qquad\Longrightarrow\qquad #$\hfil\cr
 f\in R^-(A) & f^*\in R^+(A^*)\cr
 f\in R^+(B) & f^*\in R^-(B^*)\cr}}
	\par
\end{lemma}

\proof
We verify the case for $\mathord!$.
We take an arbitrary $\mathord!g$
 from $\mathord!R^+(A)$.
Composition $g;f$ terminates by hypothesis.
So $\mathord!g;\mathord!f=\mathord!(g;f)$
 terminates as well.
Thus the first assertion follows.
The second assertion is a consequence of the
 inflation property of $(\hbox{--})^{\bot\bot}$.
The rest are similar.
For $(\hbox{--})^*$, we note that $g^*;f^*$ contracts
 to $(f;g)^*$ by $\beta$-reduction.
\endproof

	\vskip1ex

\begin{definition}\rm\label{atn84}
A {\it reducible} morphism is a morphism $A\mor fB$
 satisfying that
 $g;f;h$ terminates for every pair
 of $g\in R^+(A)$ and $h\in R^-(B)$.
\end{definition}

	\vskip1ex

\begin{lemma}\label{azj94}
For a morphism $A\mor fB$, the following are equivalent:

	\vskip0.5ex
        \hangafter0\hangindent2em
        \noindent
\llapem2{(i)}%
$f$ is reducible.

	\vskip0ex
        \noindent
\llapem2{(ii)}%
If $g\in R^+(A)$ then $g;f\in R^+(B)$.

	\vskip0ex
        \noindent
\llapem2{(iii)}%
If $h\in R^-(B)$ then $f;h\in R^-(A)$.
	\par
\end{lemma}

\proof
Straightforward as each of $R^+(A)$ and $R^-(A)$ is the complement
 of the other.
\endproof

	\vskip1ex

\begin{proposition}\label{cbw90}
Each reducible morphism terminates.
\end{proposition}

\proof
All funnels contain identity morphisms.
\endproof

	\vskip1ex
\noindent
To verify the termination property,
 therefore it suffices to show that all morphisms in the free classical
 linear category are reducible.
We start with the easy cases.

	\vskip1ex

\begin{lemma}\label{ozp36}
The following hold:

	\vskip0.5ex
	\hangafter0\hangindent2em
        \noindent
\llapem2{(i)}%
A morphism $\mathord!A\mor fB$ is reducible iff
 $\mathord!g;f$ lies in $R^+(B)$ for every
 $g$ in $R^+(A)$.

	\vskip0ex
        \noindent
\llapem2{(ii)}%
A morphism $A\otimes B\mor fC$ is reducible iff
 $(g\otimes h);f$ lies in $R^+(C)$ for
 every pair of $g$ in $R^+(A)$
 and $h$ in $R^+(B)$.

	\vskip0ex
        \noindent
\llapem2{(iii)}%
A morphism $A\mor fB\linpar C$ is reducible iff
 $f;(g\linpar h)$ lies in $R^-(A)$ for
 every pair of $g$ in $R^-(B)$
 and $h$ in $R^-(C)$.

	\vskip0ex
        \noindent
\llapem2{(iv)}%
A morphism ${\bf 1}\mor fA$ is reducible iff
 $f$ lies in $R^+(A)$.

	\vskip0ex
        \noindent
\llapem2{(v)}%
A morphism $B^*\mor fC$ is reducible iff
 $g^*;f$ lies in $R^+(C)$ for every
 $g$ in $R^-(B)$.
	\par
\end{lemma}

\proof
We prove (i) as the argument is similar.
Supposed that $f$ is reducible,
$f;h\in R^-(\mathord!A)$ for every $h\in R^-(B)$ by Lem.~\ref{azj94}.
By definition of $R^-(\mathord!A)$, then, $\mathord!g;f;h$ terminates
 for every $g\in R^+(A)$.
As $h$ is arbitrary, $\mathord!g;f$ belongs to $(R^-(B))^\bot=R^+(B)$.
The converse is also true.
\endproof

	\vskip1ex

\begin{lemma}\label{jhc51}
A morphism $\mathord!A\otimes \mathord!B\mor fC$ is reducible
 iff $(\mathord!g\otimes\mathord!h);f$ lies in $R^+(C)$
 for every pair of $g\in R^+(A)$ and
 $h\in R^+(B)$.
\end{lemma}

\proof
By Lem.~\ref{ozp36} $f$ is reducible iff,
 for every $p$ in $R^+(\mathord!A)$,
 every $q$ in $R^+(\mathord!B)$,
 and every $k$ in $R^-(C)$,
 the morphism $(p\otimes q);f;k$ terminates.
By the bending technique of Lem.~\ref{oou03}, it means termination of
$$
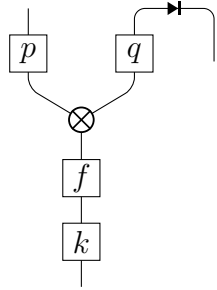
\vcenter{\hbox{\begin{tikzpicture}[xscale=0.0352778, yscale=0.0352778, thin, inner sep=0]
  \def\p#1#2{%
    \ifcase #1
      \or \ifx#2x  0       \else  0 \fi
      \or \ifx#2x  \x1     \else  \y1-24 \fi
      \or \ifx#2x  \x1     \else  \y1+22 \fi
      \or \ifx#2x  \x3+20  \else  \y3+25 \fi
      \or \ifx#2x  \x3-20  \else  \y4 \fi
      \or \ifx#2x  \x4+15  \else  \y4+17 \fi
    \fi}
  \def\x#1{\p#1x}
  \def\y#1{\p#1y}
  \path[use as bounding box] (62,-46) rectangle (-33,76);
  \draw (\x1+7,\y1+7) rectangle (\x1-7,\y1-7);
  \draw (\x2+7,\y2+7) rectangle (\x2-7,\y2-7);
  \draw (\x4+7,\y4+7) rectangle (\x4-7,\y4-7);
  \draw (\x5+7,\y5+7) rectangle (\x5-7,\y5-7);
  \draw (\x1,\y1-7) -- (\x2,\y2+7);
  \draw (\x2,\y2-7) -- (\x2,\y2-17);
  \draw (\x1,\y1+7) -- (\x3,\y3);
  \draw (\x3,\y3)[rounded corners=4] -- (\x4,\y3+12) -- (\x4,\y4-7);
  \draw (\x3,\y3)[rounded corners=4] -- (\x5,\y3+12) -- (\x5,\y5-7);
  \draw (\x4,\y4+7)[rounded corners=4] -- (\x4,\y6) -- (\x6,\y6);
  \draw (\x5,\y5+7) -- (\x5,\y5+17);
  \draw (\x6,\y6)[rounded corners=4] -- (\x6+15,\y6) -- (\x6+15,\y6-20);
  \putTensor(\x3,\y3)
  \putRightDiode(\x6,\y6)
  \node at (\x1,\y1){$f$};
  \node at (\x2,\y2){$k$};
  \node at (\x4,\y4){$q$};
  \node at (\x5,\y5){$p$};
\end{tikzpicture}}}
$$
Since $p$ is arbitrary, it amounts to that
$$
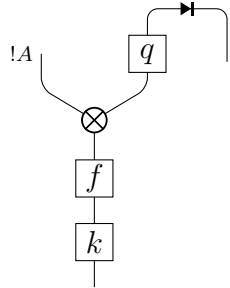
\vcenter{\hbox{\begin{tikzpicture}[xscale=0.0352778, yscale=0.0352778, thin, inner sep=0]
  \def\p#1#2{%
    \ifcase #1
      \or \ifx#2x  0       \else  0 \fi
      \or \ifx#2x  \x1     \else  \y1-24 \fi
      \or \ifx#2x  \x1     \else  \y1+22 \fi
      \or \ifx#2x  \x3+20  \else  \y3+25 \fi
      \or \ifx#2x  \x3-20  \else  \y4 \fi
      \or \ifx#2x  \x4+15  \else  \y4+17 \fi
    \fi}
  \def\x#1{\p#1x}
  \def\y#1{\p#1y}
  \path[use as bounding box] (62,-46) rectangle (-33,76);
  \draw (\x1+7,\y1+7) rectangle (\x1-7,\y1-7);
  \draw (\x2+7,\y2+7) rectangle (\x2-7,\y2-7);
  \draw (\x4+7,\y4+7) rectangle (\x4-7,\y4-7);
  \draw (\x1,\y1-7) -- (\x2,\y2+7);
  \draw (\x2,\y2-7) -- (\x2,\y2-17);
  \draw (\x1,\y1+7) -- (\x3,\y3);
  \draw (\x3,\y3)[rounded corners=4] -- (\x4,\y3+12) -- (\x4,\y4-7);
  \draw (\x3,\y3)[rounded corners=4] -- (\x5,\y3+12) -- (\x5,\y5);
  \draw (\x4,\y4+7)[rounded corners=4] -- (\x4,\y6) -- (\x6,\y6);
  \draw (\x6,\y6)[rounded corners=4] -- (\x6+15,\y6) -- (\x6+15,\y6-20);
  \putTensor(\x3,\y3)
  \putRightDiode(\x6,\y6)
  \node at (\x1,\y1){$f$};
  \node at (\x2,\y2){$k$};
  \node at (\x4,\y4){$q$};
  \node[left] at (\x5-3,\y5) {$\scriptstyle\mathord!A$};
\end{tikzpicture}}}
$$
 lies in $R^-(\mathord!A)$.
By Lem.~\ref{ozp36}, it says that
$$
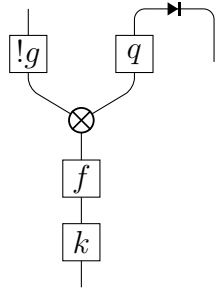
\vcenter{\hbox{\begin{tikzpicture}[xscale=0.0352778, yscale=0.0352778, thin, inner sep=0]
  \def\p#1#2{%
    \ifcase #1
      \or \ifx#2x  0       \else  0 \fi
      \or \ifx#2x  \x1     \else  \y1-24 \fi
      \or \ifx#2x  \x1     \else  \y1+22 \fi
      \or \ifx#2x  \x3+20  \else  \y3+25 \fi
      \or \ifx#2x  \x3-20  \else  \y4 \fi
      \or \ifx#2x  \x4+15  \else  \y4+17 \fi
    \fi}
  \def\x#1{\p#1x}
  \def\y#1{\p#1y}
  \path[use as bounding box] (62,-46) rectangle (-33,76);
  \draw (\x1+7,\y1+7) rectangle (\x1-7,\y1-7);
  \draw (\x2+7,\y2+7) rectangle (\x2-7,\y2-7);
  \draw (\x4+7,\y4+7) rectangle (\x4-7,\y4-7);
  \draw (\x5+7,\y5+7) rectangle (\x5-7,\y5-7);
  \draw (\x1,\y1-7) -- (\x2,\y2+7);
  \draw (\x2,\y2-7) -- (\x2,\y2-17);
  \draw (\x1,\y1+7) -- (\x3,\y3);
  \draw (\x3,\y3)[rounded corners=4] -- (\x4,\y3+12) -- (\x4,\y4-7);
  \draw (\x3,\y3)[rounded corners=4] -- (\x5,\y3+12) -- (\x5,\y5-7);
  \draw (\x4,\y4+7)[rounded corners=4] -- (\x4,\y6) -- (\x6,\y6);
  \draw (\x5,\y5+7) -- (\x5,\y5+17);
  \draw (\x6,\y6)[rounded corners=4] -- (\x6+15,\y6) -- (\x6+15,\y6-20);
  \putTensor(\x3,\y3)
  \putRightDiode(\x6,\y6)
  \node at (\x1,\y1){$f$};
  \node at (\x2,\y2){$k$};
  \node at (\x4,\y4){$q$};
  \node at (\x5,\y5){$\mathord!g$};
\end{tikzpicture}}}
$$
 terminates for every $g\in R^+(A)$.
Straightening the bend over $q$, we have succeeded in replacing
 $p$ with $\mathord!g$.
Applying the same process to the right wire as well, we obtain the lemma.
\endproof

	\vskip1ex

\begin{proposition}\label{hpc73}
The following hold:

	\vskip0.5ex
	\hangafter0\hangindent2em
        \noindent
\llapem2{(i)}%
Identities and structural isomorphisms are reducible.

\noindent
\llapem2{(ii)}%
The composition of reducible morphisms is reducible.
	\par
\end{proposition}

\proof
(i) is obvious.
(ii) is an immediate consequence of Lem.~\ref{azj94}.
\endproof

	\vskip1ex

\begin{proposition}\label{xsi96}
The following hold:

	\vskip0.5ex
        \hangafter0\hangindent2em
        \noindent
\llapem2{(i)}%
If $f$ is reducible, $\mathord!f$ is reducible.

	\vskip0ex
        \noindent
\llapem2{(ii)}%
If $f$ and $g$ are reducible, $f\otimes g$ is reducible.

	\vskip0ex
        \noindent
\llapem2{(iii)}%
If $f$ and $g$ are reducible, $f\linpar g$ is reducible.

	\vskip0ex
        \noindent
\llapem2{(iv)}%
If $f$ is reducible, $f^*$ is reducible.
	\par
\end{proposition}

\proof
(i) through (iii) are consequences of Lem.~\ref{ozp36}.
We prove (iv).
Suppose that $f:A\rightarrow B$ is reducible.
Again by Lem.~\ref{ozp36}, it suffices to show that $g^*;f^*;h$
 terminates for every pair of $g\in R^-(B)$ and $h\in R^-(A^*)$.
We note that $g^*;f^*$ contracts to $(f;g)^*$ by a duality $\beta$-reduction.
Here $f;g$ lies in $R^-(A)$, thus $(f;g)^*$ lies in $R^+(A^*)$.
So $(f;g)^*;h$ terminates.
\endproof

	\vskip1ex

\begin{proposition}\label{lqm59}
$\partial$ is reducible.
\end{proposition}

\proof
Suppose $\partial :A\otimes (B\linpar C)\rightarrow (A\otimes B)\linpar C$.
We must show that $(l\otimes f);\partial;(g\linpar h)$
 terminates for every $l\in R^+(A)$, every
 $f\in R^+(B\linpar C)$, every
 $g\in R^-(A\otimes B)$, and every
 $h\in R^-(C)$.
For every $k\in R^+(B)$, $(l\otimes k);g$ terminates.
Thus, by the bending technique of Lem.~\ref{oou03},
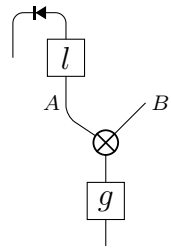
$$
\vcenter{\hbox{\begin{tikzpicture}[xscale=0.0352778, yscale=0.0352778, thin, inner sep=0]
  \def\p#1#2{%
    \ifcase #1
      \or \ifx#2x  0       \else  0 \fi
      \or \ifx#2x  \x1     \else  \y1+22 \fi
      \or \ifx#2x  \x2+15  \else  \y2+15 \fi
      \or \ifx#2x  \x2-15  \else  \y2+32 \fi
      \or \ifx#2x  \x4-10  \else  \y4+17 \fi
    \fi}
  \def\x#1{\p#1x}
  \def\y#1{\p#1y}
  \path[use as bounding box] (26,79) rectangle (-44,-22);
  \draw (\x1,\y1-17) -- (\x1,\y1-7);
  \draw (\x1,\y1+7) -- (\x2,\y2);
  \draw (\x2,\y2) -- (\x3,\y3);
  \draw (\x2,\y2)[rounded corners=4] -- (\x4,\y2+10) -- (\x4,\y4-7);
  \draw (\x4,\y4+7)[rounded corners=4] -- (\x4,\y5) -- (\x5,\y5);
  \draw (\x5,\y5)[rounded corners=4] -- (\x5-10,\y5) -- (\x5-10,\y4);
  \draw (\x1-7,\y1-7) rectangle (\x1+7,\y1+7);
  \draw (\x4-7,\y4-7) rectangle (\x4+7,\y4+7);
  \putTensor(\x2,\y2)
  \putLeftDiode(\x5,\y5)
  \node at (\x1,\y1){$g$};
  \node at (\x4,\y4){$l$};
  \node[left] at (\x4-2,\y4-17){$\scriptstyle A$};
  \node[right] at (\x3+2,\y3){$\scriptstyle B$};
\end{tikzpicture}}}$$
 lies in $R^-(B)$.
Hence
$$
\vcenter
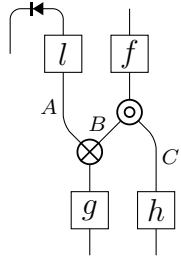
{\hbox{\begin{tikzpicture}[xscale=0.0352778, yscale=0.0352778, thin, inner sep=0]
  \def\p#1#2{%
    \ifcase #1
      \or \ifx#2x  0       \else  0 \fi
      \or \ifx#2x  \x1     \else  \y1+22 \fi
      \or \ifx#2x  \x2+15  \else  \y2+15 \fi
      \or \ifx#2x  \x3+10  \else  \y1 \fi
      \or \ifx#2x  \x3     \else  \y3+22 \fi
      \or \ifx#2x  \x2-10  \else  \y5 \fi
      \or \ifx#2x  \x6-10  \else  \y6+17 \fi
    \fi}
  \def\x#1{\p#1x}
  \def\y#1{\p#1y}
  \path[use as bounding box] (26,79) rectangle (-44,-22);
  \draw (\x1,\y1-17) -- (\x1,\y1-7);
  \draw (\x1,\y1+7) -- (\x2,\y2);
  \draw (\x2,\y2) -- (\x3,\y3);
  \draw (\x3,\y3) -- (\x5,\y5-7);
  \draw (\x5,\y5+7) -- (\x5,\y5+17);
  \draw (\x4,\y4-17) -- (\x4,\y4-7);
  \draw (\x1-7,\y1-7) rectangle (\x1+7,\y1+7);
  \draw (\x4-7,\y4-7) rectangle (\x4+7,\y4+7);
  \draw (\x5-7,\y5-7) rectangle (\x5+7,\y5+7);
  \draw (\x6-7,\y6-7) rectangle (\x6+7,\y6+7);
  \draw (\x2,\y2)[rounded corners=4] -- (\x6,\y2+10) -- (\x6,\y6-7);
  \draw (\x3,\y3)[rounded corners=4] -- (\x4,\y3-10) -- (\x4,\y4+7);
  \draw (\x6,\y6+7)[rounded corners=4] -- (\x6,\y7) -- (\x7,\y7);
  \draw (\x7,\y7)[rounded corners=4] -- (\x7-10,\y7) -- (\x7-10,\y6);
  \putTensor(\x2,\y2)
  \putPar(\x3,\y3)
  \putLeftDiode(\x7,\y7)
  \node at (\x1,\y1){$g$};
  \node at (\x4,\y4){$h$};
  \node at (\x5,\y5){$f$};
  \node at (\x6,\y6){$l$};
  \node[left] at (\x6-2,\y6-20){$\scriptstyle A$};
  \node[left] at ($ .5*(\x2,\y2)+.5*(\x3,\y3)+(-1,3) $){$\scriptstyle B$};
  \node[right] at (\x4+2,\y4+20){$\scriptstyle C$};
\end{tikzpicture}}}
$$
 terminates.
Straightening the bend over $l$, we obtain the proposition.
\endproof

	\vskip1ex

\begin{proposition}\label{dci74}
$\tau_A$ and $\gamma_A$ are reducible.
\end{proposition}

\proof
Verification is the same for both cases.
We give the proof
 for $\gamma_A:A^*\otimes A\rightarrow \bot$.
We must show that $(f\otimes g);\gamma_A$ terminates
 for every pair of $f\in R^+(A^*)$ and $g\in R^+(A)$.
Composition $f;g^*$ terminates since $g^*\in R^-(A^*)$ by Lem~\ref{aun57}.
Graphically this composition means
$$
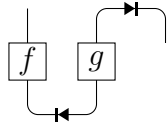
\vcenter{\hbox{\begin{tikzpicture}[xscale=0.0352778, yscale=0.0352778, thin, inner sep=0]
  \def\p#1#2{%
    \ifcase #1
      \or \ifx#2x  0       \else  0 \fi
      \or \ifx#2x  \x1+13  \else  \y1-20 \fi
      \or \ifx#2x  \x2+13  \else  \y2+20 \fi
      \or \ifx#2x  \x3+13  \else  \y3+20 \fi
    \fi}
  \def\x#1{\p#1x}
  \def\y#1{\p#1y}
  \path[use as bounding box] (66,29) rectangle (-13,-26);
  \draw (\x1+7,\y1+7) rectangle (\x1-7,\y1-7);
  \draw (\x3+7,\y3+7) rectangle (\x3-7,\y3-7);
  \draw (\x1,\y1+7) -- (\x1,\y1+20);
  \draw (\x1,\y1-7)[rounded corners=4] -- (\x1,\y2) -- (\x2,\y2);
  \draw (\x2,\y2)[rounded corners=4] -- (\x3,\y2) -- (\x3,\y3-7);
  \draw (\x3,\y3+7)[rounded corners=4] -- (\x3,\y4) -- (\x4,\y4);
  \draw (\x4,\y4)[rounded corners=4] -- (\x4+13,\y4) -- (\x4+13,\y4-13);
  \putLeftDiode(\x2,\y2)
  \putRightDiode(\x4,\y4)
  \node at (\x1,\y1){$f$};
  \node at (\x3,\y3){$g$};
\end{tikzpicture}}}$$
Straightening the bend over $g$, we obtain the proposition.
\endproof

	\vskip1ex

It remains to verify that the six algebraic morphisms are reducible.
There are technical difficulties that are unable to be
 covered by the reducibility method.
To that end, we introduce several notions.

Let $X$ range over a chosen non-empty family of objects.
Later we use the case where the family is a singleton or
 consists of two elements.
We regard $X$ as if they are atomic objects.
We consider two classes of objects
 generated by the following generative grammar.

	\vskip2ex
        \noindent\kern5em
\vbox{\halign{$#$\hfil &${}\ \ \mathrel{::=}\ \ #\ \ $\hfil
   &\hfil ${}#{}$&&${}\ \ \mathbin|\ \ #$\hfil\cr
 A && X & A\otimes A & \mathord!A\cr
 B & {\bf 1} & \mathbin|\ \ X & B\otimes B & \mathord!B\cr}}

	\vskip1ex
        \noindent

\begin{definition}\rm\label{osp47}
\noindent
	\vskip0ex
        \hangafter0\hangindent2em
        \noindent
\llapem2{(i)}%
A {\it composite algebraic morphism} $t$ is a member of the class generated
 from $1_B$ and $\varphi_0,\tilde\varphi_{B,B'},
 \delta_B,\varepsilon_B,d_B,e_B$ as well as
 $\alpha_{B,B',B''},\sigma_{B,B'},\lambda_B,\rho_B$ and their inverses,
 closed under operations $t\otimes t'$, $\mathord!t$ and composition $t;t'$.

	\vskip0.5ex
        \noindent
\llapem2{(ii)}%
A {\it strict composite algebraic morphism} $s$ is a member of the class generated
 from $1_A$ and $\tilde\varphi_{A,A'},
 \delta_A,\varepsilon_A,d_A,e_A$ as well as
 $\alpha_{A,A',A''},\sigma_{A,A'}$ and their inverses,
 closed under operations $s\otimes s'$, $\mathord!s$ and composition $s;s'$.
	\par
\end{definition}

	\vskip0ex
\noindent
Namely, composite algebraic morphisms can use everything unrelated
 to $\linpar$ or $(\hbox{--})^*$ as long as $X$ is regarded as an atomic object.
Strict composite algebraic morphisms preclude $\varphi_0$ and the isomorphisms
 involving ${\bf 1}$.
We comment that the target of $e_A:\mathord!A
 \rightarrow {\bf 1}$ is not a member of class $A$.
Save this exception, composite algebraic morphisms are between members of class $B$
 and strict composite algebraic morphisms are between members of $A$.

\begin{example}\label{cfn37}
$\mathord!X\mor{\delta_X}\mathord!\mathord!X\mor{d_{\mathord!X}}
 \mathord!\mathord!X\otimes \mathord!\mathord!X\mor{\cdot\mathord!e_X}
 \mathord!\mathord!X\otimes \mathord!{\bf 1}$
 is a strict composite algebraic morphism.
It contracts to
$\mathord!X\mor{d_X}\mathord!X\otimes \mathord!X\mor{\delta_X\delta_X}
 \mathord!\mathord!X\otimes \mathord!\mathord!X\mor{\cdot \mathord!e_X}
 \mathord!\mathord!X\otimes \mathord!{\bf 1}$, which is strict composite algebraic.
It further contracts to
$\mathord!X\mor{d_X}\mathord!X\otimes \mathord!X\mor{\delta_Xe_X}
 \mathord!\mathord!X\otimes{\bf 1}\mor{\cdot\varphi_0}
 \mathord!\mathord!X\otimes \mathord!{\bf 1}$,
 which is composite algebraic but not strictly composite algebraic
 since it contains $\varphi_0$.
Finally, it contracts to $\mathord!X\mor{\delta_X}\mathord!\mathord!X
 \mor\sim \mathord!\mathord!X\otimes {\bf 1}\mor{\cdot\varphi_0}
 \mathord!\mathord!X\otimes \mathord!{\bf 1}$, which is composite
 algebraic.
We will return to this sequence of contractions in Example~\ref{vtv72}.
\end{example}

A strict composite algebraic morphism $s$ that has
 no naturality redexes at the beginning
 may create naturality through reduction.
For example, $\mathord!X\mor{\delta}\mathord!\mathord!X
\mor{\delta}\mathord!\mathord!\mathord!X\mor{e}{\bf 1}$
 contracts to $\mathord!X\mor{\delta}
 \mathord!\mathord!X\mor{\mathord!\delta}
 \mathord!\mathord!\mathord!X\mor{e}{\bf 1}$, the latter
 containing naturality redex $\mathord!\delta;e$ while the former has none.
This happens because rule (1) produces $\mathord!\delta$ wrapped by
 $\mathord!(\hbox{--})$.
Likewise rule (9) produces $\mathord!\tilde\varphi$.
We do not have to consider rule (13) as $\varphi_0;\delta$ is 
 not allowed in strict composite algebraic morphisms.

	\vskip1ex

\begin{lemma}\label{rti32}
Let $u$ denote one of $\delta,\varepsilon,d$ and $e$.
Suppose that a strict composite algebraic morphism $s$ has no naturality redexes
 other than those of the form $\mathord!f;u_A$ where $f$
 consists of $\delta$ and $\tilde\varphi$ only.
Any morphism obtained by contraction of $s$
 satisfies the same property for naturality redexes.
\end{lemma}

\proof
Simple case analysis.
We cannot create naturality redexes when $f$ contains
 something other than $\delta$\ and $\tilde\varphi$ by contraction unless
 we have such redexes from the outset.
\endproof

	\vskip-1ex

A {\it restricted} naturality redex is $\mathord!f;u_A$ where $f$
 consists solely of $\delta$ and $\tilde\varphi$.
The above lemma asserts that if the naturality redexes of
 a strict composite algebraic morphism are restricted then
 the property is preserved under contraction.

We verify that strict composite algebraic morphisms (strongly) terminate
 if their naturality redexes are restricted.
Although $X$ may run over a family of two or more objects,
 the following argument is irrelevant to the number of distinct $X$.
So we describe the case when $X$ is unique.
If there are two or more, each $X$ should read one of them
 appropriately.
We write $A=A[X,X,\ldots,X]$ displaying each occurrence of $X$.
We further write $A=A[X^{x_1},
X^{x_2},\ldots,X^{x_n}]$.
At this stage, the $x_i$ are merely the
 labels to distinguish occurrences.
As we explain shortly, however, we assign natural numbers greater
 than or equal to $2$.

Suppose that

	\vskip2ex
        \noindent\kern5em
$s:A[X^{y_1},X^{y_2},\ldots,X^{y_n}]\rightarrow
 B[X^{x_1},X^{x_2},\ldots,X^{x_m}]$.

	\vskip2ex
        \noindent
Each $y_i$ is computed by applying a function $|s|$ determined
 by the shape of $s$ to some of $x_1,x_2,\ldots,x_m$.
We show that if $s$ is contracted by applying a certain type of
 reduction rules, then $y_1+y_2+\cdots +y_n$ strictly decreases.
To define $|s|$ we need some auxiliary data given below.

	\vskip1ex

\begin{definition}\rm\label{lfq82}
Let $x$ denote an occurrence of $X$ in $A$.
We define $\theta_A(x)$ recursively as follows:
(i) If $A=X$ then we set $\theta_X(x)=x$.
(ii) For the exponential, we set $\theta_{\mathord!A}(x)=2\theta_A(x)$.
(iii) For the tensor, we set $\theta_{A\otimes A'}(x)=b+\theta_A(x)$ and
 symmetrically $\theta_{A'\otimes A}(x)=b+\theta_A(x)$ where
 $b$ denotes the number of occurrences of $X$ in $A'$.
\end{definition}

	\vskip1ex
	\noindent
This recursive definition is applied to each occurrence of $X$.
For example if $A=\mathord!(X^{x_1}\otimes \mathord!\mathord!X^{x_2})$ then
 $\theta_A(x_1)=2(1+x_1)$ and $\theta_A(x_2)=2(1+4x_2)$.
Observe that $\theta_A$ is not a single function, the shape of which changes per
 occurrence.
We remark that a structural isomorphism does not affect
 $\theta_A$ since it does not change the number of occurrences of $X$.
For example $\theta_{(A\otimes B)\otimes C}(x)=\theta_{A\otimes(B\otimes C)}(x)$.

Let $A[X^x]$ denote a specific occurrence of $X$ in $A$.
Suppose that $s:A\rightarrow B$ is a strict composite algebraic morphism.
With each occurrence $A[X^y]$ of $X$ in $A$, we can naturally associate
 a finite number of occurrences $B[X^{x_1},X^{x_2},\ldots,X^{x_n}]$
 of $X$.
The number $n$ depends on the shape of $s$.
If $s=d_A$ then $n=2$ and we set
 $d_A:\bang A[X^y]\rightarrow\bang A[X^{x_1}]\otimes
\bang A[X^{x_2}]$, where $A[X^{x_i}]$ signifies
 the occurrence of $X$ at the same position as $A[X^y]$.
If $s=e_A$ then $n=0$.
For $\tilde\varphi_{A,A'},\delta_A,\varepsilon_A,d_A,e_A,
 \alpha_{A,A',A''}$, and $\sigma_{A,A'}$, we have $n=1$ and the
 association is straightforward.
This association is naturally extended to the composition $s;t$:
if $X^z$ associates with $X^{y_1},
 X^{y_2},\ldots,X^{y_n}$ in $s$, and if each $X^{y_i}$
 associates with $X^{x_{i1}},X^{x_{i2}},\ldots,X^{x_{im_i}}$ in
 $t$, then $z$ associates with
 all of $X^{x_{11}}$ through $X^{x_{nm_n}}$ in $s;t$.
The tensor $s\otimes t$ and the exponentiation $\mathord!s$ do
 not alter the association.

	\vskip1ex

\begin{definition}\rm\label{ibg60}
We define the arithmetic expression $y=|s|(x_1,x_2,\ldots,
x_n)$ for each
strict composite algebraic morphism $s:A[X^y]\rightarrow
 B[X^{x_1},X^{x_2},\ldots,X^{x_n}]$ and each occurrence $A[X^y]$ of
 $X$ in $A$.

	\vskip.5ex

\begingroup
	\hangafter0\hangindent2em

\noindent
\llapem2{(i)}%
For $\delta_A:\mathord!A[X^y]\rightarrow
 \mathord!\mathord!A[X^x]$ we associate $y=|\delta_A|(x)$
 where $|\delta_A|(x)=2^{\theta_A(x)}$.

\noindent
\llapem2{(ii)}%
For $d_A:\mathord!A[X^y]\rightarrow
 \mathord!A[X^{x}]\otimes \mathord!A[X^{x'}]$ we set $y=|d_A|(x,x')$
 where $|d_A|(x,x')=\theta_A(x)+\theta_A(x')$.

\noindent
\llapem2{(iii)}%
For $\varepsilon_A:\mathord!A[X^y]\rightarrow
 A[X^x]$ we set $y=|\varepsilon_A|(x)$ where
 $|\varepsilon_A|(x)=\theta_A(x)$.

\noindent
\llapem2{(iv)}%
For $e_A:\mathord!A[X^y]\rightarrow{\bf 1}$
 we set $y=|e_A|()$ where
 $|e_A|()=\theta_A(2)$.

\noindent
\llapem2{(v)}%
For the other components $1_A,\tilde\varphi_{A,A'},
 \alpha_{A,A',A''}$, and $\sigma_{A,A'}$, we set $y=x$.

\noindent
\llapem2{(vi)}%
For the composition, $|s;t|$ is defined by the composition
 of the reverse order, $|s|\scirc |t|$.
Namely, if $z=|s|(y_1,y_2,\ldots,y_n)$ and
$y_i=|t|(x_{i1},x_{i2},\ldots,x_{im_i})$, then $z=|s;t|(x_{11},\ldots,
	\penalty-3000
x_{nm_n})$ is obtained by substitutions for $y_i$.
For the tensor, $|s\otimes t|$ is either $|s|$ or $|t|$, depending
 on which side of tensor $X^y$ lies in.
Finally, we set $|\mathord!s|=|s|$.

	\vskip0ex
\endgroup
\end{definition}

	\vskip1ex
	\noindent
The definition is applied to each occurrence separately.
For instance, when $A=\mathord!(X\otimes \mathord!\mathord!X)$ and
 $d_A:\mathord!(X^{y_1}\otimes \mathord!\mathord!X^{y_2})\rightarrow
 \mathord!(X^{x_1}\otimes \mathord!\mathord!X^{x_2})\otimes
 \mathord!(X^{x'_1}\otimes \mathord!\mathord!X^{x'_2})$ then
 $y_1=|d_A|(x_1,x'_1)=2(1+x_1)+2(1+x'_1)$ and
 $y_2=|d_A|(x_2,x'_2)=2(1+4x_2)+2(1+4x'_2)$.
As usual, we can interpret the arithmetic expression
 $|s|(x_1,x_2,\ldots,x_n)$ as a function.
For the reason explained in Lem.~\ref{jod26},
 we assume $x_i$ are natural numbers
 greater than or equal to $2$.
As is clear from the definition, the functions are increasing.

\begin{example}\label{vtv72}
We consider the morphisms in Example~\ref{cfn37}.
If we label the first morphism as in

	\vskip2ex
        \noindent\kern5em
$\mathord!X^y\mor{\delta_X}\mathord!\mathord!X
 \mor{d_{\mathord!X}}\mathord!\mathord!X\otimes \mathord!\mathord!X
 \mor{\cdot\mathord!e_X}\mathord!\mathord!X^x\otimes \mathord!{\bf 1}$,

	\vskip2ex
        \noindent
 we have $y=2^{\theta_{\mathord!X}(x)+\theta_{\mathord!X}(2)}=2^{2x+4}$.
It contracts to

	\vskip2ex
        \noindent\kern5em
 $\mathord!X^{y'}\mor{d_X}
 \mathord!X\otimes \mathord!X\mor{\delta_X\delta_X}
 \mathord!\mathord!X\otimes \mathord!\mathord!X\mor{\cdot \mathord!e_X}
 \mathord!\mathord!X^x\otimes \mathord!{\bf 1}$,

	\vskip2ex
        \noindent
 for which
 we have $y'=2^x+2^2$.
It further contracts to

	\vskip2ex
        \noindent\kern5em
 $\mathord!X^{y''}\mor{d_X}
 \mathord!X\otimes \mathord!X\mor{\delta_Xe_X}\mathord!\mathord!X
 \otimes {\bf 1}\mor{\cdot\varphi_0}
 \mathord!\mathord!X^x\otimes \mathord!{\bf 1}$,

	\vskip2ex
        \noindent
 for which
 $y''=2^x+2$.
Finally it contracts to

	\vskip2ex
        \noindent\kern5em
 $\mathord!X^{y'''}\mor{\delta_X}\mathord!\mathord!X
 \mor\sim \mathord!\mathord!X\otimes {\bf 1}\mor{\cdot\varphi_0}
 \mathord!\mathord!X^x\otimes \mathord!{\bf 1}$,

	\vskip2ex
        \noindent
 for which $y'''=2^x$.
We observe $2^{2x+4}>2^x+2^2>2^x+2>2^x$.
We will verify that this is universally true.
\end{example}

\begin{remark}
Tranquilli
 assigns natural numbers to show the termination of net rewriting~\cite{tran}.
Exact correspondence to our assignment is not immediate.
We stop here by commenting that the assignment to diagonal $d$ has similarity.
\end{remark}

	\vskip1ex

\begin{lemma}\label{jod26}
Algebraic reduction in a strict composite algebraic
 morphism decreases the natural numbers involved in the redex.
\end{lemma}

\proof
As $\varphi_0$-type contraction never occurs,
 it suffices to consider rule (1) through (12).
Since ${\bf 1}$ is not involved,
 in the definition of $\theta_{A\otimes B}(x)=b+
 \theta_A(x)$, the number $b$ is greater than or equal to $1$.
We verify several subtle cases, leaving the others to the reader.

Rule (1).
Suppose $\mathord!\mathord!\mathord!A[X^x]$ where
 $X^x$ denotes an arbitrary occurrence in $A$.
Then

	\vskip2ex
        \noindent\kern5em
\vbox{\halign{$#$\hfil &${}\ =\ #$\hfil\cr
 |\delta_A;\mathord!\delta_{A}|(x) & 2^{\theta_A(2^{\theta_A(x)})}\cr
	\noalign{\vskip.5ex}
 |\delta_A;\delta_{\mathord!A}|(x) & 2^{\theta_A(2^{2\theta_A(x)})}\cr}}

	\vskip2ex
        \noindent
 as $\theta_{\mathord!A}(x)=2\theta_A(x)$.
Since $\theta_A(x)<2\theta_A(x)$ we have $|\delta_A;\mathord!\delta_{A}|(x)
 <|\delta_A;\delta_{\mathord!A}|(x)$.

Rule (3).
Suppose that $\mathord!(A[X^x]\otimes A[X^{x'}])$ displays two
 occurrences at the corresponding same positions in $A$.
Then

	\vskip2ex
        \noindent\kern5em
\vbox{\halign{$#$\hfil &${}\ =\ #$\hfil\cr
 |d_A;(\delta_A\otimes\delta_A);
  \tilde\varphi_{\mathord!A,\mathord!A}|(x,x')
  & \theta_A(2^{\theta_A(x)})+\theta_A(2^{\theta_A(x')})\cr
	\noalign{\vskip.5ex}
 |\delta_A;\mathord!d_A|(x,x') & 2^{\theta_A(\theta_A(x)+\theta_A(x'))}.\cr}}

	\vskip2ex
        \noindent
So, putting $u=\theta_A(x)$ and $v=\theta_A(x')$, we must show that
 $\theta_A(2^u)+\theta_A(2^v)<2^{\theta_A(u+v)}$.
This inequality is verified by induction.
If $\theta_A$ is an identity function, the inequality amounts
 to $2^u+2^v<2^{u+v}$, which is correct as we have $u,v\geq 2$ since
 we assumed that $x_i\geq 2$.
If $\theta_A=2\theta_B$ the inequality amounts
 to $2\theta_B(2^u)+2\theta_B(2^v)<2^{2\theta_B(u+v)}$.
By the induction hypothesis (LHS)${}<2\cdot 2^{\theta_B(u+v)}$.
By $1<\theta_B(u+v)$ this is smaller than (RHS).
If $\theta_A=b+\theta_B$ the inequality amounts
 to $2b+\theta_B(2^u)+\theta_B(2^v)<2^{b+\theta_B(u+v)}$.
By the induction hypothesis (LHS)${}<2b+2^{\theta_B(u+v)}$.
This is less than or equal to $2^b+2^{\theta_B(u+v)}\leq{}$(RHS).
For the last inequality we use $1\leq b$.

$\tilde\varphi$-type rules.
These are manipulated uniformly.
For example, let us consider rule (9).
Suppose $\mathord!\mathord!(A[X^x]\otimes B)$.
Then

	\vskip2ex
        \noindent\kern5em
\vbox{\halign{$#$\hfil &${}\ =\ #$\hfil\cr
 |(\delta_A\otimes\delta_B);\tilde\varphi_{\mathord!A,\mathord!B};
  \mathord!\tilde\varphi_{A,B}|(x) & |\delta_A|(x)=2^{\theta_A(x)}\cr
	\noalign{\vskip.5ex}
 |\tilde\varphi_{A,B};\delta_{A\otimes B}|(x)
  & |\delta_{A\otimes B}|(x)=2^{b+\theta_A(x)}\cr
}}

	\vskip2ex
        \noindent
 where
 $b$ is the number of occurrences of $X$ in $B$.
Since $1\leq b$ the former is smaller than the latter.
The case when the specified $X$ occurs in $B$ is similar.
\endproof

	\vskip1ex

Next, we verify that restricted naturality reductions decrease
 the assigned natural numbers.
We give several lemmata towards it.

	\vskip1ex

\begin{lemma}\label{cyf24}
The inequality $1+\theta_A(x)\leq \theta_A(1+x)$ holds.
\end{lemma}

\proof
By induction on the construction of $A$.
\endproof

	\vskip1ex

\begin{lemma}\label{mtc69}
The inequality $2\theta_A(x)\leq \theta_A(2b^-+2x)$ holds where
 $b^-$ is one less than the number of occurrences of $X$ in $A$.
\end{lemma}

\proof
If we integrate serial applications of tensor, we have
$$\theta_A(x)=2^{k_0}b_0+2^{k_0+k_1}b_1+\cdots+
 2^{k_0+k_1+\cdots+k_{q-1}}b_{q-1}+2^{k_0+k_1+\cdots +k_q}x.$$
This is the case, for example, if
$A=\mathord!^{k_0}(A_0\otimes \mathord!^{k_1}(A_1\otimes\cdots
 \mathord!^{k_{q-1}}(A_{q-1} \otimes \mathord!^{k_q}X)\cdots))$
 and each $A_i$ contains $b_i$ occurrences of $X$.
We have $b_i\geq 1$.
The two numbers $k_0$ and $k_q$ in both ends are non-negative while
 the other $k_i$ are strictly positive.
If $q=0$, i.e., when $A$ is $\mathord!^{k_0}X$,
 we have $\theta_A(x)=2^{k_0}x$.
The equality holds in this case as $b^-=0$.
If $q>0$ we note $\mathop{\rm max}\{b_0,b_1,\ldots,b_{q-1}\}\leq b^-$.
We observe that $2^{k_0}+2^{k_0+k_1}+\cdots+ 2^{k_0+k_1+\cdots+k_{q-1}}
 <2^{k_0+k_1+\cdots+k_{q-1}+1}$ holds,
 which is clear if regarded as numbers in base $2$.
Therefore we have
 $2^{k_0}b_0+2^{k_0+k_1}b_1+\cdots+ 2^{k_0+k_1+\cdots+k_{q-1}}b_{q-1}
 <2^{k_0+k_1+\cdots+k_{q-1}+1}b^-\leq 2^{k_0+k_1+\cdots+k_{q-1}+k_q}(2b^-)$.
Thus $2\theta_A(x)<2^{k_0+k_1+\cdots+k_{q}}(2b^-)+\theta_A(2x)
 =\theta_A(2b^-+2x)$.
\endproof

	\vskip1ex

\begin{lemma}\label{zrv33}
Let $b^-$ be one less than the number of occurrences of $X$ in $A$.
Then $b^-<\theta_A(x)$ holds.
\end{lemma}

\proof
The definition of $\theta_A(x)$ sums up
 all occurrences of $X$ through recursive calls.
\endproof

	\vskip1ex

\begin{lemma}\label{ddp27}
Restricted naturality reduction in a strict composite algebraic morphism
 decreases the natural numbers involved in the redex.
\end{lemma}

\proof
Consider the redex $\mathord!f;u_A$.
It suffices to prove the case where $f$ consists of
 a single $\delta$ or of a single $\tilde\varphi$.
Since the latter is simpler, we prove it first.
Namely, suppose that $f=F(\tilde\varphi_{A\otimes A'}):F(\mathord!A
 \otimes \mathord!A')\rightarrow F(\mathord!(A\otimes A'))$ for a
 functor $F$.
For example, let us suppose $u=\delta$.
We have

	\vskip2ex
        \noindent\kern5em
\vbox{\halign{$#$\hfil &${}\ =\ #$\hfil\cr
 |\delta_{F(\mathord!A\otimes \mathord!A')};\mathord!\mathord!F(\tilde
  \varphi_{A,A'})|(x) & 2^{\theta_{F(\mathord!A\otimes \mathord!A')}(x)}\cr
	\noalign{\vskip.5ex}
 |\mathord!F(\tilde\varphi_{A,A'});\delta_{F(\mathord!(A\otimes A'))}|(x)
  & 2^{\theta_{F(\mathord!(A\otimes A'))}(x)}.\cr
}}

	\vskip2ex
        \noindent
If $X^x$ occurs in $A$ and if $b$ is the nubmer of occurrences
 of $X$ in $A'$,
 we have $\theta_{\mathord!A\otimes \mathord!A'}(x)=b+2\theta_A(x)<
 2(b+\theta_A(x))=\theta_{\mathord!(A\otimes A')}(x)$ as $1\leq b$.
Therefore the former is smaller than the latter.
We note that this relies only on comparison
 between $\theta_{\mathord!A\otimes \mathord!A'}$ and
 $\theta_{\mathord!(A\otimes A')}$.
Hence the same argument applies to all the naturality rules.

Next we deal with the case when $f=F(\delta_A):F(\mathord!A)
 \rightarrow F(\mathord!\mathord!A)$.

Rule (18).

	\vskip2ex
        \noindent\kern5em
\vbox{\halign{$#$\hfil &${}\ =\ #$\hfil\cr
 |\delta_{F(\mathord!A)};\mathord!\mathord!F(\delta_A)|
 (x) & 2^{\theta_F(2\theta_A(2^{\theta_A(x)}))}\cr
	\noalign{\vskip.5ex}
 |\mathord!F(\delta_A);\delta_{F(\mathord!\mathord!A)}|(x)
  & 2^{\theta_A(2^{\theta_F(4\theta_A(x))})}.\cr
}}

	\vskip2ex
        \noindent
So, putting $u=\theta_A(x)$, we must show $\theta_F(2\theta_A(2^u))
 <\theta_A(2^{\theta_F(4u)})$.
This is verified by induction on $F$.
We start with the case $\theta_F=2\theta_G$.
By the induction hypothesis and Lem.~\ref{mtc69}, (LHS)${}=2\theta_G
 (2\theta_A(2^u))<\theta_A(2b^-+2\cdot 2^{\theta_G(4u)})$ where
 $b^-$ is one less than the number of occurrences of $X$ in $A$.
On the other hand (RHS)${}=\theta_A(2^{2\theta_G(4u)})$.
So, if we put $t=\theta_G(4u)$, it suffices to show that
 $2b^-+2\cdot 2^t\leq 2^{2t}$.
As $u\leq t$ we can assume that $0\leq b^-<t$ by Lem.~\ref{zrv33}.
If $b^-=0$ and $t=1$ the inequality is directly checked.
Assume $t\geq 2$.
Then $2b^-+2\cdot 2^t<2t+2\cdot 2^t<2\cdot 2^t+2\cdot 2^t=
 2^{2+t}\leq 2^{2t}$ holds.
The next case is $\theta_F=c+\theta_G$.
By the induction hypothesis and Lem.~\ref{cyf24},
 (LHS)${}=c+\theta_G(2\theta_A(2^u))<\theta_A(c+2^{\theta_G(4u)})$.
Applying $1+2^t<2^{1+t}$ repeatedly, we conclude that
 it is smaller than $\theta_A(2^{c+\theta_G(4u)})={}$(RHS).
The base case is that $\theta_F$ is an identity function.
By Lem.~\ref{mtc69}, (LHS)${}=2\theta_A(2^u)\leq \theta_A(2b^-+2\cdot 2^u)$
 where $b^-$ is one less than the number of $X$ in $A$.
On the other hand (RHS)${}=\theta_A(2^{4u})$.
So it suffices to show that $2b^-+2\cdot 2^u<2^{4u}$, for which a sharper result
 has been verified in the first case.

Rule (19).

	\vskip2ex
        \noindent\kern5em
\vbox{\halign{$#$\hfil &${}\ =\ #$\hfil\cr
 |\varepsilon_{F(\mathord!A)};F(\delta_A)|(x)
  & \theta_F(2\theta_A(2^{\theta_A(x)}))\cr
	\noalign{\vskip.5ex}
 |\mathord!F(\delta_A);\varepsilon_{F(\mathord!\mathord!A)}|(x)
  & 2^{\theta_A(\theta_F(4\theta_A(x)))}.\cr
}}

	\vskip2ex
        \noindent
So we must prove that $\theta_F(2\theta_A(2^u))<2^{\theta_A(\theta_F(4u))}$ with
 $u=\theta_A(x)$.
It is verified by induction on $F$.
If $\theta_F$ is an identity, we show that $2\theta_A(2^u)<2^{\theta_A(4u)}$
 by induction on $A$.
If $\theta_A$ is an identity, then obviously $2\cdot 2^u<2^{4u}$.
If $\theta_A=b+\theta_B$, the inner induction hypothesis implies
 (LHS)${}<2b+2^{\theta_B(4u)}$.
So, for $t=\theta_B(4u)$, we show $2b+2^t\leq 2^{b+t}$.
We can assume $1\leq b<t$ by Lem.~\ref{zrv33}.
Then the inequality is justified as $2b+2^t<2^b+2^t<2^{b+t}$.
If $\theta_A=2\theta_B$ then by the inner induction hypothesis
 (LHS)${}<2\cdot 2^{\theta_B(4u)}\leq 2^{2\theta_B(4u)}={}$(RHS).
This finishes the base case.
If $\theta_F=c+\theta_G$ then by the induction hypothesis
 (LSH)${}<c+2^{\theta_A(\theta_G(4u))}$, which is, by $1+2^t<2^{1+t}$
 and Lem.~\ref{cyf24}, smaller than $2^{\theta_A(c+\theta_G(4u))}={}$(RHS).
If $\theta=2\theta_G$ then by the induction hypothesis
 (LSH)${}<2\cdot2^{\theta_A(\theta_G(4u))}$, which is, by Lem.~\ref{cyf24},
 smaller than $2^{\theta_A(1+\theta_G(4u))}\leq 2^{\theta_A(2\theta_G(4u))}={}$(RHS).

Rule (20).

	\vskip2ex
        \noindent\kern5em
\vbox{\halign{$#$\hfil &${}\ =\ #$\hfil\cr
 |d_{F(\mathord!A)};\mathord!F(\delta_A)\otimes
 \mathord!F(\delta_A)|(x,x') & \theta_F(2\theta_A(2^{\theta_A(x)}))
 +\theta_F(2\theta_A(2^{\theta_A(x')}))\cr
	\noalign{\vskip.5ex}
 |\mathord!F(\delta_A);
 d_{F(\mathord!\mathord!A)}|(x,x')
  & 2^{\theta_A(\theta_F(4\theta_A(x))+\theta_F(4\theta_A(x')))}.\cr
}}

	\vskip2ex
        \noindent
We must show that
 $\theta_F(2\theta_A(2^{u}))+\theta_F(2\theta_A(2^{v}))<
 2^{\theta_A(\theta_F(4u)+\theta_F(4v))}$,
 for $u=\theta_A(x)$ and $v=\theta_A(x')$. 
It is verified as in the previous case.
In this time, meanwhile, $4b+2^t\leq 2^{b+t}$
 appears as the inequality that must be shown.
This is valid for $1\leq b<t$.

Rule (21).

	\vskip2ex
        \noindent\kern5em
\vbox{\halign{$#$\hfil &${}\ =\ #$\hfil\cr
 |e_{F(\mathord!A)}|() & \theta_F(2\theta_A(2))\cr
	\noalign{\vskip.5ex}
 |\mathord!F(\delta_A);e_{F(\mathord!\mathord!A)}|()
  & 2^{\theta_A(\theta_F(4\theta_A(2)))}.\cr
}}

	\vskip2ex
        \noindent
Trivially the former is smaller than the latter.
\endproof

	\vskip1ex

\begin{lemma}\label{jex61}
Consider a strict composite algebraic morphism
 $s:A[X^{y_1},X^{y_2},\ldots, X^{y_n}]
	\penalty-3000
\rightarrow B[X^{x_1},X^{x_2}, \ldots,X^{x_m}]$.
Suppose that $s$ contracts to $t:A[X^{y'_1},X^{y'_2},\ldots,X^{y'_n}]\rightarrow
 B[X^{x_1},
	\penalty-3000
 X^{x_2},\ldots,X^{x_m}]$ by a reduction sequence where
 naturality reductions are restricted.
Then $y_i\leq y'_i$ for all $i$ and strictly $y_i<y'_i$ for one or more $i$,
 provided $x_j\geq 2$ for all $j$.
\end{lemma}

\proof
Since the composition of morphisms is realized by the composition of functions
 and all involving functions are strictly increasing, the local arguments
 proved in Lem.~\ref{jod26} and \ref{ddp27} imply the lemma.
\endproof

	\vskip1ex

We are interested in morphisms of the shape $f=s;t;h$ where $s$ is a strict
 composite algebraic morphism, $t$ is a composite algebraic morphism,
 and $h$ is normal.
Two punctuations are called {\it frontiers}.
For reference, $s$ is called {\it demesne} and $t$ {\it fief}.
The frontiers are not absolute since
 definitions
 (i) and (ii) in Def.~\ref{osp47} are not exclusive.

We consider the following condition:
 $f=s;t;h$ contains no naturality redexes except restricted ones
 and $t;h$ contains no redexes other than $\varphi_0$-type.
We also assume that $s$ is a strict
 composite algebraic morphism, $t$ is a composite algebraic morphism,
 and $h$ is normal.
We denote this condition by $\circledast$.

	\vskip1ex

\begin{lemma}\label{afd41}
Suppose that $f=s;t;h$ fulfills the condition $\circledast$
 above and it contracts to $f'$.
Then there is a decomposition $f'=s';t';h'$ satisfying $\circledast$.
Moreover, for an arbitrary decomposition
 $f'=s';t';h'$ subject to the condition $\circledast$,
 there is a decomposition $f=\tilde s;\tilde t;\tilde h$
 satisfying $\circledast$
 such that $s'$ is a contractum of $\tilde s$ or a part of it.
\end{lemma}

\proof
The first assertion means that contraction creates no
 redexes in the fief except of $\varphi_0$-type.
The second means that contraction creates
 no fresh part that can be added to the demesne beyond the area
 attached thereto at the outset.
The following are crucial cases.
Suppose that ${\bf 1}\mor{\varphi_0}\mathord!{\bf 1}\mor\delta
\mathord!\mathord!{\bf 1}$ contacts to ${\bf 1}\mor{\varphi_0}
 \mathord!{\bf 1}\mor{\mathord!\varphi_0}\mathord!\mathord!{\bf 1}$ in the fief.
When they are followed by $\mathord!\mathord!{\bf 1}
 \mor \delta\mathord!\mathord!\mathord!{\bf 1}$ for instance,
 contraction creates a naturality redex
 $\mathord!\varphi_0;\delta$ in the fief that violates the condition $\circledast$.
This situation is precluded, since
 the fief then contained
 a prohibited $\delta$-type redex $\delta;\delta$ beforehand.
Next suppose that ${\bf 1}\otimes\mathord!A\mor{\varphi_0\cdot}\mathord!{\bf 1}
 \otimes \mathord!A\mor{\tilde\varphi}\mathord!({\bf 1}\otimes A)$
 contracts to ${\bf 1}\otimes \mathord!A\mor\sim\mathord!A
 \mor\sim \mathord!({\bf 1}\otimes A)$ in the fief.
Two sides encircling the part may
 form a new redex,
 or the right side of the part may be newly attached to the demesne $s$,
 as the intervening $\varphi_0$ and $\tilde\varphi$ vanish.
For example, if the right side is $\mathord!({\bf 1}\otimes A)\mor\delta
 \mathord!\mathord!({\bf 1}\otimes A)$ such a problem may happen.
However, the fief then had a redex $\tilde\varphi;\delta$, which was prohibited
 by the condition $\circledast$.
Next, consider the naturality of $\tilde\varphi$.
As typical in the
 equivalence of $\mathord!\mathord!C\otimes \mathord!A\mor{\tilde\varphi}
 \mathord!(\mathord!C\otimes A)\mor{\mathord!(e\cdot)}
 \mathord!({\bf 1}\otimes A)$ and
 $\mathord!\mathord!C\otimes \mathord!A\mor{\mathord!e\cdot}
 \mathord!{\bf 1}\otimes \mathord!A\mor{\tilde\varphi}
 \mathord!({\bf 1}\otimes A)$, the former $\tilde\varphi$ can be a part of the
 demesne while the latter must be in the fief.
If they are followed by, for example, $\delta$, we have a $\tilde\varphi$-type
 redex in the fief, violating the hypothesis.
However, this is precluded since non-restricted
 naturality redex $\mathord!(e\cdot);\delta$ is
 forbidden by the condition $\circledast$.
We remark that no redex contains $e$ in
 its left half except such forbidden naturality redexes.
Hence all redex crossing the frontier
 can be engulfed in the demesne by extending $s$.
For example, if $\delta;\varepsilon$ crosses
 the frontier, $\delta$ is in the demesne while
 $\varepsilon$ is in the fief, we may enlarge
 the demesne so that $\varepsilon$ is a part of it.
So if we take sufficiently large $\tilde s$ then $s'$ is its
 contractum or a part of it.
\endproof

	\vskip1ex

\begin{lemma}\label{tbq29}
If $f=s;t;h$ satisfies the condition $\circledast$ introduced
 before Lem.~\ref{afd41}, $f$ satisfies the (strong) termination property.
\end{lemma}

\proof
By Lem.~\ref{afd41}, we can enlarge the demesne $s$ at the outset so that subsequent
 reductions in the demesne are all done in the descendants of $s$.
We assign natural numbers to each $X$ in the demesne as in
 Def.~\ref{ibg60}.
By Lem.~\ref{jex61}, the sum of associated natural numbers $y_i$ strictly decreases
 by contractions in the demesne.
Occasionally $\varphi_0$-type reductions occur in the fief, but
 they do not alter the associated natural numbers.
So the sum is eventually constant.
Thereafter only $\varphi_0$-type reductions can occur.
The sequence of $\varphi_0$-type reductions must be finite
 since they reduce the number of algebraic morphisms other than $\varphi_0$.
For example $\varphi_0;\delta$ contracts to $\varphi_0;\mathord!\varphi_0$, where
 $\delta$ disappears.
\endproof

	\vskip1ex

\begin{proposition}\label{ick83}
Algebraic morphisms
$\delta,\varepsilon,d,e$, and $\tilde\varphi,\varphi_0$ are reducible.
\end{proposition}

\proof
We describe the case of $\delta_A:\mathord!A\rightarrow
 \mathord!\mathord!A$.
By Lem.~\ref{ozp36} it suffices to prove that, for any $f:X\rightarrow
 A$ in $R^+(A)$
 and any $g:\mathord!\mathord!A\rightarrow Y$ in
 $R^-(\mathord!\mathord!A)$, $\mathord!f;\delta_A;g$ terminates.
By a naturality reduction, it contracts to $\delta_X;\mathord!\mathord!f
 ;g$.
By Lem.~\ref{aun57}, $\mathord!\mathord!f$ lies in $R^+(
\mathord!\mathord!A)$.
Hence $\mathord!\mathord!f;g$ contracts to a normal form $h$.
Now $\delta_X;h$ satisfies the hypothesis of Prop.~\ref{tbq29}
 if we set the demesne to be $\delta_X$ and the fief to be empty
 (an identity).
So it terminates.
The same argument applies to $\tilde\varphi_{X,Y}$, which
 is a strict composite algebraic morphism in $X,Y$.
We use Lem.~\ref{jhc51}.
For $\varphi_0$, we take the demesne to be
 empty and the fief to be $\varphi_0$.
\endproof

	\vskip1ex

All atomic morphisms are reducible by Prop.\ref{hpc73}, \ref{lqm59},
 \ref{dci74}, and \ref{ick83}.
Moreover, Prop.~\ref{xsi96} shows that all constructions preserve
 the property of being reducible.
With Prop.~\ref{cbw90}, we can conclude the weak termination property:

	\vskip3ex

\begin{theorem}\label{kbc55}
Every morphism terminates.
\end{theorem}

	\vskip0ex

\begin{remark}\label{cdo79}
Each morphism terminates under the following specific strategy.
First, any $\beta$-redexes are contracted.
If no $\beta$-redexes remain, naturality redexes are contracted.
Finally, if there are neither $\beta$-redexes nor naturality redexes, the
 rightmost redexes are contracted.
The verification of the theorem remains applicable if we interpret
 ``terminate'' as termination under this strategy.
Prop.~\ref{xsi96} (iv), Prop.~\ref{aun57}, and
 Lem.~\ref{ick83} depend on this particular strategy.
The rightmost redexes are not unique in general,
 as both $f$ and $g$ may have redexes in $f\otimes g$ for example.
 
We conjecture that strong normalizability is fulfilled.
We define strong termination to hold if
 all infinite reduction sequence repeats
 only reversible reductions after some finite number
 of reduction steps.
\end{remark}

\begin{remark}\label{lho86}
We obtain a cartesian closed category if we enforce $\mathord!$ to
 be an identity functor.
The tensor turns out to be the cartesian product
 $\times$ and the unit object is a terminal object.
Accordingly, we obtain a reduction system for a free cartesian closed category.
It contains reduction
 $e_{A\times B}\Rightarrow e_A\times e_B$ and $e_{{\bf 1}}\Rightarrow 1$
 among others.
Note that reductions depend on subscripts, i.e., the shape of
 objects.
This system has a looping reduction sequence
 $e_{A\times {\bf 1}}\Rightarrow e_A\times e_{{\bf 1}}
 \Rightarrow e_A\times 1\cong e_{A\times{\bf 1}}$.
In our system, rule (17) blocks this to happen.
Moreover, all reduction rules make sense if we omit subscripts.
\end{remark}

\section{Conclusion}

We define a rewriting system
 on the categorical semantics of
 the linear logic.
Namely, the free (intuitionistic or classical)
 linear category can be regarded
 as a calculus.
In this paper, we verify that
 the calculus on the free classical linear category
 satisfies the weak termination property.
In a forthcoming paper, we will verify that
 it is almost confluent
(we say ``almost'' since we cannot properly
 deal with the tensor/cotensor units, which are
 difficult to handle.)
These two results together imply that each morphism has a unique
 normal form as far as no units are involved.

A reward brought about by introducing a calculus
 is the mechanization
 of diagram chasing.
Given two morphisms, we first convert them into normal forms.
We can replace the judgment of equality between morphisms
 by comparison between normal forms.
If the tensor/cotensor units are not involved,
 we can automatically check whether they are equal.
In this sense, our result will give a kind of (partial) coherence
 result.
This paper provides the first step towards this purpose.

\section*{Acknowledgements}
The author is supported by JSPS Kakenhi Grant Number JP15500003.
We thank an anonymous referee for a number of helpful suggestions to improve
 the presentation.
Of course, it is completely our responsibility if
 there remain poor presentations yet.

\begin{references*}

\bibitem{aspe}
A.~Asperti, {\rm Linear logic, comonads and optimal reductions},
 {\it Fundamenta Informaticae}, 22(1-2):3--22, 1993.

\bibitem{asgu}
A.~Asperti and G.~Guerrini, {\it The Optimal Implementation of Functional
 Programming Languages}, Cambridge University Press, 1998.

\bibitem{barb}
A.~Barber, {\rm Dual intuitionistic linear logic}, preprint,
 Laboratory for Foundations of Computer Science, The University of
 Edinburgh, U.K., 1996.

\bibitem{barr}
M.~Barr, {\it $*$-Autonomous Categories}, Lecture Notes in Mathematics,
 vol.~752, Springer Verlag, 1979.

\bibitem{bent}
N.~Benton, {\rm A mixed linear and non-linear logic: proofs,
 terms, models}, in {\it Computer Science Logic, CSL 1994},
 L.~Pacholski, J.~Tiuryn, eds., Lecture Notes in Computer Science, vol.~933,
 pp.121--135, Springer Verlag, 1994.

\bibitem{bbph}
N.~Benton, G.~Bierman, V.~de~Paiva, and M.~Hyland, {\rm Linear
 $\lambda$-calculus and categorical models revisited}, in
 {\it Computer Science Logic. CSL 1992}, E.~B\"orger, G.~J\"ager,
 H.~Kleine B\"uning, S.~Martini, M.~M.~Richter, eds.,
 Lecture Notes in Computer Science, vol.~702, pp.61--84,
 Springer Verlag, 1993.

\bibitem{bcst}
R.~F.~Blute, J.~R.~B.~Cockett, R.~A.~G.~Seely, and T.~H.~Trimble,
{\rm Natural deduction and coherence for weakly distributive
 categories}, {\it Journal of Pure and Applied Algebra}, 113(3):229--296,
 1996.

\bibitem{cose}
J.~R.~B.~Cockett and R.~A.~G.~Seely, {\rm Weakly distributive categories},
 {\it Journal of Pure and Applied Algebra}, 114(2):133--173, 1997.

\bibitem{ccm}
G.~Cousineau, P.-L.~Curien, and M.~Mauny, {\rm The categorical abstract
 machine}, {\it Science of Computer Programming}, 8(2):173--202, 1987.

\bibitem{curi}
P.-L.~Curien, {\rm Categorical combinators}, {\it Information and Control},
 69(1-3):188--254, 1986.

\bibitem{epst}
D.~B.~A.~Epstein, {\rm Functors between tensored categories}, {\it
 Inventiones Mathematcae}, 1(3):221--228, 1966.

\bibitem{ghan}
N.~Ghani, {\rm Adjoint rewriting}, preprint,
 Laboratory for Foundations of Computer Science, The University of
 Edinburgh, U.K., 1995.

\bibitem{gira}
J.-Y.~Girard, {\rm Linear logic}, {\it Theoretical Computer Science},
 50(1):1--101, 1987.

\bibitem{gal}
G.~Gonthier, M.~Abadi, and J.-J.~L\'evy, {\rm The geometry
 of optimal lambda reduction}, in {\it Proceedings of the 19th ACM
 SIGPLAN-SIGACT Symposium on Principles of programming languages, POPL '92},
 pp.15--26, ACM, 1992.

\bibitem{heho}
W.~Heijltjes and R.~Houston, {\rm No proof nets for MLL with units:
 Proof equivalence in MLL is PSPACE complete}, in {\it 
Proceedings of the Joint Meeting of the Twenty-Third EACSL
 Annual Conference on Computer Science Logic 
 and the Twenty-Ninth Annual ACM/IEEE Symposium on Logic in Computer Science,
 CSL-LICS '14}, ACM, 2014.

\bibitem{hugh}
D.~J.~D.~Hughes, {\rm Simple free star-autonomous categories
 and full coherence}, {\it Journal of Pure and Applied Algebra},
 216(11):2386--2410, 2012.

\bibitem{jaco}
B.~Jacobs, {\it Categorical Logic and Type theory}, Elsevier, 2001.

\bibitem{jay}
C.~B.~Jay, {\rm Modelling reduction in confluent categories}, preprint,
 Laboratory for Foundations of Computer Science, The University of
 Edinburgh, U.K., 1991.

\bibitem{kell}
G.~M.~Kelly, {\rm On MacLane's conditions for coherence of natural
 associativities, commutativities, etc.}, {\it Journal of
 Algebra}, 1(4):397--402, 1964.

\bibitem{kemc}
G.~M.~Kelly and S.~MacLane, {\rm Coherence in closed categories},
 {\it Journal of Pure and Applied Algebra}, 1(1):97--140, 1971.

\bibitem{lasc}
J.~Lambek and P.~J.~Scott, {\it Introduction to Higher-Order Categorical
 Logic}, Cambridge Universit Press, 1988.

\bibitem{macl}
S.~Mac~Lane, {\it Categories for the Working Mathematicien},
 Springer, 1978.

\bibitem{mmpr}
M.~E.~Maietti, P.~Maneggia, V.~de~Paiva, and E.~Ritter,
 {\rm Relating categorical semantics for intuitionsitic linear logic},
 {\it Applied Categorical Structures}, 13(1):1--36, 2005.

\bibitem{mell}
P.-A.~Melli\`es, {\rm Categorical models of linear logic revisited}, 
 preprint, HAL-Inria, France, 2002.

\bibitem{mel2}
P.-A.~Melli\`es, {\rm Categorical semantics of linear logic},
 manuscript, CNRS and Universit\'e Paris 7, France, 2007.

\bibitem{ryst}
D.~E.~Rydeheard and J.~G.~Stell, {\rm Foundations of equational
 deduction: A categorical treatment of equational proofs and
 unification algorithm}, in {\it Category Theory and Computer Science},
 D.~H.~Pitt, A.~Poign\'e, eds., Lecture Notes in Computer Science, vol.~283,
 pp.114--139, 1987.

\bibitem{scha}
A.~Schalk, {\rm What is a categorical model for linear
 logic?}, manuscript, Univ. of Manchester, U.K., 2004.

\bibitem{seel}
R.~A.~G.~Seely, {\rm Modelling computations: a 2-categorical framework},
 {\it Proceedings of the Symposium on Logic in Computer Science, LICS '87},
 pp.65--71, IEEE, 1987.

\bibitem{tran}
P.~Tranquilli, {\rm Confluence of pure differential nets with promotion},
 in {\it Computer Science Logic, CSL 2009}, E.~Gr\"adel, R.~Kahle, eds.,
 Lecture Notes in Computer Science, vol.~5771, pp.504--514, 2009.

\end{references*}

\end{document}